\documentclass[11pt]{article}

\usepackage[latin1]{inputenc}    
\usepackage[OT1]{fontenc}
\usepackage[english]{babel}
\usepackage{natbib}
\usepackage{fullpage}
\usepackage{amsmath}
\usepackage{amssymb}
\usepackage{amsthm}
\usepackage{mathrsfs}
\usepackage{graphicx}
\usepackage{dsfont}
\usepackage{wrapfig}
\usepackage{comment}
\usepackage{rotating}
\usepackage{hyperref} 
\usepackage{color} 
\usepackage[all]{xy}
\newcommand{\bert}[1]{\noindent \textcolor{blue}{\textsf{[B: #1]}}}

\newcommand{\R}{\mathbb{R}}
\newcommand{\p}{\mathbb{P}}
\newcommand{\E}{\mathbb{E}}
\newcommand{\Rop}{\mathbb{R}^{\text{op}}}
\newcommand{\Rext}{\mathbb{R}_{\text{Ext}}}
\newcommand{\N}{\mathbb{N}}

\newcommand{\X}{\mathbb{X}}
\newcommand{\Y}{\mathbb{Y}}

\newcommand{\I}{\mathcal{I}}

\newcommand{\Reeb}{\mathrm{R}}
\newcommand{\Ord}{\mathrm{Ord}}
\newcommand{\Ext}{\mathrm{Ext}}
\newcommand{\Rel}{\mathrm{Rel}}
\newcommand{\Dg}{\mathrm{Dg}}
\newcommand{\Mapper}{\mathrm{M}}
\newcommand{\MMapper}{\overline{\mathrm{M}}}

\newcommand{\Shift}{\mathrm{Shift}}
\newcommand{\Merge}{\mathrm{Merge}}
\newcommand{\Split}{\mathrm{Split}}
\newcommand{\Crit}{\mathrm{Crit}}
\newcommand{\Rips}{\mathrm{Rips}}

\newcommand{\Map}{\Mapper}
\newcommand{\disth}{d_{\rm H}}

\newcommand{\dimension}{b}
\newcommand{\const}{13}
\newcommand{\proba}[1]{\mathbb{P}\left(#1\right)}
\newcommand{\TV}{\operatorname{TV}}
\newcommand{\freeb}[1]{#1_{\rm R}}
\newcommand{\fmapp}[1]{#1_{\I}}
\newcommand{\frips}[1]{#1^{\rm PL}}
\newcommand{\End}{\mathrm{End}}
\newcommand{\Xset}{\mathcal{X}}
\newcommand{\Yset}{\mathcal{Y}}
\newcommand{\Xs}{\mathbb{X}}
\newcommand{\Ys}{\mathbb{Y}}
\newcommand{\distb}{d_{\Delta}}
\newcommand{\dper}{d_{\Delta}}

\newtheorem{thm}{Theorem}[section]
\newtheorem{cor}[thm]{Corollary}
\newtheorem{prop}[thm]{Proposition}
\newtheorem{lem}[thm]{Lemma}
\newtheorem{defin}[thm]{Definition}

\newtheorem{rmq}[thm]{Remark}

\title{Statistical analysis and parameter selection for Mapper}
\author{Mathieu Carri\`ere\footnote{mathieu.carriere@inria.fr}, Bertrand Michel\footnote{bertrand.michel@ec-nantes.fr} and Steve Oudot\footnote{steve.oudot@inria.fr}}

\begin{document}

\maketitle

\begin{abstract}

In this article, we study the question of the statistical convergence of the 1-dimensional Mapper
to its continuous analogue, the Reeb graph. We show that the Mapper
is an optimal estimator of the Reeb graph, which gives, as a byproduct, a method to automatically tune
its parameters and compute confidence regions on its topological features, 
such as its loops and flares. This allows to circumvent the issue of testing a large grid of parameters
and keeping the most stable ones in the brute-force setting, which is widely used in visualization, clustering
and feature selection with the Mapper. \\

\end{abstract}

\section{Introduction}



In statistical learning, a large class of problems can be categorized into supervised or unsupervised problems. 
For supervised learning problems, an output quantity $Y$ must be predicted or explained from the  
input measures $X$. On the contrary, for unsupervised problems there is no output quantity $Y$ to 
predict and the aim is to explain and model the underlying structure or distribution in the data. 
In a sense, unsupervised learning can be thought of as extracting features from the data, assuming 
that the latter come with unstructured noise. Many methods in data sciences can be qualified as 
unsupervised methods, among the most popular examples are association methods, clustering methods, 
linear and non linear dimension reduction methods and matrix factorization to cite a few 
(see for instance Chapter~14 in~\cite{Friedman01}). Topological Data Analysis (TDA) 
has emerged in the recent years as a new field whose aim is to uncover, understand and exploit the 
topological and geometric structure underlying complex.
and possibly high-dimensional data. 
Most of TDA methods can thus be 
qualified  as unsupervised. In this paper, we study a recent TDA algorithm called Mapper 
which was first introduced in~\cite{Singh07}. 

Starting from a point cloud $\Xs_n$ sampled from a metric space $\Xset$, 
the idea of Mapper is to study the topology of the sublevel sets of a function $f:\Xs_n\rightarrow\R$ 
defined on the point cloud. The function $f$ is called a filter function and it has to be chosen by the user. 
The construction of Mapper depends on the choice of a cover $\I$ of the image of $f$ by open sets. 
Pulling back $\I$ through $f$ gives an open cover of the domain $\Xs_n$. 
It is then refined into a connected cover by splitting each
element into its various clusters using a clustering algorithm whose choice is left to the user. 
Then, the Mapper is defined as the nerve of the connected
cover, having one vertex per element, one edge per pair of intersecting elements, and more generally, one
$k$-simplex per non-empty $(k+1)$-fold intersection. 

In practice, the Mapper has two major applications. The first one is data visualization and clustering.
Indeed, when the cover $\I$ is minimal, the Mapper provides a visualization of the data in the form of a graph whose topology reflects
that of the data. As such, it brings additional information to the usual clustering algorithms 
by identifying {\em flares} and {\em loops} that outline potentially remarkable subpopulations
in the various clusters. See e.g.~\cite{Yao09, Lum13, Sarikonda14, Hinks15} for examples of applications. 
The second application of Mapper deals with feature selection. Indeed,
each feature of the data can be evaluated on its ability to discriminate  
the interesting subpopulations mentioned above (flares, loops) from the rest of the data, 
using for instance Kolmogorov-Smirnov tests.
See e.g.~\cite{Lum13, Nielson15, Rucco15} for examples of applications.

\begin{figure}
\begin{tabular}{cc}
\includegraphics[width=6cm]{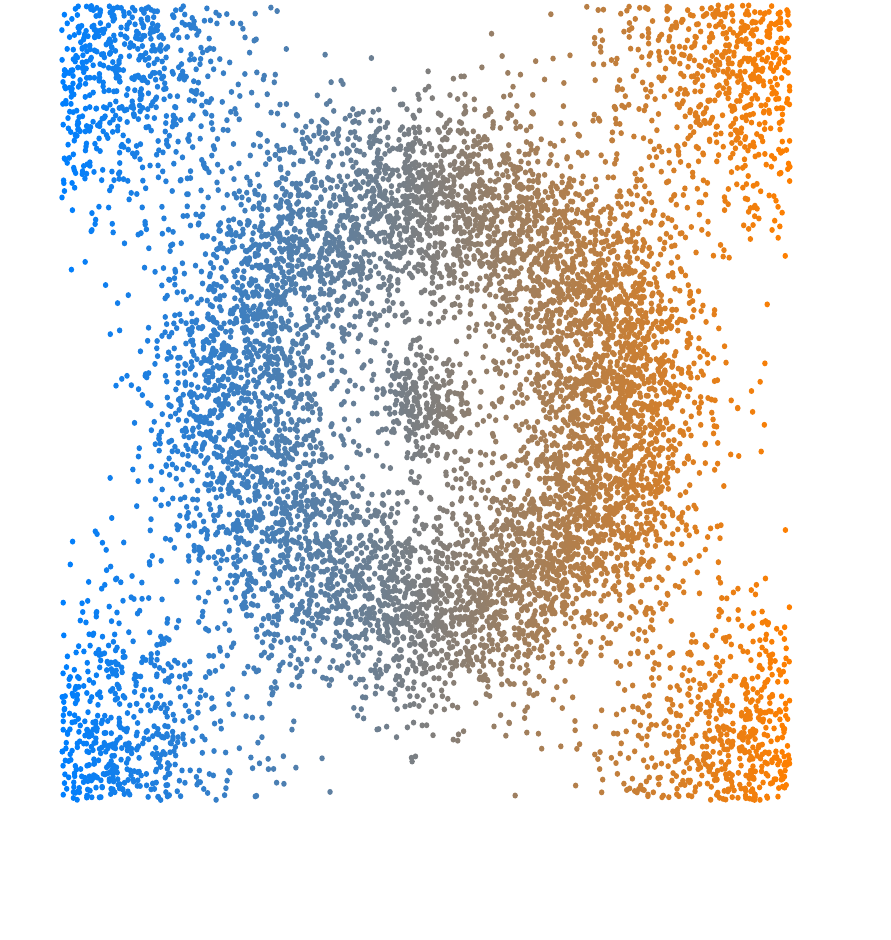} & \includegraphics[width=10cm]{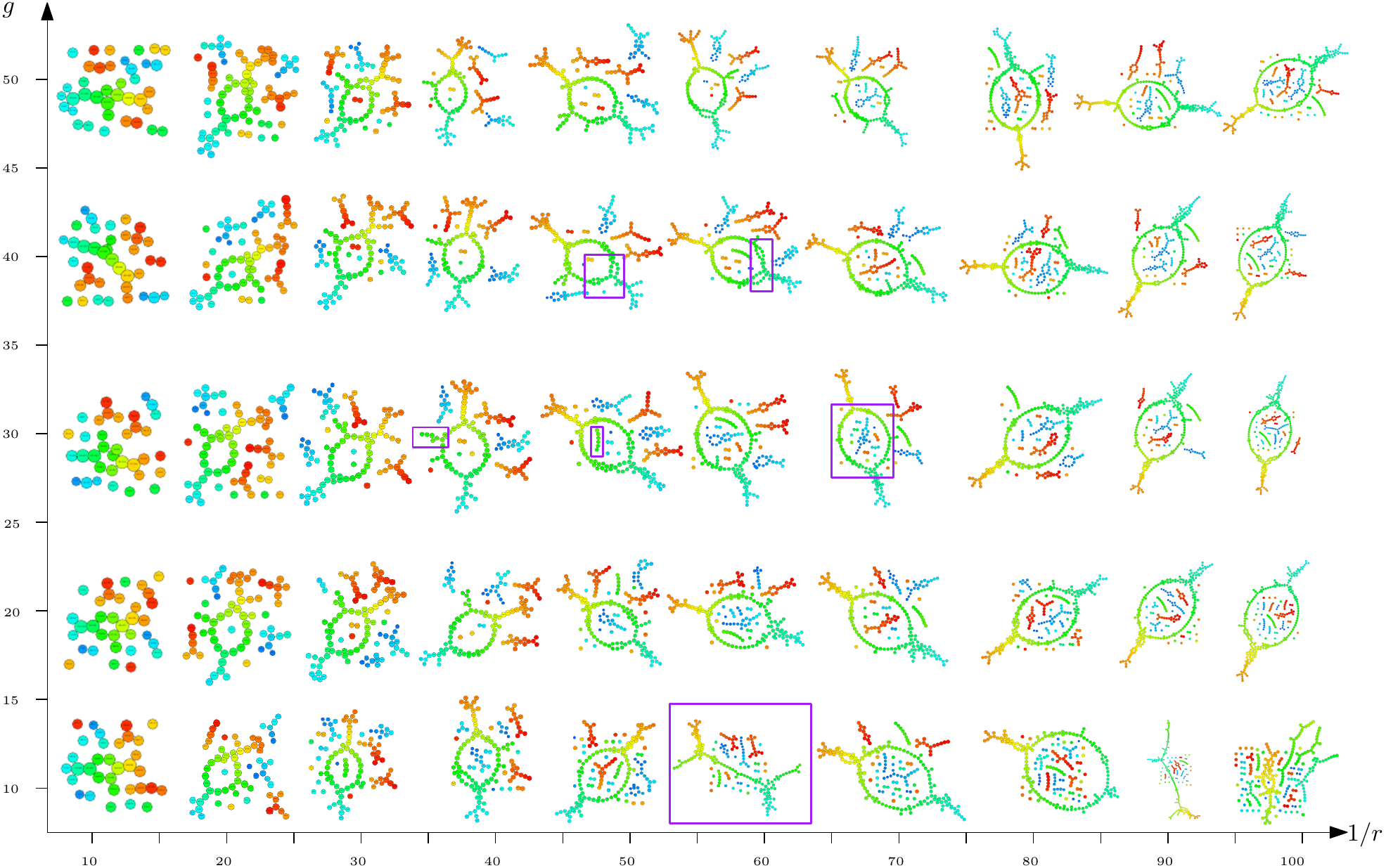}
\end{tabular}
\caption{\label{fig:instab}
A collection of Mappers computed with various parameters. Left: crater dataset. Right: outputs of Mapper with various parameters. One can see that for some Mappers,
(the ones with purple squares), topological features suddenly appear and disappear. These
are discretization artifacts, that we overcome in this article by appropriately tuning the parameters.
} 
\end{figure}

Unsupervised methods generally depend on parameters that 
need to be chosen by the user. For instance, the number of selected dimensions for dimension reduction 
methods or the number of clusters for clustering methods have to be chosen. 
Contrarily to supervised problems, it is tricky to evaluate the output of unsupervised methods and thus to select parameters. This situation is highly problematic with Mapper since, as for many TDA methods, it is not robust to outliers. This major drawback of Mapper is an important obstacle to its use  
in Exploratory Data Analysis with non trivial datasets. This phenomenon is illustrated for instance in 
Figure~\ref{fig:instab} on a dataset that we study further in Section~\ref{sec:appli}. The only answer proposed to 
this drawback in the literature consists in selecting parameters in a range of values for which 
the Mapper seems to be stable---see for instance~\cite{Nielson15}. 
We believe that such an approach is not satisfactory because it does not provide statistical guarantees on the inferred Mapper.

Our main goal in this article is to provide a statistical method to tune the parameters of Mapper automatically. 
To select parameters for Mapper, or more generally to evaluate the significance of topological features provided by Mapper, 
we develop a rigorous statistical framework for the convergence of the Mapper. This contribution  
is made possible by the recent work (in a deterministic setting) of~\cite{Carriere17b} about the structure and the stability of the Mapper. 
In this article, the authors explicit a way to go from the input space to the Mapper using small perturbations. We build on this 
relation between the input space and its Mapper to show that the Mapper is itself a measurable construction.
In~\cite{Carriere17b}, the authors also show that the topological structure of the Mapper can actually be predicted from 
the cover $\I$ by looking at appropriate {\em signatures}
that take the form of {\em extended persistence diagrams}. In this article, we use this observation, together with an approximation inequality, 
to show that the Mapper, computed with a specific set of parameters, is actually
an optimal estimator of its continuous analogue, the so-called {\em Reeb graph}. Moreover, these specific parameters act as natural candidates 
to obtain a reliable Mapper with no artifacts,
avoiding the computational cost of testing millions of candidates and selecting the most stable ones in the brute-force setting of many practitioners.
Finally, we also provide methods to assess the stability and compute confidence regions for the topological features of the Mapper.
We believe that this set of methods open the way to an accessible and intuitive 
utilization of Mapper for non expert researchers in applied topology.

Section~\ref{sec:ApproxReeb} presents the necessary background 
on the Reeb graph and Mapper, and it also gives an approximation 
inequality---Theorem~\ref{thm:geomineq}---for the Reeb graph with the Mapper. 
From this approximation result, we derive rates of 
convergences as well as candidate parameters in Section~\ref{sec:ConfConv}, 
and we show how to get confidence regions in Section~\ref{sec:confR}. 
Section~\ref{sec:appli} illustrates the validity of our parameter tuning and
confidence regions with numerical experiments on smooth and noisy data.


\section{Approximation of a Reeb graph with Mapper} \label{sec:ApproxReeb} 

\subsection{Background on the Reeb graph and Mapper} \label{subsec:Backg}

We start with some background on the Reeb graph and Mapper. 
In particular, we present the specific Mapper algorithm that we study in this article.

\paragraph{Reeb graph.} Let $\Xset$ be a topological space and let $f: \Xset \rightarrow \R$
be a continuous function. Such a function on $\Xset$  is called a 
{\it filter function} in the following. 
Then, we define the equivalence relation $\sim_f$ as follows: 
for all $x$ and $x'$ in $\Xset$,  $x$ and $x'$ are in the same class ($x \sim_f x'$) if and only if  $x$ and $x'$ belong to the same connected component of  
$f^{-1}(y)$, for some $y$ in the image of $f$. 

\begin{defin} The {\em Reeb graph} $\Reeb_f(\Xset)$ of $\Xset$ computed with the filter function $f$ is 
the quotient space $\Xset/\sim_f$ endowed with the quotient topology.
\end{defin}

\begin{figure}[h]\centering
\includegraphics[width=6cm]{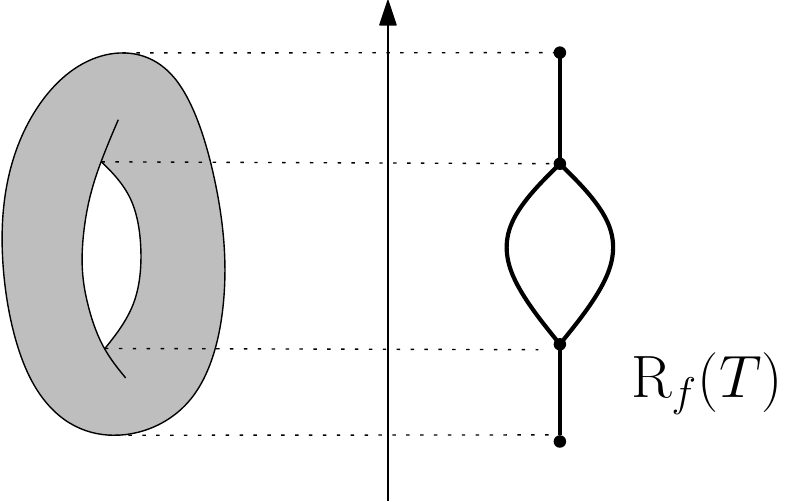}
\caption{\label{fig:ReebGraphExample} Example of Reeb graph computed on the torus $T$ with the height function $f$.}
\end{figure}

See Figure~\ref{fig:ReebGraphExample} for an illustration.
Note that, since $f$ is constant on equivalence classes, there is an induced map $\freeb{f}:\Reeb_f(\Xset)\rightarrow\R$
such that $f = \freeb{f} \circ \pi$, where $\pi$ is the quotient map $\Xset\rightarrow\Reeb_f(\Xset)$. 

The topological structure of a Reeb graph can be described if the pair ($\Xset,f)$ is regular enough. 
From now on, we will assume that 
the filter function $f:\Xset\rightarrow\R$ is {\it Morse-type}. 
Morse-type functions are generalizations of classical Morse functions that share some of their properties 
without having to be differentiable (nor even defined over a smooth manifold):

\begin{defin}\label{def:Morse-type}
A continuous real-valued function $f$ on a compact space $\Xset$ is \emph{of Morse type} if:
\begin{itemize}
\item[{\rm (i)}] There is a finite set $\emph{Crit}(f)=\{a_1<...<a_n\}$, called the set of \emph{critical values},
such that over every open interval $(a_0=-\infty,a_1),...,(a_i,a_{i+1}),...,(a_n,a_{n+1}=+\infty)$
there is a compact and locally connected space $\Yset_i$ and a homeomorphism $\mu_i:\Yset_i\times(a_i,a_{i+1})\rightarrow \Xset ^{(a_i,a_{i+1})}$ s.t. $\forall i=0,...,n, f|_{\Xset^{(a_i,a_{i+1})}}=\pi_2\circ\mu_i^{-1}$, where
$\pi_2$ is the projection onto the second factor;

\item[{\rm (ii)}]$\forall i=1,...,n-1,\mu_i$ extends to a continuous function $\bar{\mu}_i:\Yset_i\times[a_i,a_{i+1}]\rightarrow \Xset^{[a_i,a_{i+1}]}$
-- similarly $\mu_0$ extends to $\bar{\mu}_0:\Yset_0\times(-\infty,a_1]\rightarrow \Xset^{(-\infty,a_1]}$
and $\mu_n$ extends to $\bar{\mu}_n:\Yset_n\times[a_n,+\infty)\rightarrow \Xset^{[a_n,+\infty)}$;

\item[{\rm (iii)}]Each levelset $\Xset^t$ has a finitely-generated homology.
\end{itemize}
\end{defin}


\medskip

\noindent {\bf Key fact 1a:} For 
$f:\Xset\rightarrow\R$ a Morse-type function, 
the Reeb graph $\Reeb_f(\Xset)$ is a multigraph. 

\medskip

For our purposes, in the following we further assume that $\Xset$ is a smooth and compact submanifold of $\R^D$.
The space of Reeb graphs computed with Morse-type functions over such spaces is denoted $\mathcal R$ in this article. 

\begin{figure}\centering
\includegraphics[width=12cm]{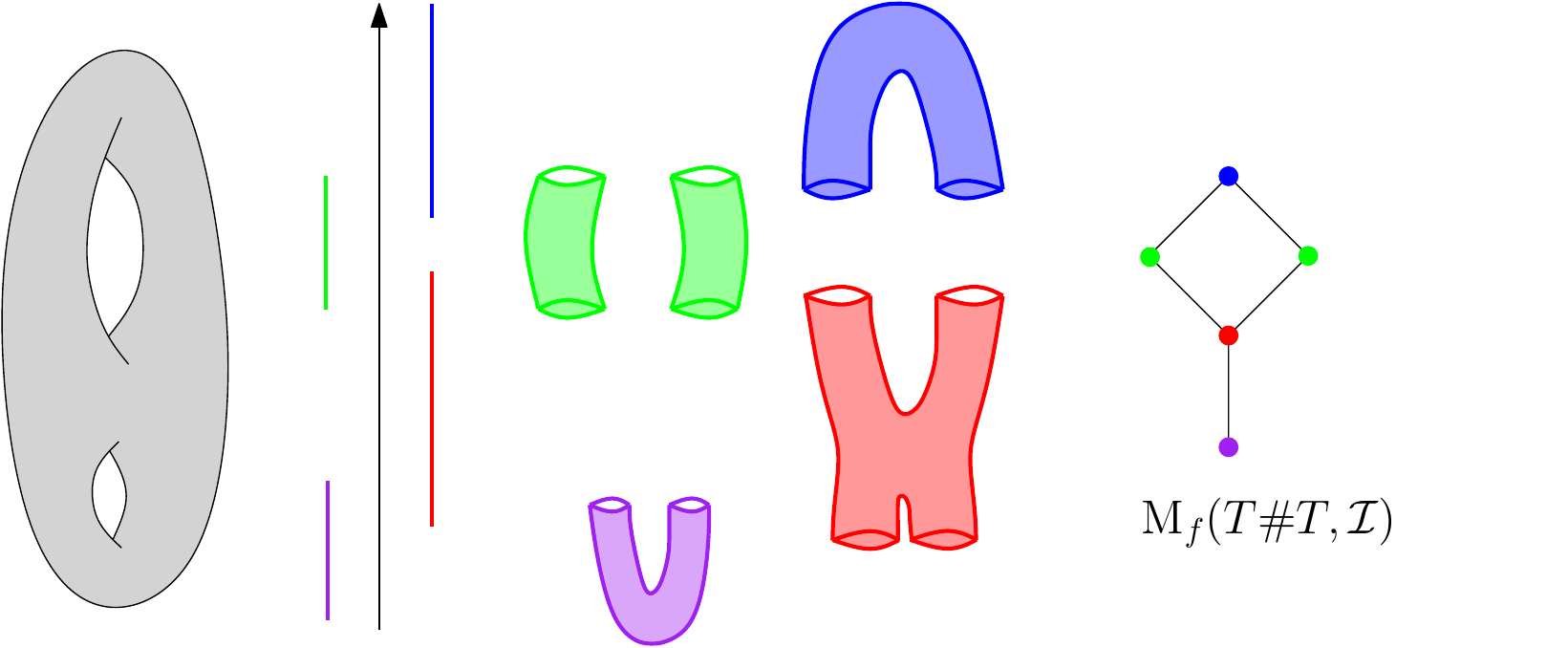}
\caption{\label{fig:MapperExample} Example of Mapper computed on the double torus $T\# T$ with the height function $f$ and a cover $\I$ of its range
with four open intervals.}
\end{figure}

\paragraph{Mapper.} The Mapper is introduced in \cite{Singh07} as a statistical version of the Reeb graph $\Reeb_f(\Xset)$ in the sense 
that it is a discrete and computable approximation of the Reeb graph computed with some filter function. 
Assume that we observe a point cloud 
$\Xs_n= \{X_1,\dots,X_n\}\subset \Xset$ with known pairwise dissimilarities.  
A filter function is chosen and can be computed on each point of  $\Xs_n$.  
The generic version of the Mapper algorithm on $\Xs_n$ computed with the filter function $f$ can be summarized as follows:
\begin{enumerate}
\item Cover the range of values $\Ys_n = f(\Xs_n)$ with a set of consecutive intervals $I_1,\dots, I_S$  which overlap.
\item Apply a clustering algorithm to each pre-image $f^{-1}(I_s)$, $s\in\{1,...,S\}$. This defines a {\em pullback cover}
$\mathcal{C}=\{\mathcal{C}_{1,1},\dots,\mathcal{C}_{1,k_1},\dots,\mathcal C_{S,1},\dots,\mathcal{C}_{S,k_S}\}$ of the point cloud $\Xs_n$.
\item The Mapper is then the {\em nerve} of $\mathcal{C}$. Each vertex $v_{s,k}$ of the Mapper corresponds to one element $\mathcal{C}_{s,k}$ 
and two vertices $v_{s,k}$ and $v_{s',k'}$ are connected if and only if  $\mathcal{C}_{s,k} \cap \mathcal{C}_{s',k'} $ is not empty.
\end{enumerate}
Even for one given filter function, many versions of the Mapper algorithm can be proposed depending on how one 
chooses the intervals that cover the image of $f$, and which method is used to cluster the pre-images.
Moreover, note that the Mapper can be defined as well for continuous spaces. The definition is strictly the same except for the clustering step,
since the connected components of each pre-image $f^{-1}(I_s)$, $s\in\{1,...,S\}$ are now well-defined. 
See Figure~\ref{fig:MapperExample}.

\paragraph{Our version of Mapper.} In this article, we focus on a Mapper algorithm that uses neighborhood graphs. 
Of course, more sophisticated versions of Mapper can be used in practice but then the statistical analysis is more tricky.  
We assume that there exists a distance on $\Xs_n$ and that the matrix of pairwise distances is available. 
First, from the distance matrix we compute 
the 1-skeleton of a Rips complex with parameter $\delta$, i.e. the $\delta$-neighborhood graph built on top of $\Xs_n$.
This objet plays the role of an approximation of the underlying and unknown metric space $\Xset$ on which the data are sampled. 
Second,  given $\Ys_n = f(\Xs_n)$  the set of filter values, we choose a regular cover of $\Ys_n $ with open intervals,
where no more than two intervals can intersect at a time.  
More precisely, we use open intervals with same length $r$ (apart from the first and the last one, which can have any positive length): 
$\forall s \in  \{2,  \dots,  S-1 \}$, 
$$ r = \ell(I_s)   $$  
where $\ell$ is the Lebesgue measure on $\R$. 
The overlap $g$ between two consecutive intervals is also a fixed constant: $\forall s \in \{1,\dots, S-1\},$
$$ 0 <g =  \frac{\ell(I_s \cap I_{s+1})}{\ell(I_s)} < \frac{1}{2} .  $$  
The parameters $g$ and $r$ are generally called the {\it gain} and the  {\it resolution}  
in the literature on the Mapper algorithm.
Finally, for the clustering step, we simply consider the connected components of the pre-images $f^{-1}(I_s)$ 
that are induced by the 1-skeleton of the Rips complex. 
The corresponding Mapper is denoted $\Map_{r,g,\delta}(\Xs_n,\Y_n)$  or $\Map_n$ for short in the following.
When dealing with a continuous space $\Xset$, there is no need to compute a neighborhood graph since the
connected components are well-defined, so we let $\Map_{r,g}(\Xset,f)$ denote such a Mapper. 

\medskip

\noindent {\bf Key fact 1b:}  The Mapper $\Map_{r,g,\delta}(\Xs_n,\Ys_n)$ 
is a combinatorial graph.

\medskip

Moreover, following~\cite{Carriere17b}, we can define a function on the nodes of $\Map_n$ as follows. 

\begin{defin}\label{def:arbitfunc}
Let $v$ be a node of $\Map_n$, i.e. $v$ represents a connected component of $f^{-1}(I_s)$
for some $s\in\{1,\dots, S\}$. Then, we let $$\fmapp{f}(v) = {\rm mid}(\tilde I_s),$$ where $\tilde I_s= I_s\setminus(I_s\cap I_{s-1})\cup(I_s\cap I_{s+1})$
 and ${\rm mid}(\tilde I_s)$ denotes the midpoint of the interval $\tilde I_s$.
\end{defin}

\paragraph{Filter functions.}
In practice, it is common to choose filter functions that are coordinate-independent, in order to avoid
depending on solid transformations of the data like rotations or translations. The two most common
filters that are used in the literature are:
\begin{itemize}
\item the {\em eccentricity}: $x\mapsto {\rm sup}_{y\in\Xset} d(x,y)$,
\item the eigenfunctions given by a Principal Component Analysis of the data.
\end{itemize}

\subsection{Extended persistence signatures and the persistence metric $\dper$ }


In this section, we introduce {\em extended persistence} and its associated metric, the {\em bottleneck distance},
which we will use later to compare Reeb graphs and Mappers.

\paragraph{Extended persistence.} Given any graph $G=(V,E)$ and a function attached to its nodes $f:V\rightarrow\R$,
the so-called {\em extended persistence diagram} $\Dg(G,f)$ is a multiset of points in the
Euclidean plane $\R^2$ that can be computed with {\em extended persistence theory}.
Each of the diagram points has a specific {\em type}, which is either $\Ord_0$, $\Rel_1$, $\Ext_0^+$ or $\Ext_1^-$.
We refer the reader to Appendix~\ref{sec:ExtPers} for formal definitions and further details about extended persistence.
A rigorous connexion between the Mapper and the Reeb graph was drawn recently by~\cite{Carriere17b}, 
who show how extended persistence 
provides a relevant and efficient framework to compare a Reeb graph with a Mapper.  
We summarize below the main points of this work in the perspective of the present article. 

\begin{figure}\centering
\includegraphics[width=15cm]{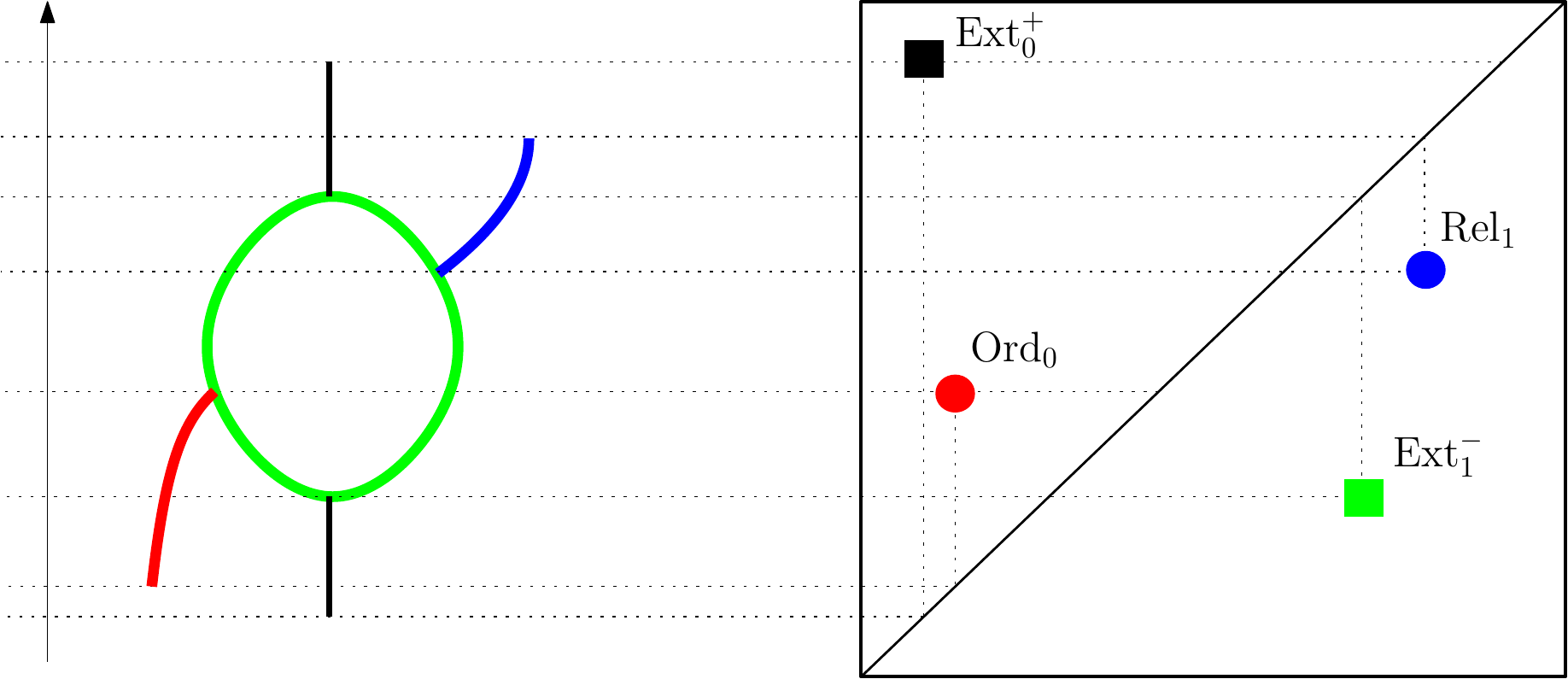}
\caption{\label{fig:topodict} Example of correspondences between topological features of a graph
and points in its corresponding extended persistence diagram. Note that ordinary persistence
is unable to detect the blue upwards branch.}
\end{figure}

\paragraph{Topological dictionary.} Given a topological space $\Xset$ and a Morse-type function $f:\Xset\rightarrow\R$,
there is a nice interpretation of $\Dg(\Reeb_f(\Xset),\freeb{f})$ in terms of the
structure of $\Reeb_f(\Xset)$. 
Orienting the Reeb graph vertically so $\freeb{f}$ is the height function,
we can see each connected component of the graph as a trunk with
multiple branches (some oriented upwards, others oriented downwards)
and holes.  Then, one has the following correspondences, where the
{\em vertical span} of a feature is the span of its image by~$\freeb{f}$:
\begin{itemize}
\item The vertical spans of the trunks are given by the points in $\Ext_0^+(\Reeb_f(\Xset),\freeb{f})$;
\item The vertical spans of the branches that are oriented downwards are given by the points in $\Ord_0(\Reeb_f(\Xset),\freeb{f})$;
\item The vertical spans of the branches that are oriented upwards are given by the points in $\Rel_1(\Reeb_f(\Xset),\freeb{f})$;
\item The vertical spans of the holes are given by the points in $\Ext_1^-(\Reeb_f(\Xset),\freeb{f})$. 
\end{itemize}
%
These correspondences
provide a dictionary to read off the structure of the Reeb graph from
the corresponding extended persistence diagram. See Figure~\ref{fig:topodict} for an illustration.

Note that it is a bag-of-features type descriptor, taking an inventory of all
the features (trunks, branches, holes) together with their vertical
spans, but leaving aside the actual layout of the features. As a
consequence, it is an incomplete descriptor: two Reeb graphs with the
same persistence diagram may not be isomorphic.



\paragraph{Bottleneck distance.} We now define the commonly used metric between persistence diagrams.

\begin{defin}
Given two persistence diagrams $D,D'$, a \emph{partial matching} between $D$ and $D'$ is a subset $\Gamma$ of $D\times D'$
such that:
\[\forall p\in D, \text{ there is at most one }p'\in D'\text{ s.t. }(p,p')\in\Gamma,\]
\[\forall p'\in D', \text{ there is at most one }p\in D\text{ s.t. }(p,p')\in\Gamma.\]
Furthermore, $\Gamma$ must match points of the same type (ordinary, relative, extended) and of the same homological dimension only.  
Let $\Delta$ be the diagonal $\Delta=\{(x,x)\ |\ x\in\R\}$. 
The \emph{cost} of $\Gamma$ is:
\[\text{cost}(\Gamma)=\max\left\{\max_{p\in D}\ \delta_D(p),\ \max_{p'\in D'}\ \delta_{D'}(p')\right\},\]
where
\[\delta_D(p)=\|p-p'\|_\infty \text{ if } \exists p'\in D'\text{ s.t. }(p,p')\in\Gamma \text{, otherwise } \delta_D(p)=\inf_{q\in\Delta} \|p-q\|_\infty,\]
\[\delta_{D'}(p')=\|p-p'\|_\infty \text{ if } \exists p\in D\text{ s.t. }(p,p')\in\Gamma \text{, otherwise } \delta_{D'}(p')=\inf_{q\in\Delta} \|p'-q\|_\infty.\]
\end{defin}

\begin{defin}
\label{def:bottleneck}
Let $D,D'$ be two persistence diagrams.
The \emph{bottleneck distance} between $D$ and $D'$ is:
\[d_\Delta(D,D')=\inf_{\Gamma}\ \emph{cost}(\Gamma),\]
where $\Gamma$ ranges over all partial matchings between $D$ and $D'$.
\end{defin}

Note that $d_\Delta$ is only a pseudometric, not a true metric, because points lying on $\Delta$ can be left unmatched at no cost.
\begin{defin}
Let $G_1=(V_1,E_1)$ and $G_2=(V_2,E_2)$ be two combinatorial graphs with real-valued functions  
$f_1:V_1\rightarrow\R$ and $f_2:V_2\rightarrow\R$ attached to their nodes.
The {\em persistence metric $\dper$} between the pairs $(G_1,f_1)$ and $(G_2,f_2)$ is:
$$\dper(G_1,G_2)=d_\Delta\left(\Dg(G_1,f_1),\Dg(G_2,f_2)\right).$$ 
\end{defin}

For a Morse-type function $f$ defined on $\Xset$ and for a finite point cloud $\Xs_n \subset \Xset$, we can thus consider 
$\Dg(\Reeb_f(\Xset)) = \Dg(\Reeb_f(\Xset),\freeb{f})$ and $\Dg(\Map_n) = \Dg(\Map_n, \fmapp f)$, with $\fmapp f$
as in Definition~\ref{def:arbitfunc}. 
In this context the bottleneck distance $\dper(\Reeb_f(\Xset), \Map_n)=d_\Delta(\Dg(\Reeb_f(\Xset)), \Dg(\Map_n))$ 
is well defined and we use this quantity to assess if the Mapper $\Map_n$ is a good approximation of the Reeb graph $\Reeb_f(\Xset)$. 
Moreover, note that, even though $d_\Delta$ is only a pseudometric, it has been shown to a be 
a true metric {\em locally} for Reeb graphs by~\cite{Carriere17a}.

As noted in~\cite{Carriere17b}, the choice of $f_\I$ is in some sense arbitrary since any function 
defined on the nodes of the Mapper that respects the ordering of the intervals of $\I$
carries the same information in its extended persistence diagram. To avoid this issue,~\cite{Carriere17b}
define a pruned version of $\Dg(\Reeb_f(\Xset), \freeb f)$ as a canonical descriptor for the Mapper.
The problem is that computing this canonical descriptor requires to know the critical values of $\freeb f$ beforehand.
Here, by considering $\Dg(\Map_n, \fmapp f)$ instead, the descriptor becomes computable.
Moreover, one can see from the proofs in the Appendix that the canonical descriptor and its arbitrary 
version actually enjoy the same rate of convergence, up to some constant.

\subsection{An approximation inequality for Mapper}\label{subsec:confreg}

We are now ready to give the key ingredient of this paper to derive a statistical analysis of the Mapper in Euclidean spaces. 
The ingredient is an upper bound on the bottleneck distance between the Reeb graph of a pair $(\Xset,f)$ and the Mapper 
computed with the same filter function $f$ and a specific cover $\I$ of a sampled point cloud $\Xs_n \subset \Xset$.  
From now on, it is assumed that the underlying space $\Xset$ is a smooth and compact submanifold embedded in $\R^D$,
and that the filter function $f$ is Morse-type on $\Xset$.

\paragraph{Regularity of the filter function.} 
Intuitively, approximating  a Reeb graph computed with a filter function $f$ that has large variations is more difficult than for a smooth filter function, 
for some notion of regularity that we now specify. Our result is given in a general setting by considering the modulus of continuity of $f$. 
In our framework, $f$ is assumed to be Morse-type and thus uniformly continuous on the compact set $\Xset$. Following for instance Section~6 in \cite{DeVore93},  we define the (exact) modulus of continuity of $f$ by 
\begin{equation*}
\omega_f(\delta) = \sup_{\|x - x'\| \leq \delta}  |f(x) -  f(x')|
\end{equation*} 
for any $\delta >0$, where $\|\cdot\|$ denotes the Euclidean norm in $\R^D$. Then $\omega_f$  satisfies :
\begin{enumerate}
\item $ \omega_f(\delta)  \rightarrow \omega(0) = 0$  as $ \delta \rightarrow 0$ ;
\item $\omega_f$ is non negative and non-decreasing on $\R^+$ ;
\item $\omega_f$ is subadditive : $\omega_f(\delta_1 + \delta_2) \leq \omega_f(\delta_1) +\omega_f(\delta_2) $ for any $\delta_1$, $\delta_2 >0$;
\item $\omega_f$  is continous on $\R^+$.
\end{enumerate}
In this paper we say that a function $\omega$ defined on $\R^+$ is \emph{a modulus of continuity} if it satisfies the four properties above 
and we say that  it is \emph{a modulus of continuity for $f$} if, in addition, we have
\begin{equation*}
 |f(x) -  f(x')| \leq \omega(\|x - x'\|),
\end{equation*}
for any $x,x' \in \Xset$.

\begin{thm}\label{thm:geomineq}
Assume that $\Xset$ has 
positive reach $\mathop{rch}$ and convexity radius $\rho$. 
Let $\Xs_n$ be a point cloud of $n$ points, all lying in $\Xset$. Assume that the filter function $f$ is Morse-type on $\Xset$. 
Let $\omega$ be a modulus of continuity for $f$.
If the three following conditions hold:
\begin{align}
&\delta \leq \frac 1 4 \min\left\{\mathop{rch}, \rho\right\}, \label{CdtA} \\ 
&\max\{|f(X)-f(X')|\, :\, X,X'\in\Xs_n, \|X-X'\|\leq \delta\}    <  g r, \label{CdtB} \\ 
&4  \disth(\Xset ,\Xs_n) \leq\delta ,  \label{CdtD}
\end{align}
then the Mapper $\Map_n = \Map_{r,g,\delta}(\Xs_n,\Ys_n)$ with parameters $r$, $g$ and $\delta$ is such that:
\begin{equation} \label{ApproxBound}
\dper\left(\Reeb_f(\Xset), \Map_n \right)\leq r +2\omega(\delta).
\end{equation}
\end{thm}


\begin{rmq} \label{rq:metrics}
Using the edge-based MultiNerve Mapper---as defined in~Section~8 of~\cite{Carriere17b}---allows to 
weaken Assumption~(\ref{CdtB}) since $gr$ can be replaced by $r$ in the corresponding equation,
and $r$ can be replaced by $r/2$ in Equation~(\ref{ApproxBound}).
\end{rmq}

\paragraph{Analysis of the hypotheses.} 
On the one hand, the scale parameter of the Rips complex could not be smaller than the approximation error corresponding to 
the Hausdorff distance between the sample and the underlying space $\Xset$ (Assumption~\eqref{CdtD}). 
On the other hand, it must be smaller than the reach and convexity radius to provide a correct estimation of 
the geometry and topology of $\Xset$ (Assumption~\eqref{CdtA}). 
The quantity $g r$ corresponds to the minimum scale at which the filter's codomain is analyzed. This minimum resolution 
has to be compared with the regularity of the filter at scale $\delta$ (Assumption~\eqref{CdtB}). Indeed the pre-images of a filter with 
strong variations will be more difficult to analyze than when the filter does not vary too fast. 

\paragraph{Analysis of the upper bound.} The upper bound given in \eqref{ApproxBound} makes sense in that the 
approximation error is controlled by the resolution level in the 
codomain and by the regularity of the filter. If one uses a filter with strong variations, or if the grid in the codomain has a too 
rough resolution, then the approximation will be poor. On the other hand, a sufficiently dense sampling is required in order to take $r$ small, 
as prescribed in the assumptions.

\paragraph{Lipschitz filters.} A large class of filters used for Mapper are actually Lipschitz functions and of course, 
in this case, one can take $\omega(\delta) = c \delta$ 
for some positive constant $c$. In particular, $c=1$ for linear projections (PCA, SVD, Laplacian or coordinate filter for instance). 
The distance to a measure (DTM) is also a 1-Lipschitz function, see~\cite{Chazal11}. On the other hand, 
the modulus of continuity of filter functions defined from 
estimators, e.g. density estimators, is less obvious although still well-defined. 


\paragraph{Filter approximation.} In some situations, the filter function $\hat f$ used to compute the Mapper is only an approximation of 
the filter function $f$ with which the Reeb graph is computed. 
In this context, 
the pair $(\Xs_n, \hat f)$ appears as an approximation of the pair $(\Xset,f)$.
The following result is directly derived from Theorem~\ref{thm:geomineq} and 
Theorem~5.1 in~\cite{Carriere17b} (that derives stability for Mappers building on
the stability theorem of extended persistence diagrams proved by \cite{Cohen09}):
\begin{cor} \label{cor:approxfilter}
Let $\hat f: \Xset \rightarrow \R$ be a Morse-type filter function approximating $f$.
Assume that Assumptions~(\ref{CdtA}) and~(\ref{CdtD}) of Theorem~\ref{thm:geomineq} are satisfied, 
and assume moreover that
\begin{equation}
\max\{\max\{|f(X)-f(X')|,|\hat f(X)-\hat f(X')|\}\,:\,X,X'\in\Xs_n,\|X-X'\|\leq\delta\} < gr.
\end{equation} 
Then, the Mapper $\hat{\Map}_n  =   \Map_{r,g,\delta} (\Xs_n, \hat f(\Xs_n))$ built on $\Xs_n$ with filter function $\hat f$  
and parameters $r,g,\delta$ satisfies:
$$\dper(\Reeb_f(\Xset),  \hat{\Map}_n )\leq 2r+2\omega(\delta)  + \max_{1\leq i\leq n} |f(X_i)-\hat f(X_i)|. $$
\end{cor}
%


\section{Statistical Analysis of Mapper}
\label{sec:ConfConv}


From now on, the set of observations $\Xs_n$  
is assumed to be composed of $n$ independent points $X_1,...,X_n$ sampled from a probability distribution 
$\mathbb{P}$ in $\R^D$ (endowed with its Borel algebra). We assume that each point $X_i$ comes with a 
filter value which is represented by a random variable $Y_i$. Contrarily to the $X_i$'s, 
the filter values $Y_i$'s are not necessarily independent. In the following, we consider two different settings: 
in the first one, $Y_i = f(X_i)$, where the filter $f$ is a deterministic function, in the second one, $Y_i = \hat f(X_i)$ 
where $\hat f$ is an estimator of the filter function $f$. In the latter case, the $Y_i $'s are obviously dependent.
We first provide the following Proposition, whose proof is deferred to Appendix~\ref{sec:proofMeasurability}, which states that
computing probabilities on the Mapper makes sense:

\begin{prop}\label{prop:Measurability}
For any fixed choice of parameters $r,g,\delta$ and for any fixed $n\in \mathbb{N}$, the function 
$$\Phi:\left\{\begin{array}{ccc}(\R^D)^n\times \R^n & \rightarrow & \mathcal{R} \\ (\Xs_n,\Ys_n) & \mapsto & \Map_{r,g,\delta}(\Xs_n,\Ys_n)\end{array}\right.$$ 
is measurable, where $\mathcal R$ denotes the set of Reeb graphs computed from Morse-type functions.
\end{prop}

\subsection{Statistical Model for Mapper}

In this section, we study the convergence of the Mapper for a general generative model and a class of filter functions. 
We first introduce the generative model and next we present different settings depending on the nature of the filter function.

\paragraph{Generative model.} The set of observations $\Xs_n$  is assumed to be composed of $n$ independent points 
$X_1,...,X_n$ sampled from a probability distribution $\mathbb{P}$ in $\R^D$. The support of $\p$ is denoted 
$\Xset_\p$ and is assumed to be a smooth and compact  submanifold of $\R^D$ with positive reach and positive convexity radius, 
as in the setting of Theorem~\ref{thm:geomineq}. We also assume that $0 <{\rm diam}(\mathcal{X}_\p)\leq L$. 
Next, the probability distribution $\mathbb{P}$ is assumed to be 
$(a,\dimension)$-standard for some constants $a>0$  and $\dimension\geq D$, that is for any Euclidean ball $B(x,t)$ centered on $x\in \Xset$ with radius $t$ :
 $$ \proba{B(x,t)} \geq \min (1,a t^\dimension). $$
This assumption is popular in the literature about set estimation  \citep[see for instance][]{Cuevas09,Cuevas04}. 
It is also widely used in the TDA literature \citep{Chazal15c,Fasy14,Chazal15a}. 
For instance, when $\dimension=D$, this assumption is satisfied when the distribution is absolutely continuous with 
respect to the Hausdorff measure on $\Xset_\p$. 
We introduce the set  $\mathcal{P}_{a,\dimension}=\mathcal{P}_{a,\dimension, \kappa, \rho, L  }$ 
which is composed of all  the $(a,\dimension)$-standard probability distributions for which the support $\Xset_\p$ 
is a smooth and compact submanifold of $\R^D$ with reach larger than $\kappa$, convexity radius larger than $\rho $
and diameter less than $L$. 
 
\paragraph{Filter functions in the statistical setting.} The filter function  $f:\Xset_\p \rightarrow \R$ 
for the Reeb graph is assumed as before to be a Morse-type function. Two different settings have to be 
considered regarding how the filter function is defined. In the first setting, the same filter function is used 
to define the Reeb graph  and the Mapper. The Mapper can be defined by taking the exact values of the filter function 
at the observation points $ f(X_1),\dots,f(X_n)$. Note that this does not mean that the function $f$ is completely known 
since, in our framework, knowing $f$ would imply to know its domain and thus $\Xset_\p$ would be known which is of course 
not the case in practice. This first setting is referred to as the {\it exact filter setting} in the following. It corresponds 
to the situations where the Mapper algorithm is used with coordinate functions for instance. In the second setting, the 
filter function used for the Mapper is not available and an estimation of this filter function has to be computed from the data. 
This second setting is referred to as the {\it inferred filter setting} in the following. It corresponds to  
PCA or Laplacian eigenfunctions, distance functions (such as the DTM), or regression and density estimators.

\paragraph{Risk of Mapper.} We study, in various settings, the problem of inferring a Reeb graph using Mappers and we use 
the metric $\dper$ to assess the performance of the Mapper, seen as an estimator of the Reeb graph:
$$\mathbb{E} \left[ \dper\left(\Map_{n},\Reeb_f(\Xset_\p)  \right)\right],$$
where $\Map_n$ is computed with the exact filter $f$ or the inferred filter $\hat f$, depending on the context.

\subsection{Reeb graph inference with exact filter and known generative model }
\label{subsec:knownab}

We first consider the exact filter setting in  the simplest situation where the parameters  
$a$ and $\dimension$ of the generative model are known. In this setting, for given Rips parameter $\delta$, 
gain $g$ and resolution $r$, the Mapper $\Map_n = \Map_{r,g,\delta}(\Xs_n,\Ys_n)$  
is computed with $\Ys_n = f(\Xs_n)$.  We now tune the triple of parameters $(r,g,\delta)$ 
depending on the parameters $a$ and $b$.  
Let $V_n(\delta_n)=\max\{|f(X)-f(X')|\, :\, X,X'\in\Xs_n, \|X-X'\|\leq\delta_n\}$.
We choose for $g$ a fixed value in $\left(\frac13,\frac12\right)$ and we take:
$$\delta_n=8\left(\frac{2{\log}(n)}{an}\right)^{1/\dimension}
\quad 
\textrm{ and }
\quad 
r_n=  \frac{V_n(\delta_n)^+} {g},$$
where $V_n(\delta_n)^+$ denotes a value that is strictly larger but arbitrarily close to $V_n(\delta_n)$.
We  give below a general upper bound on the risk of $\Map_n$ with these parameters, which depends on the regularity 
of the filter function and on the parameters of the generative model.  We show a uniform convergence over a class of possible filter functions. 
This class of filters necessarily depends on the support of $\p$, so we define the class of filters for each probability measure in 
$\mathcal{P}_{a,\dimension}$. For any $\p \in \mathcal{P}_{a,\dimension}$, we let 
$\mathcal F (\p,\omega)$ 
denote the set of filter functions 
$f: \Xset_\p \rightarrow \R$ such that $f$ is Morse-type on $ \Xset_\p$ with $\omega_f \leq \omega$. 

\begin{prop} \label{prop:UpBdRestr}
Let $\omega$ be a   modulus of continuity for $f$ such that  $\omega(x) / x $ is a non-increasing function on $\R^+$. 
For $n$ large enough, the Mapper computed with parameters $(r_n,g,\delta_n)$ defined before satisfies
$$
\sup_{\p \in \mathcal P_{a,b}} \mathbb{E}\left[   \sup_{f \in \mathcal F(\p,\omega) } \dper\left(\Reeb_f(\Xset_\p), \Map_n\right)\right] \leq C\,\omega\left(\frac{2\cdot 8^{\dimension}}{a}\frac{{\rm log}(n)}{n}\right)^{1/\dimension} 
$$
where the constant $C$ only depends on $a$, $b$, and on the geometric parameters of the model.
\end{prop}

Assuming that $\omega(x) / x $ is non-increasing is not a very strong assumption. 
This property is satisfied in particular when $\omega$ is concave, as in the case of
concave majorant (see for instance Section~6 in~\cite{DeVore93}).  As expected, we see that the rate of convergence of the Mapper to the Reeb graph directly depends on 
the regularity of the filter function and on the parameter $\dimension$ which roughly represents the intrinsic dimension of the data. 
For Lipschitz filter functions, the rate is similar to the one for persistence diagram inference \cite{Chazal15c}, 
namely it corresponds to the one of support estimation for the Hausdorff metric (see for instance \cite{Cuevas04}) 
and \cite{Genovese12a}). In the other cases where the filters only admit a concave modulus of continuity, we see that 
the ``distortion" created by the filter function slows down the convergence of the Mapper to the Reeb graph.

We now give a lower bound that matches with the upper bound of Proposition~\ref{prop:UpBdRestr}.


\begin{prop}\label{prop:lecam}
Let  $\omega$ be a  modulus of continuity for $f$. Then, for any estimator ${\hat \Reeb}_n$  of  $\Reeb_f(\Xset_\p)$, 
we have
\begin{equation*}
\underset{\mathbb{P} \in\mathcal{P}_{a,\dimension}}{\rm sup}\  \mathbb{E} \left[   \sup_{f \in \mathcal F(\p,\omega) }  \dper\left( 
\Reeb_f(\Xset_\p),{\hat \Reeb}_n \right)
\right]
\geq 
 C\,\omega\left( \frac{1}{an}\right)^{\frac 1 \dimension}, 
 \end{equation*}
where the constant $C$ only depends on $a$, $b$ and on the geometric parameters of the model.
\end{prop}

Propositions~\ref{prop:UpBdRestr} and~\ref{prop:lecam} together show that, with the choice of parameters  given before, 
$\Map_n$ is minimax optimal up to a logarithmic factor $\log(n)$ inside the modulus of continuity. Note that the 
lower bound is also valid whether or not the coefficients $a$ and $b$ and the filter function $f$ and its modulus of continuity are given.

%
%
%

\subsection{Reeb graph inference with exact filter and unknown generative model}

We still assume that the exact values $\Ys_n = f(\Xs_n)$ of the filter on the point could can be computed and 
that at least an upper bound on the modulus of continuity for the filter is known. 
However, the parameters $a$ and $b$ are not assumed to be known anymore. We adapt a subsampling approach 
proposed by~\cite{Fasy14}. As before, for given Rips parameter $\delta$,  gain $g$ and resolution $r$, 
the Mapper $\Map_n = \Map_{r,g,\delta}(\Xs_n,\Ys_n)$  is computed with $\Ys_n = f(\Xs_n)$.

We introduce the sequence  $s_n  =\frac n { (\log n)^{1+\beta}}$ for some fixed value $\beta > 0$. 
Let $\hat \Xs_n^{s_n}$ be an arbitrary subset of $\Xs_n$ that contains $s_n$ points. 
We tune the triple of parameters $(r,g,\delta)$ as follows: we choose for $g$ a fixed value in $\left(\frac13,\frac{1}{2}\right)$ and we take:
\begin{equation}
\label{ref:coeff_ab_inconnus}
\delta_n=d_{\rm H}(\hat  \Xs_n^{s_n},\Xs_n)
\quad 
\textrm{ and }
\quad 
r_n=  \frac{V_n(\delta_n)^+} {g},
\end{equation}
where $V_n$ is defined as in Section~\ref{subsec:knownab}.
\begin{prop}\label{prop:subs}
Let $\omega$ be a modulus of continuity for $f$ such that 
$x\mapsto \omega(x)/x$ is a non-increasing function. Then, using the same notations as in the previous section,  
the Mapper $\Map_n$ computed with parameters $(r_n,g,\delta_n)$ defined before satisfies
\begin{equation*}
\sup_{\p \in \mathcal P_{a,b}} \mathbb{E}\left[   \sup_{f \in \mathcal F(\p,\omega) } \dper\left(\Reeb_f(\Xset_\p), \Map_n\right)\right]
\leq C\,\omega\left(\frac{C'{\log}(n)^{2+\beta}}{n}\right)^{1/\dimension},
\end{equation*}
where the constants $C,C'$ only depends on $a$, $\dimension$, and on the geometric parameters of the model.
\end{prop}
Up to logarithmic factors inside the modulus of continuity, 
we find that this Mapper is still minimax optimal over the class $\mathcal P_{a,b}$ by Proposition~\ref{prop:lecam}.

\subsection{Reeb graph inference with inferred filter and unknown generative model}

One of the nice properties of Mapper is that it can easily be computed
with any filter function, including estimated filter functions such as
PCA eigenfunctions, eccentricity functions, DTM functions, Laplacian
eigenfunctions, density estimators, regression estimators, and many
other filters directly estimated from the data. In this section, we
assume that the {\em true filter} $f$ is unknown but can be estimated
from the data using an estimator $\hat f$. Without loss of generality  we assume that both $f$ and $\hat f$ are defined on $\R^D$.
As before, parameters $a$ and
$b$ are not assumed to be known and we have to tune the triple of
parameters $(r_n,g,\delta_n)$.

In this context, the quantity $V_n$ cannot be computed as before because there is no direct access to the values of $f$: we 
only know an estimation $\hat f$ of it. 
However, in many cases, an upper bound $\omega_1$ on the modulus of continuity of $f$ is known, which makes possible the 
tuning of the parameters. For instance, PCA (and kernel) projectors, eccentricity functions, DTM functions 
(see~\cite{Chazal11}) are all 1-Lipschitz functions, and Corollary~\ref{cor:rate_filtre_inconnu} below can be applied.

Let $\hat V_n(\delta_n)=\max\{|\hat f(X)-\hat f(X')|\, :\, X,X'\in\Xs_n, \|X-X'\|\leq\delta_n\}$, and let
$\omega_1$ be a modulus of continuity for $f$. Let 
\begin{equation}\label{eq:coeff_ab_filtre_inconnus}
r_n=\frac{\max\{\omega_1(\delta_n),\hat V_n(\delta_n)\}^+}{g}.
\end{equation}

Following the lines of the proof of Proposition~\ref{prop:subs} and applying Corollary~\ref{cor:approxfilter}, we obtain:

\begin{cor}\label{cor:rate_filtre_inconnu}
Let $f:\R^D \rightarrow \R $ be a Morse-type filter function  and let $\hat f: \R^D \rightarrow \R$ be a Morse-type estimator of $f$. 
Let  $\omega_1$  (resp. $\omega_2$) be a modulus of continuity for $f$ (resp. $\hat f$).
Let $\omega=\max\{\omega_1,\omega_2\}$ such that 
$x\mapsto \omega(x)/x$ is a non-increasing function. 
Let also  $\hat{\Map}_n  =   \Map_{r_n,g,\delta_n} (\Xs_n, \hat f(\Xs_n))$  be the Mapper built on $\Xs_n$ 
with function $\hat f$  and parameters $g,\delta_n$  as in Equation~\eqref{ref:coeff_ab_inconnus},
and $r_n$ as in Equation~\eqref{eq:coeff_ab_filtre_inconnus}. 
Then, $\hat{\Map}_n$ satisfies
\begin{equation*}
\label{risk_estimated_filter}
 \mathbb{E}\left[  \dper\left(\Reeb_f(\Xset_\p),  \hat{\Map}_n  \right)\right]
\leq C\omega\left(\frac{C'{\log}(n)^{2+\beta}}{n}\right)^{\frac 1 \dimension} 
+  \mathbb{E} \left[\max_{1\leq i \leq n } | f(X_i) - \hat f (X_i)| \right],
\end{equation*}
where  the constants $C,C'$ only depends on $a$, $\dimension$, and on the geometric parameters of the model.
\end{cor}

Note that $\omega_1$ has to be known to compute $\hat{\Map}_n$ in Corollary~\ref{cor:rate_filtre_inconnu} 
since it appears in the definition of $r_n$. On the contrary, $\omega_2$---and thus $\omega$---are not required to tune the parameters.



\paragraph{PCA  eigenfunctions.} In the setting of this article, the measure $\mu$ has a finite second moment. 
Following~\cite{Biau12}, we define the covariance operator $\Gamma(\cdot) =\E ( \langle  X ,  \cdot \rangle X )$ and  
we let $\Pi_k$ denote the orthogonal projection onto the space spanned by the $k$-th  eigenvector of $\Gamma$.
 In practice, we consider the empirical version of the covariance operator 
$$ \hat  \Gamma_n (\cdot)  = \frac 1n \sum_{i=1}^n \langle  X_i ,   \cdot \rangle X_i  $$
and the empirical projection  $\hat \Pi_k$ onto the space spanned by the $k$-th eigenvector of $\hat  \Gamma_n $.  
According to  \cite{Biau12}(see also \cite{Blanchard07,Shawe05}), we have
$$\E \left[\| \Pi_k  - \hat \Pi_k  \|_\infty\right]  = O\left(\frac1 {\sqrt n}\right).$$ 
This, together with Corollary~\ref{risk_estimated_filter} and the fact that both $\Pi_k$ and $\hat \Pi_k$ are 1-Lipschitz, gives that
the rate of convergence of the Mapper of $\hat \Pi_k(\Xs_n)$ computed with parameters $\delta_n$ and $g$
as in Equation~(\ref{ref:coeff_ab_inconnus}), and
$r_n$ as in Equation~(\ref{eq:coeff_ab_filtre_inconnus}), i.e. $r_n=g^{-1}\delta_n^+$, satisfies
$$
 \mathbb{E} \left[  \dper \left( \Reeb_{\Pi_k} (\Xset_\p), \Map_{r_n,g,\delta_n} (\Xs_n, \hat \Pi_k(\Xs_n)) \right) \right]
\lesssim  \left(\frac{{\log}(n)^{2+\beta}}{n}\right)^{1/\dimension}  \vee \frac1 {\sqrt n}.
$$
Hence, the rate of convergence of Mapper 
is not deteriorated by using $\hat \Pi_k$ instead of $\Pi_k$ if
the intrinsic dimension $b$ of the support of $\mu$ is at least 2.

\paragraph{The distance to measure.}  It is well known that TDA methods may fail completely in the presence of outliers. 
To address this issue, \citet{Chazal11} introduced an alternative distance function which is robust to noise,  
the {\em distance-to-a-measure} (DTM).  A similar analysis  as  with the PCA filter can be carried out with the DTM filter 
using the rates of convergence proven in \cite{Chazal16b}.


\section{Confidence sets for Reeb signatures} \label{sec:confR}
\subsection{Confidence sets for extended persistence diagrams}

In practice,  computing a Mapper $\Map_n$ and its signature $\Dg(\Map_n,\fmapp{f})$ is not sufficient: we need to know how accurate these estimations are. 
One natural way to answer this problem is to provide a confidence set for the Mapper using the bottleneck distance. 
For $\alpha \in (0,1)$, we look for some value $\eta_{n,\alpha}$ such that
 $$ \proba{\dper(\Map_n,\Reeb_f(\Xset_\p))\geq \eta_{n,\alpha} } \leq  \alpha $$
or at least such that 
 $$ \underset{n \rightarrow \infty}{\limsup}  \proba{\dper(\Map_n,\Reeb_f(\Xset_\p))\geq \eta_{n,\alpha}}  \leq  \alpha .$$
Let $$\mathcal M_\alpha = \left\{ \Reeb  \in \mathcal R  \, : \,  \dper(\Map_n, \Reeb )  \leq \alpha  \right\}$$ 
be the closed ball of radius $\alpha$ in the bottleneck distance and centered at the Mapper $\Map_n$ in the space of Reeb graphs $\mathcal R$. 
Following~\cite{Fasy14}, we can visualize the signatures of the points belonging to this ball in various ways. 
One first option is to center a box of side length $2\alpha$ at each point of the extended persistence diagram 
of $\Map_n$---see the right columns of Figure~\ref{fig:synthdata} and Figure~\ref{fig:realapp} for instance. 
An alternative solution is to visualize the confidence set by adding a band at (vertical) distance $2\alpha$ from the diagonal (the bottleneck distance being defined for the $\ell_\infty$ norm).
The points outside the band are then considered as significant topological features, see~\cite{Fasy14} for more details.  
 
Several methods have been proposed in~\cite{Fasy14} and~\cite{Chazal14b}  
to define confidence sets for persistence diagrams. We now adapt these ideas to provide confidence sets for Mappers. 
Except for the bottleneck bootstrap (see further), all the methods proposed in these two articles rely on the 
stability results for persistence diagrams, which say that persistence diagrams equipped with the bottleneck distance 
are stable under Hausdorff or Wasserstein perturbations of the data. Confidence sets for diagrams are then directly 
derived from confidence sets in the sample space. Here, we follow a similar strategy using Theorem~\ref{thm:geomineq}, as explained in the next section.

\subsection{Confidence sets derived from Theorem~\ref{thm:geomineq}}

In this section, we always assume that an upper bound $\omega$ on the exact modulus of continuity $\omega_f$ of the filter function is known. 
We start with the following remark: if we can take $\delta$ of the order of $d_{\rm H}(\Xset_\p,\Xs_n)$ in Theorem~\ref{thm:geomineq} 
and if all the conditions of the theorem are satisfied, then $\dper(\Map_n,\Reeb_f(\Xset_\p))$ can be bounded in terms of 
$\omega(d_{\rm H}(\Xset_\p,\Xs_n))$.  This means that we can adapt the methods of \cite{Fasy14} to Mappers.

\paragraph*{Known generative model.}  Let us first consider the simplest situation where  
the parameters $a$ an $b$ are also known. Following Section~\ref{subsec:knownab}, we choose for $g$ a fixed value in $\left(\frac13,\frac 12\right)$ and we take 
$$\delta_n=8\left(\frac{2{\log}(n)}{an}\right)^{1/\dimension}
\quad 
\textrm{ and }
\quad 
r_n=  \frac{V_n(\delta_n)^+} {g} ,$$
where $V_n$ is defined as in Section~\ref{subsec:knownab}.
Let $\varepsilon_n=d_{\rm H}(\Xset_\p,\Xs_n)$.
As shown in the proof of Proposition~\ref{prop:UpBdRestr} (see Appendix~\ref{sec:proofdeter}), for $n$ large enough, 
Assumption~\eqref{CdtA} and \eqref{CdtB} 
are always satisfied and then 
$$ \proba{\dper(\Map_n,\Reeb_f(\Xset_\p))\geq\eta}  \leq  \proba{ \delta_n \geq  \omega^{-1}\left(\frac{\eta}{g^{-1} + 2}\right) } .$$
Consequently,
\begin{align}
\proba{\dper(\Map_n,\Reeb_f(\Xset_\p))\geq\eta} &\leq \proba{\dper(\Map_n,\Reeb_f(\Xset_\p))\geq\eta \cap \varepsilon_n \leq 4\delta_n} + \proba{\varepsilon_n > 4\delta_n}\nonumber \\
&\leq \mathbb{I}_{\omega(\delta_n)\geq \frac{g}{1+2g}\eta} +  \min\left\{1,\frac{2^\dimension}{2{\log}(n)n}\right\} \nonumber \\
& = \Phi_n(\eta). \nonumber
\end{align}
where $ \Phi_n$ depends on the parameters of the model (or some bounds on these parameters)  which are here assumed to be known. 
Hence, given a probability level $\alpha$, one has:
$$\proba{\dper(\Map_n,\Reeb_f(\Xset_\p))\geq\Phi_n^{-1}(\alpha)} \leq \alpha.$$

\paragraph*{Unknown generative model.} We now assume that $a$ and $b$ are unknown.  
To compute confidence sets for the Mapper in this context, we approximate the distribution of 
$\disth(\Xset_\p,  \Xs_n )$ using the distribution of  $\disth (\hat \Xs_n^{s_n}, \Xs_n)$ conditionally to $\Xs_n$.  
There are $N_1=\binom{n}{s_n}$ subsets of size $s_n$ inside $\Xs_n$, so we let $\Xs^1 _{s_n}, \dots ,  \Xs^{N_1}_{s_n}$
denote all the possible configurations. Define
$$L_n(t)=\frac{1}{N_1}\sum_{k=1}^{N_1} \mathbb{I}_{\disth\left(\Xs^k_{s_n}, \Xs_n \right) > t}.$$ 
Let $s$ be the function on $\N$ defined by $s(n) = s_n$ and let $s^2_n =s (s(n)).$  
There are $N_2=\binom{n}{s^2_n}$ subsets of size $s^2_n$ inside $\Xs_n$. Again, we let $\Xs^k_{s^2_n}$, $1\leq k \leq N_2$, 
denote these configurations and we also introduce 
$$F_n(t)=\frac{1}{N_2}\sum_{k=1}^{N_2} \mathbb{I}_{\disth\left(\Xs^k_{s^2_n} , \Xs_{s_n}\right) > t} . $$
\begin{prop}\label{prop:conf1}
Let $\eta >0$.
Then, one has the following confidence set:
$$\proba{\dper(\Reeb_f(\Xset_\p),\Map_n)\geq \eta}\leq F_n\left(\frac{1}{4}\omega^{-1}\left(\frac{g}{1+2g}\eta\right)\right) 
+ L_n\left(\frac{1}{4}\omega^{-1}\left(\frac{g}{1+2g}\eta\right)\right) + o\left(\frac{s_n}{n}\right)^{1/4}.$$
\end{prop}
Both $F_n$ and $L_n$ can be computed in practice, or at least approximated using Monte Carlo procedures.  
The upper bound on $ \proba{ \dper(\Reeb_f(\Xset_\p),\Map_n)\geq \eta}$  
then provides an asymptotic confidence region for the persistence diagram of the Mapper $\Map_n$, 
which can be explicitly computed in practice. See the green squares in the first row of Figure~\ref{fig:synthdata}.
The main drawback of this approach is that it requires to know an upper bound on the modulus of continuity $\omega$
and, more importantly, the number of observations has to be very large, which is not the case on our examples in Section~\ref{sec:appli}.

\paragraph{Modulus of continuity of the filter function.} As shown in Proposition~\ref{prop:conf1}, the modulus of continuity 
of the filter function is a key quantity to describe the confidence regions. 
Inferring the modulus of continuity of the filter from the data is a tricky problem. 
Fortunately, in practice, even in the inferred filter setting, 
the modulus of continuity of the function can be upper bounded in many situations. For instance, projections  such as PCA eigenfunctions and
DTM functions are 1-Lipschitz. 

\subsection{Bottleneck Bootstrap}\label{sec:bootstrap}

The two methods given before both require an explicit upper bound on the modulus of continuity of the filter function.  
Moreover, these methods both rely on the approximation result Theorem~\ref{thm:geomineq}, which often leads to conservative confidence sets. 
An alternative strategy is the bottleneck bootstrap introduced in \cite{Chazal14b},
and which we now apply to our framework.  
 
The bootstrap is a general method for estimating standard errors and computing confidence intervals. 
Let $\p_n$ be the empirical measure defined from the sample $(X_1,Y_1), \ldots, (X_n,Y_n)$. Let  \\ $(X_1^*,Y_1^*)\dots, (X_n^*,Y_n^*)$ 
be a sample from $\p_n$ and let also $\Map_n^*$ be the random Mapper defined from this sample. We then take for 
$\hat \eta_{n,\alpha}$ the quantity  $\hat \eta_{n,\alpha}^*$ defined by
\begin{equation}
\proba{\dper ( \Map_n^*, \Map_n) > \hat \eta_{n,\alpha}^* \, |\, X_1,\ldots, X_n} = \alpha.
\end{equation}
Note that $\hat \eta_{n,\alpha}^*$ can be easily estimated with Monte Carlo procedures. 
It has been shown in~\cite{Chazal14b} that the bottleneck bootstrap is valid when computing 
the sublevel sets of a density estimator. The validity of the bottleneck bootstrap has not been proven
for the extended persistence diagram of any distance function. For Mapper, it would require to write  
$\dper ( \Map_n^*, \Map_n)$ in terms of the distance between the extrema of the filter function and 
the ones of the interpolation of the filter function on the Rips. We leave this problem open in this article.

\paragraph{Extension of the analysis.} As pointed out in Section~\ref{subsec:Backg}, many versions of the discrete Mapper
exist in the literature. One of them, called the {\em edge-based MultiNerve Mapper} $\MMapper^\triangle_{r,g,\delta}(\Xs_n,\Ys_n)$, 
is described in Section~8 of~(\cite{Carriere17b}). The main advantage of this version is that it allows for 
finer resolutions than the usual Mapper while remaining fast to compute.
Our analysis can actually handle this  version as well by replacing $gr$ by $r$ in Assumption~\eqref{CdtB} 
of Theorem~\ref{thm:geomineq}---see Remark~\ref{rq:metrics},
and changing constants accordingly in the proofs. In particular, this improves the resolution $r_n$ in Equation~(\ref{ref:coeff_ab_inconnus})
since $g^{-1}V_n(\delta_n)^+$ becomes $V_n(\delta_n)^+$. Hence, we use this edge-based version in Section~\ref{sec:appli},
where this improvement on the resolution $r_n$ allows us to compensate for the low number of observations in some datasets.

\section{Numerical experiments}\label{sec:appli}

In this section, we provide few examples of parameter selections and confidence regions (which are union of squares in the extended persistence diagrams) 
obtained with bottleneck bootstrap. The interpretation of these regions is that squares that intersect the diagonal, 
which are drawn in pink color, represent topological features in the Mappers that may be horizontal or artifacts due to the cover,
and that may not be present in the Reeb graph. 
We show in Figure~\ref{fig:synthdata} various Mappers (in each node of the Mappers, the left number is the cluster ID and the right number is the 
number of observations in that cluster) 
and 85 percent confidence regions computed on various datasets.
All $\delta$ parameters and resolutions were computed with Equation~(\ref{ref:coeff_ab_inconnus}) 
(the $\delta$ parameters were also averaged over $N=100$ subsamplings with $\beta=0.001$), 
and all gains were set to $40\%$. The code we used is expected to be added in the next release of~\cite{gudhi},
and should then be available soon. 
The confidence regions were computed by bootstrapping data 100 times. 
Note that computing confidence regions with Proposition~\ref{prop:conf1} is possible, but the numbers of observations in all of our datasets
were too low, leading to conservative confidence regions that did not allow for interpretation.   

\subsection{Mappers and confidence regions}

\begin{figure}
\begin{tabular}{ccc}
\includegraphics[width=4cm]{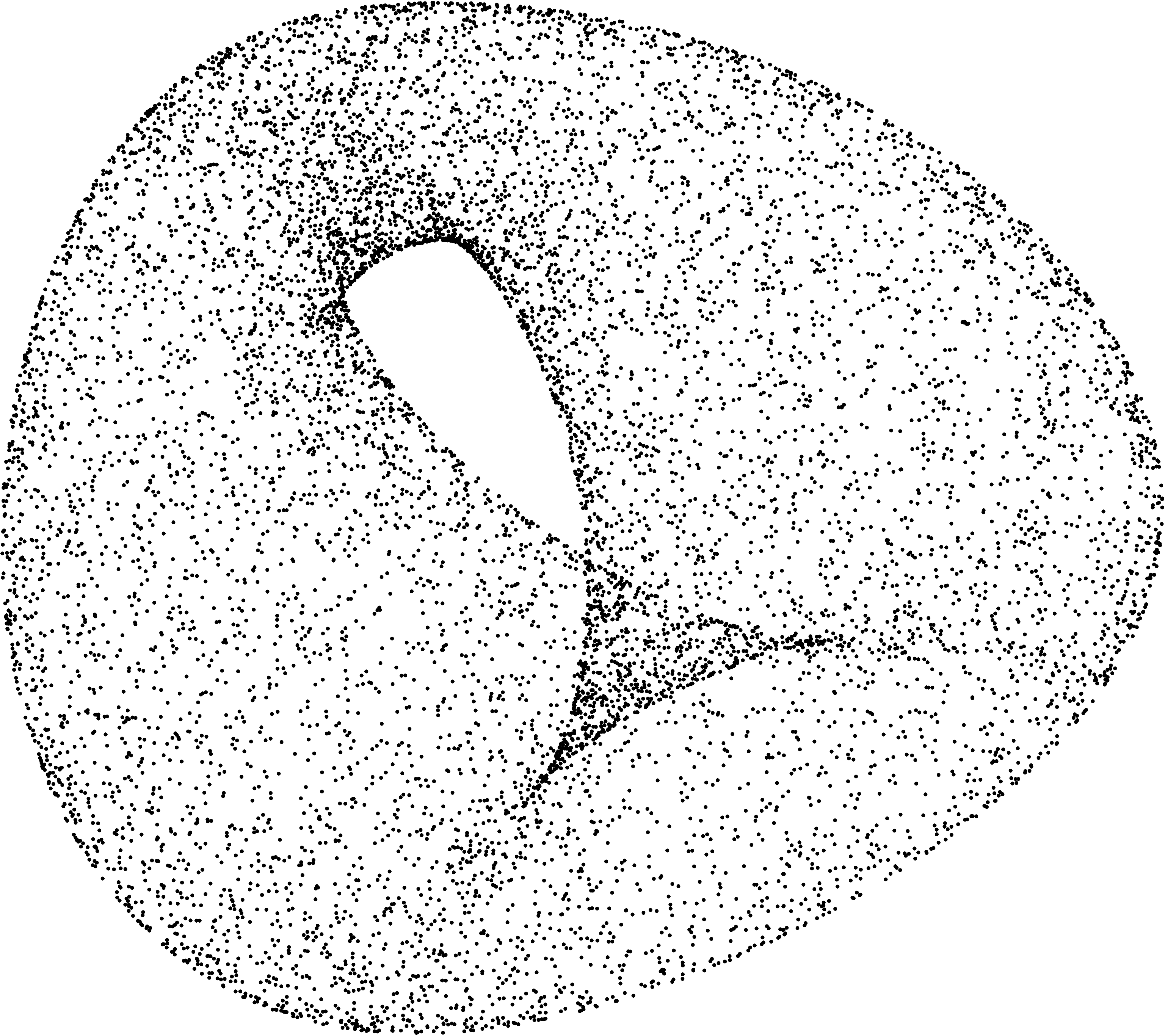} & \includegraphics[width=3cm]{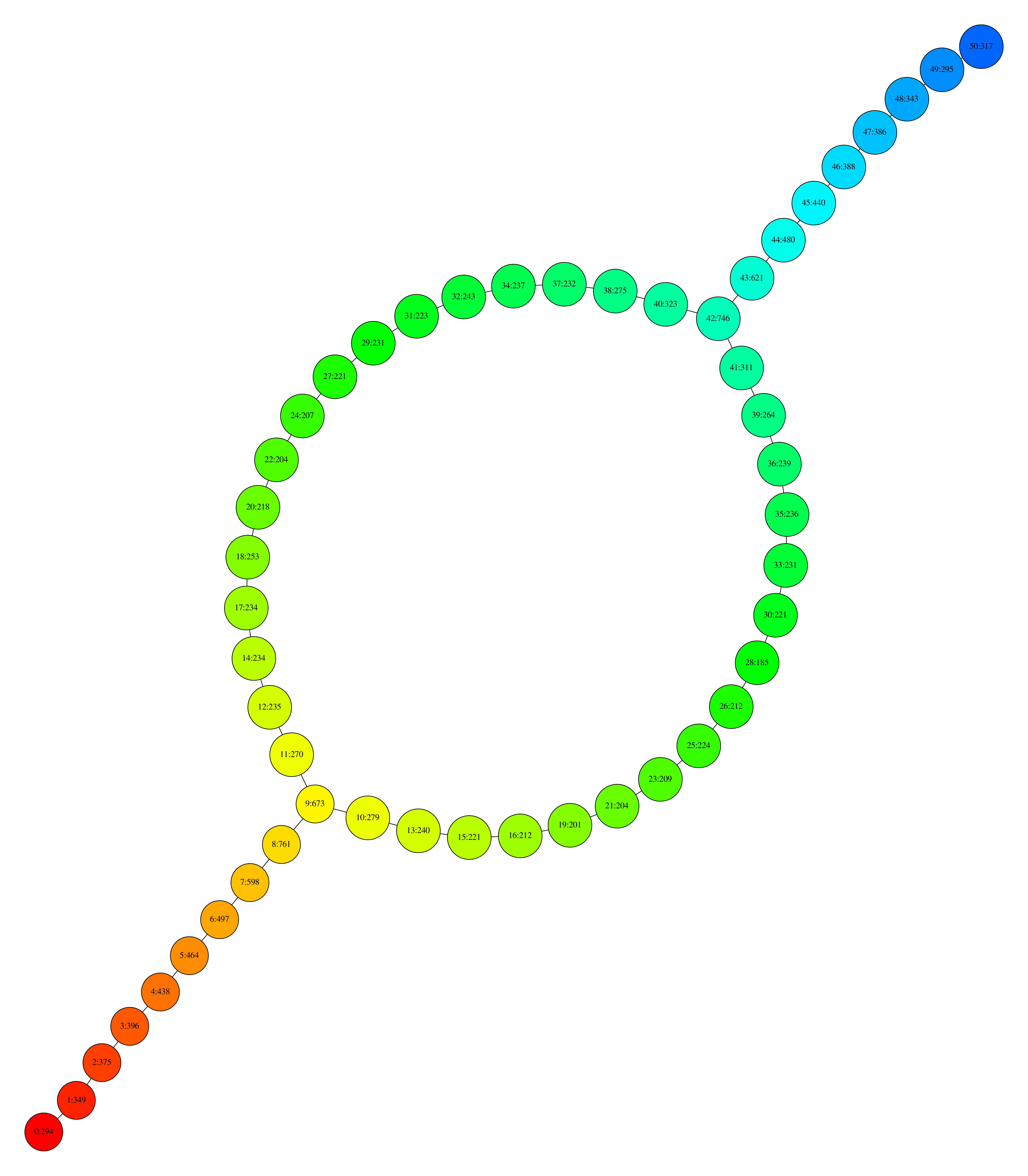} &  \includegraphics[width=5cm]{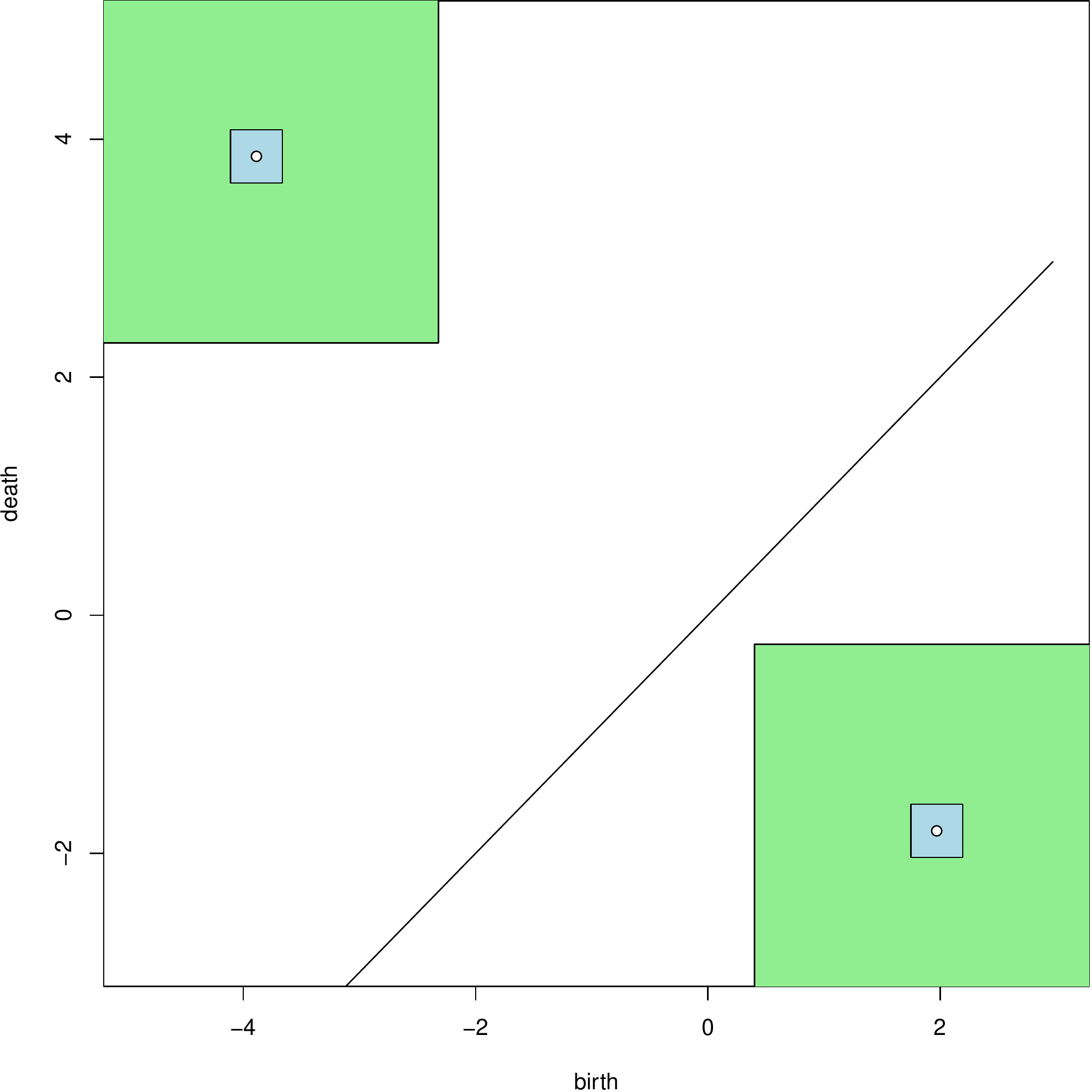} \\
\includegraphics[width=2.5cm]{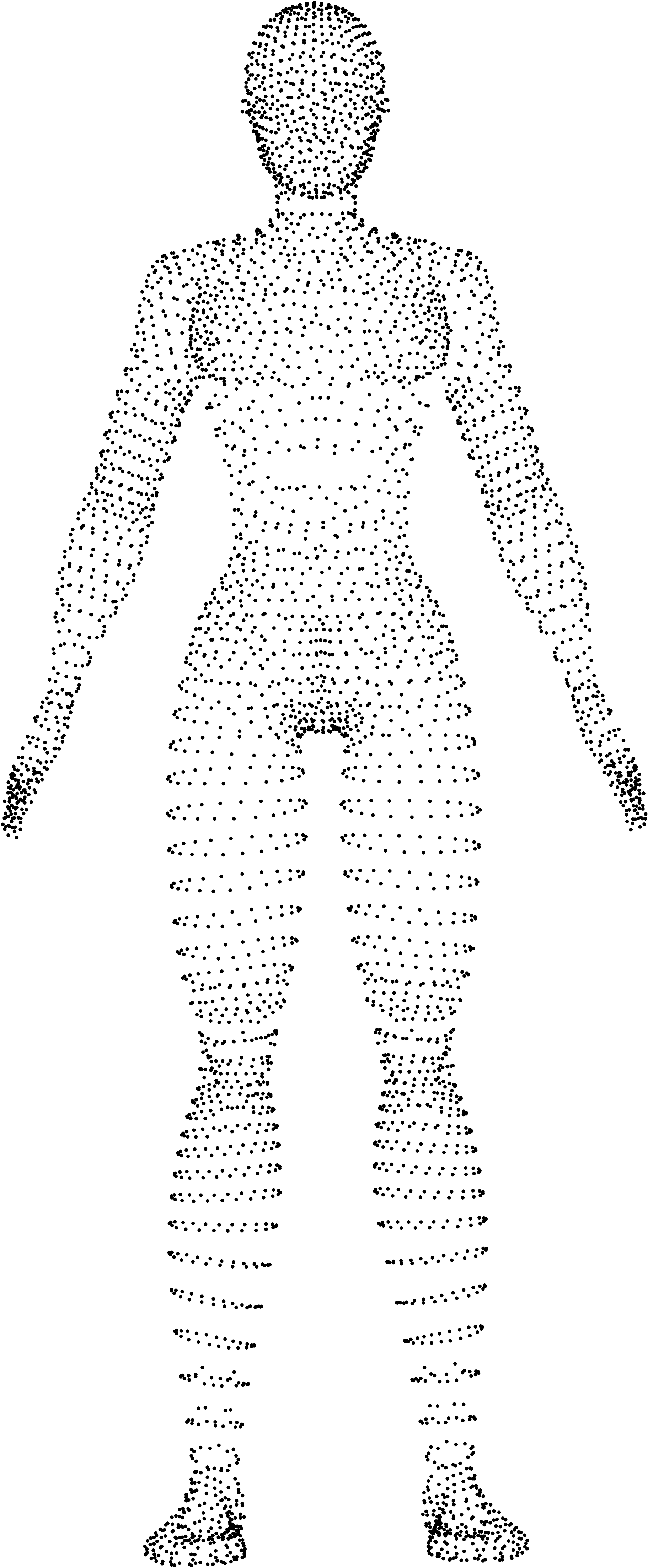} & \includegraphics[width=4.5cm]{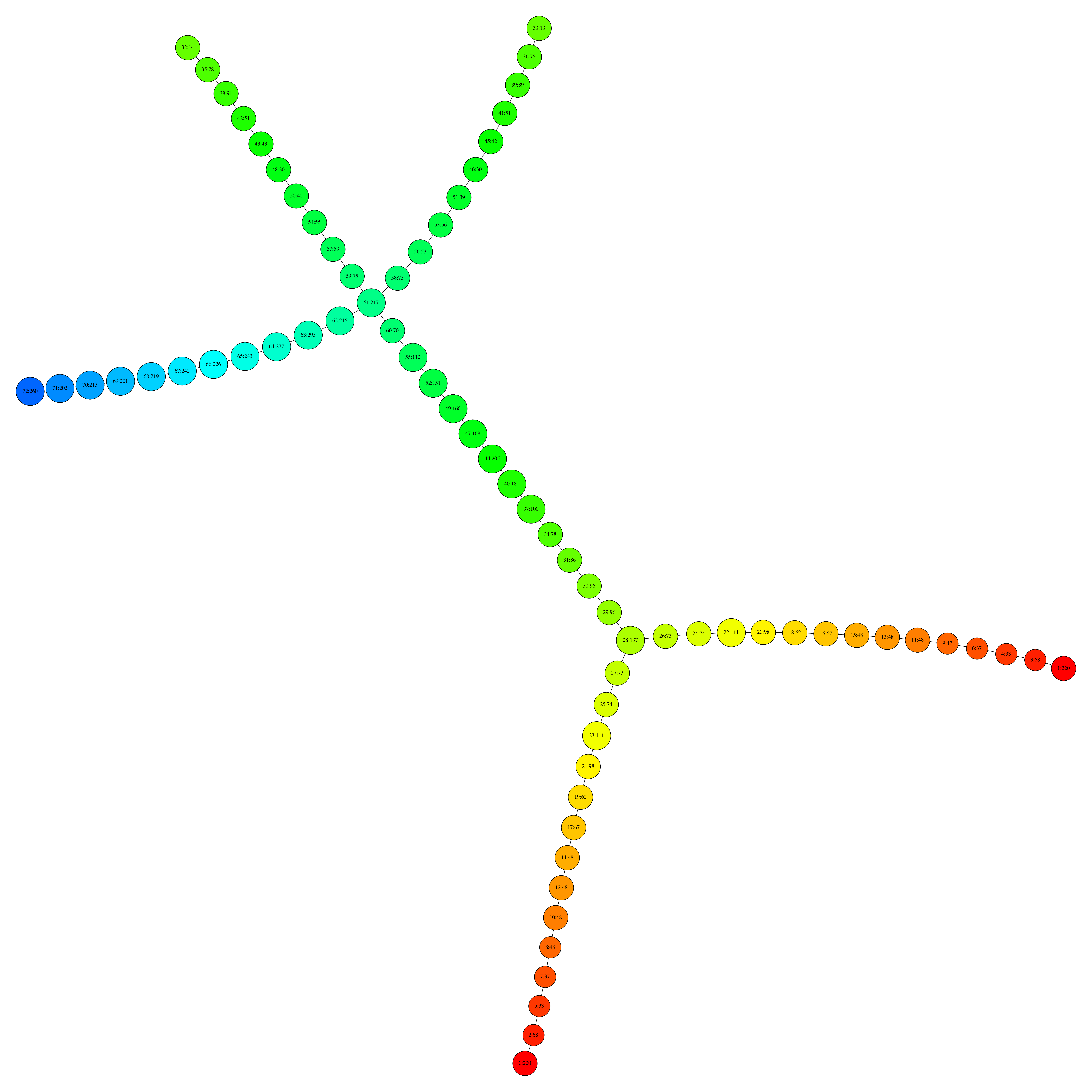} & \includegraphics[width=5cm]{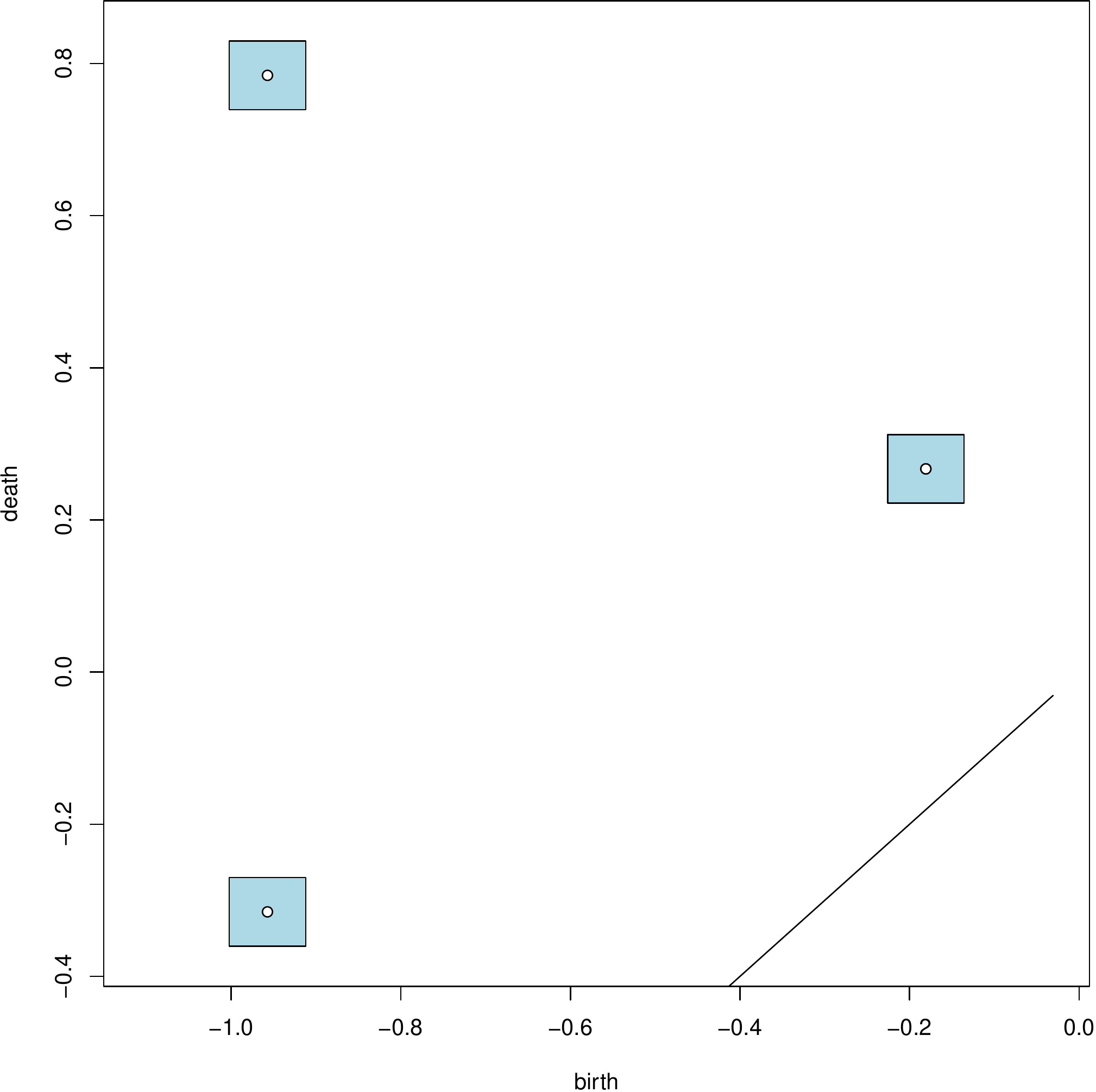} \\
\includegraphics[width=6.5cm]{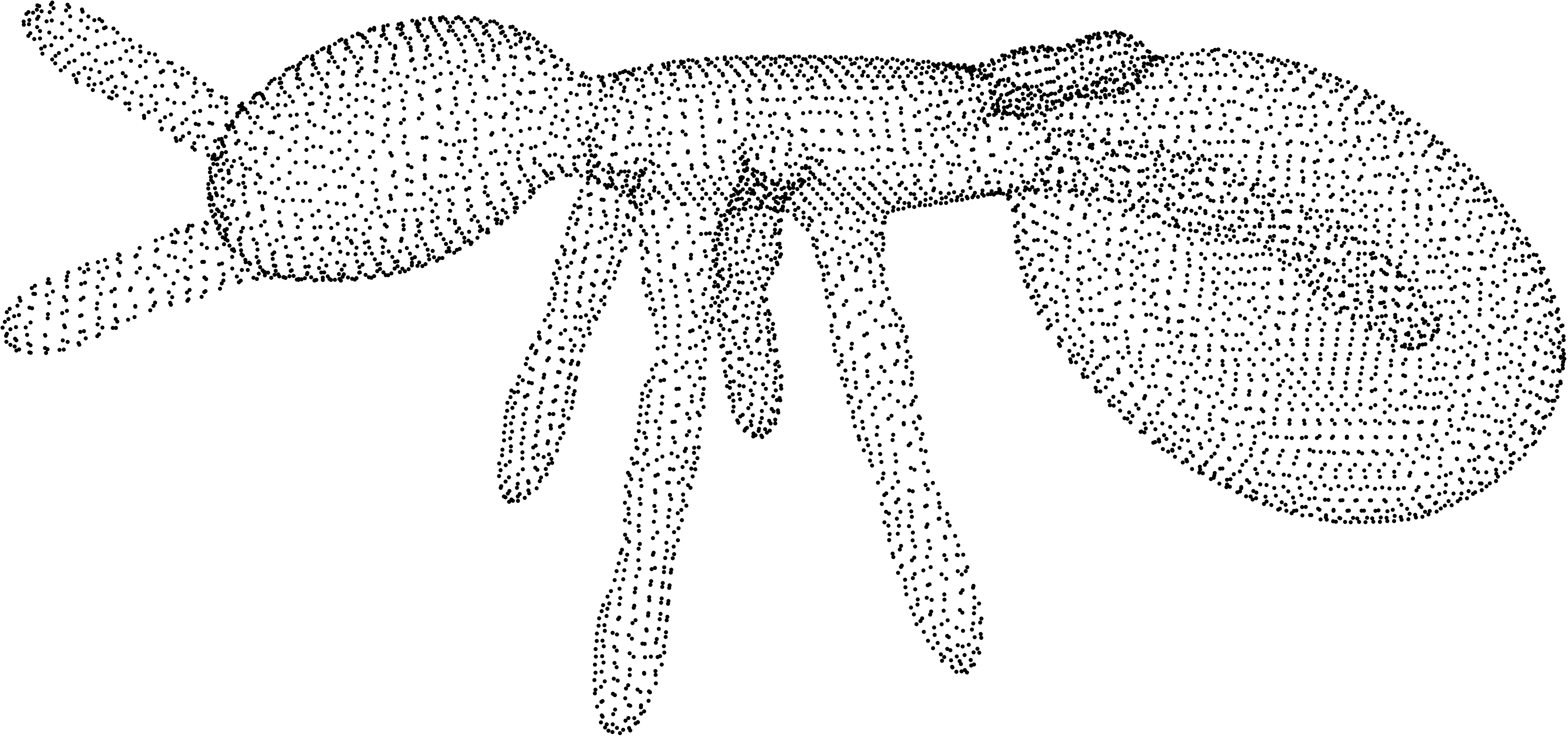} & \includegraphics[width=4.5cm]{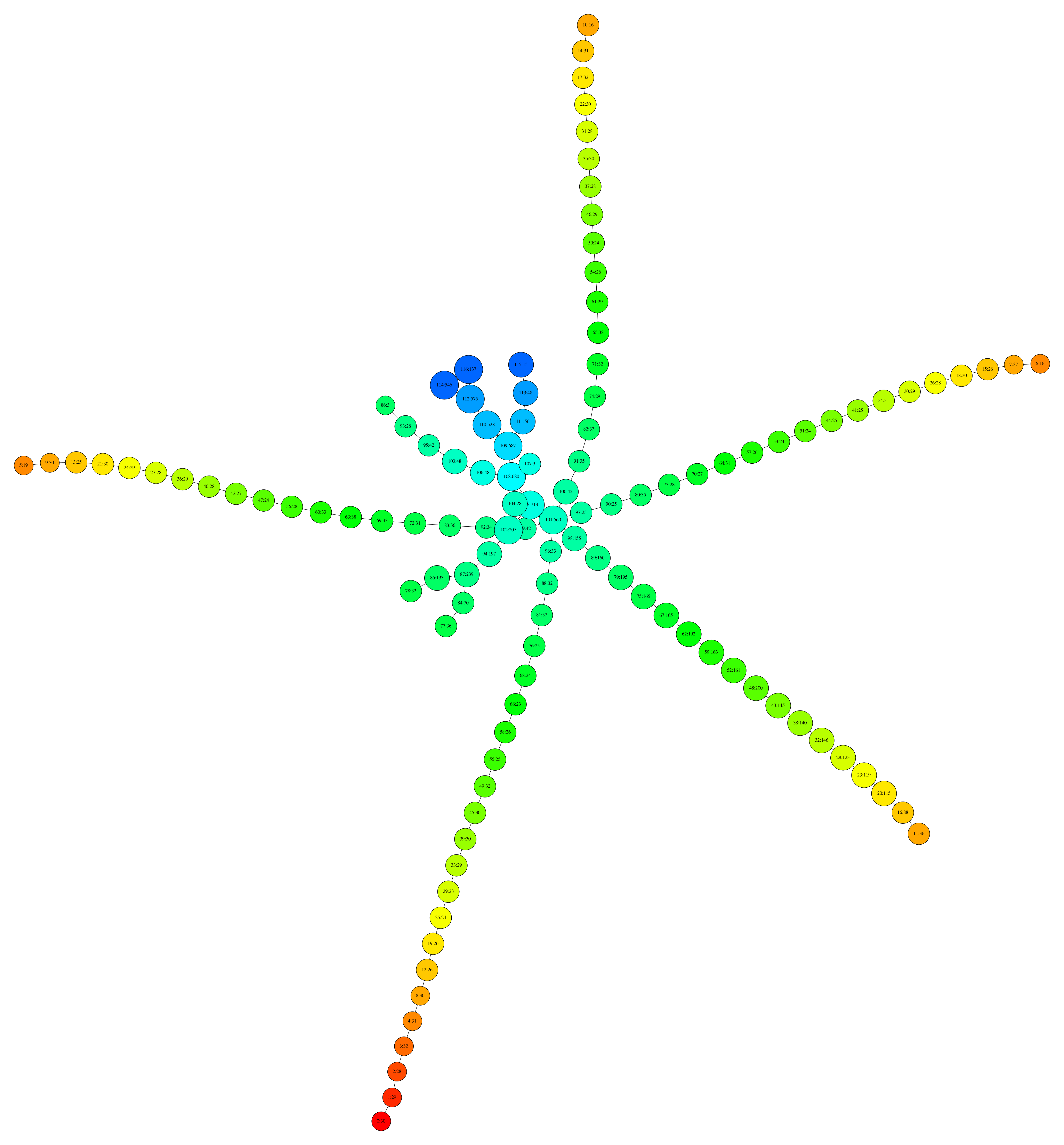} & \includegraphics[width=5cm]{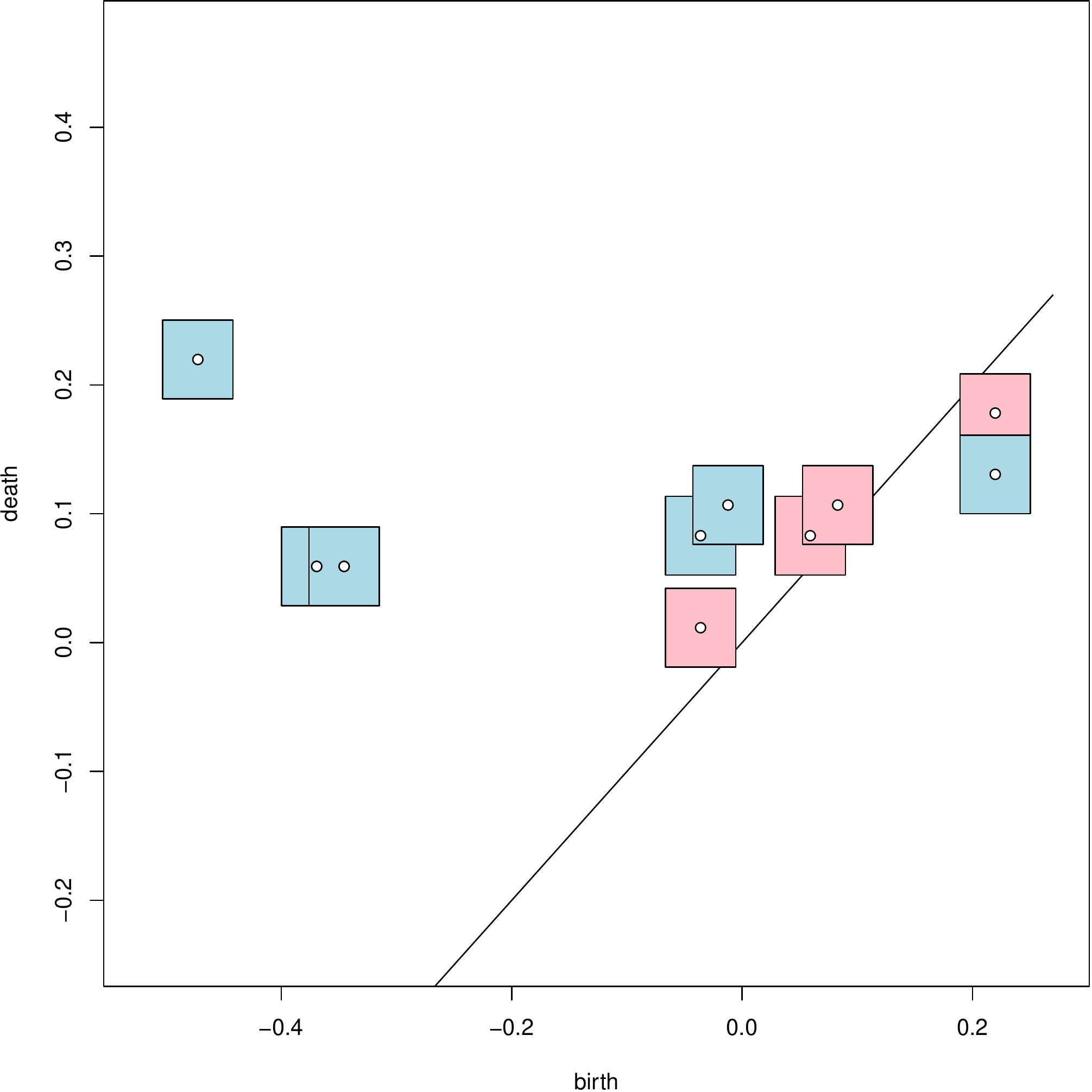} \\
\end{tabular}
\caption{\label{fig:synthdata} Mappers computed with automatic tuning (middle) and 85 percent confidence regions for their topological features (right) are provided for
an embedding of the Klein Bottle into $\R^4$ (first row), a 3D human shape (second row) and a 3D ant shape (third row). }
\end{figure}

\paragraph{Synthetic example.} We computed the Mapper of
an embedding of the Klein bottle into $\R^4$ with 10,000 points with the height function.
In order to illustrate the conservativity of confidence regions computed with Proposition~\ref{prop:conf1},
we also plot these regions for an embedding with 10,000,000 points using the fact that the height function is 1-Lipschitz. 
Corresponding squares are drawn in green color.
Their very large sizes show that Proposition~\ref{prop:conf1} requires a very large number of observations in practice.
See the first row of Figure~\ref{fig:synthdata}.

\paragraph{3D shapes.}
We computed the Mapper of an ant shape and a human shape from~\cite{Chen09} embedded in $\R^3$ (with 4,706 and 6,370 points respectively) 
Both Mappers were computed with the height function.
One can see that the confidence squares for the 
features that are almost horizontal (such as the 
small branches in the Mapper of the ant) intersect indeed the diagonal. 
See the second and third rows of Figure~\ref{fig:synthdata}.

\begin{figure}
\begin{tabular}{ccc}
\includegraphics[width=6.5cm]{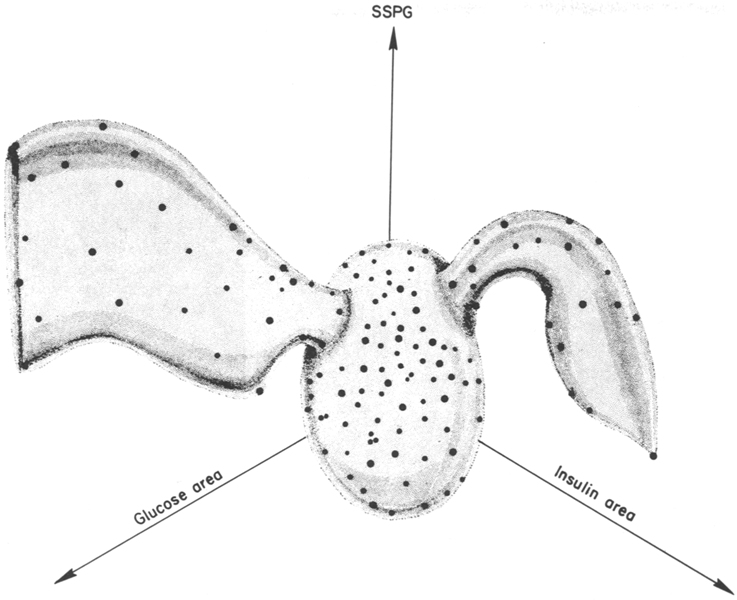} & \includegraphics[width=3cm]{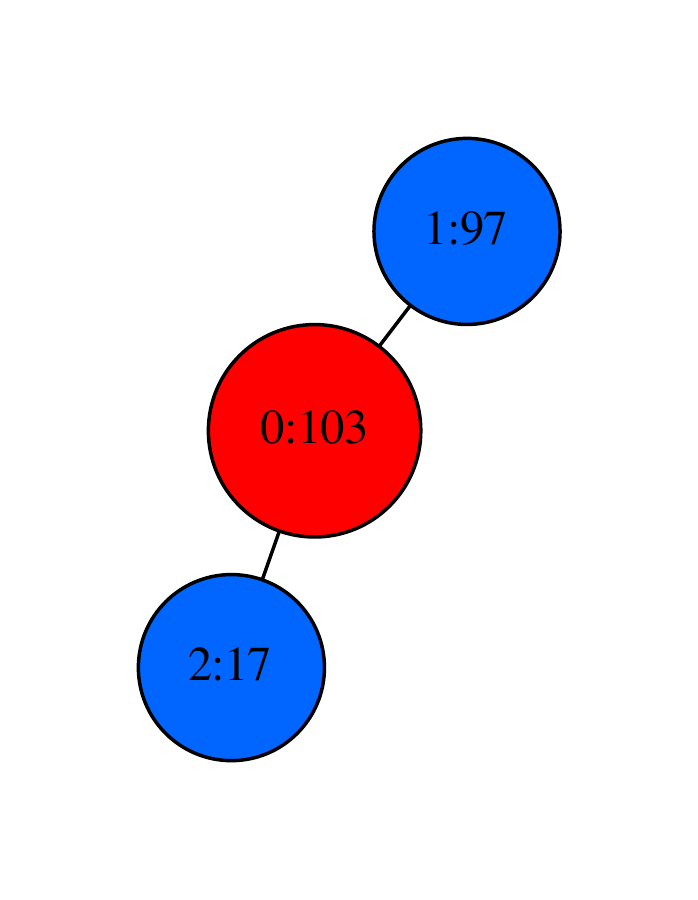} & \includegraphics[width=5cm]{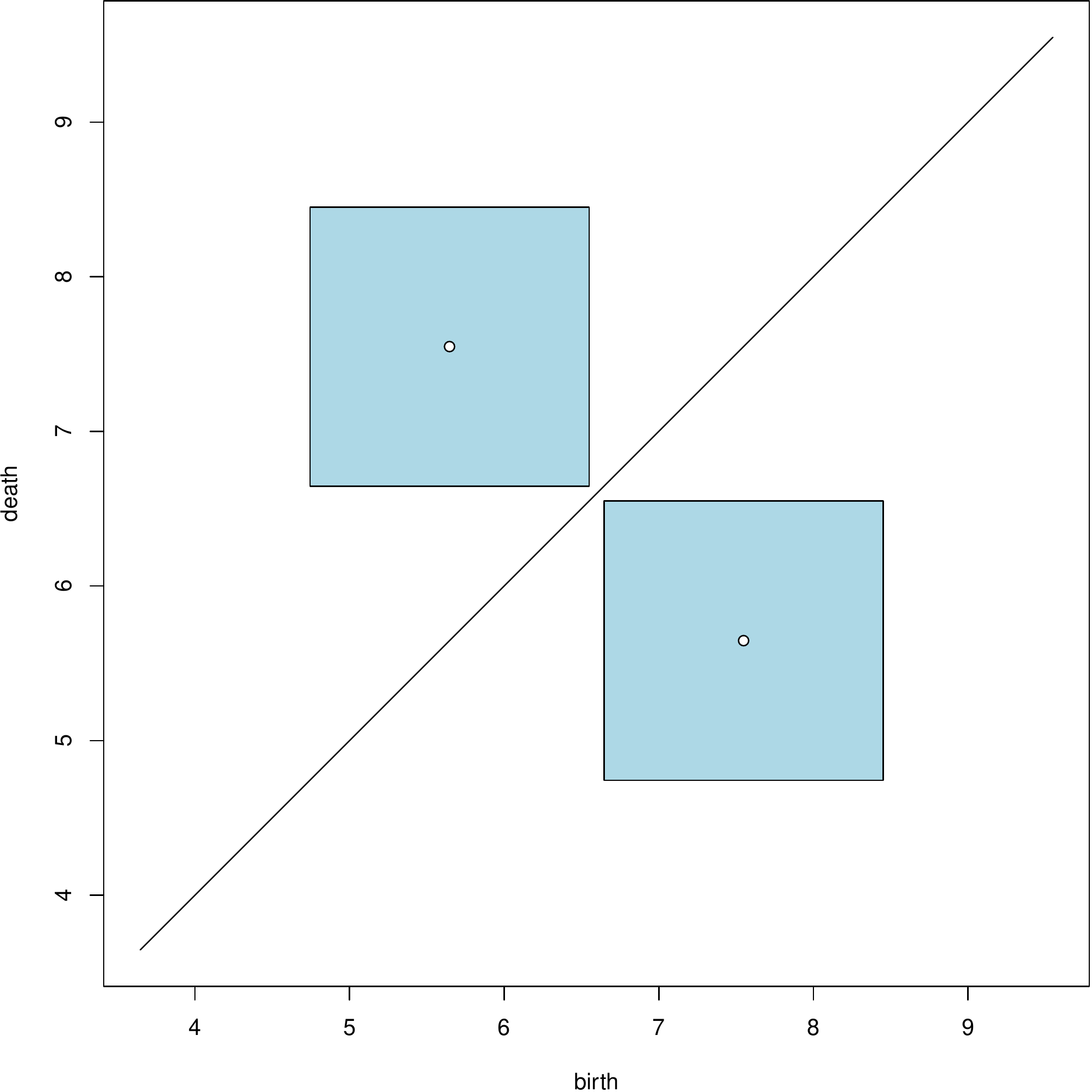} \\
\includegraphics[width=5cm]{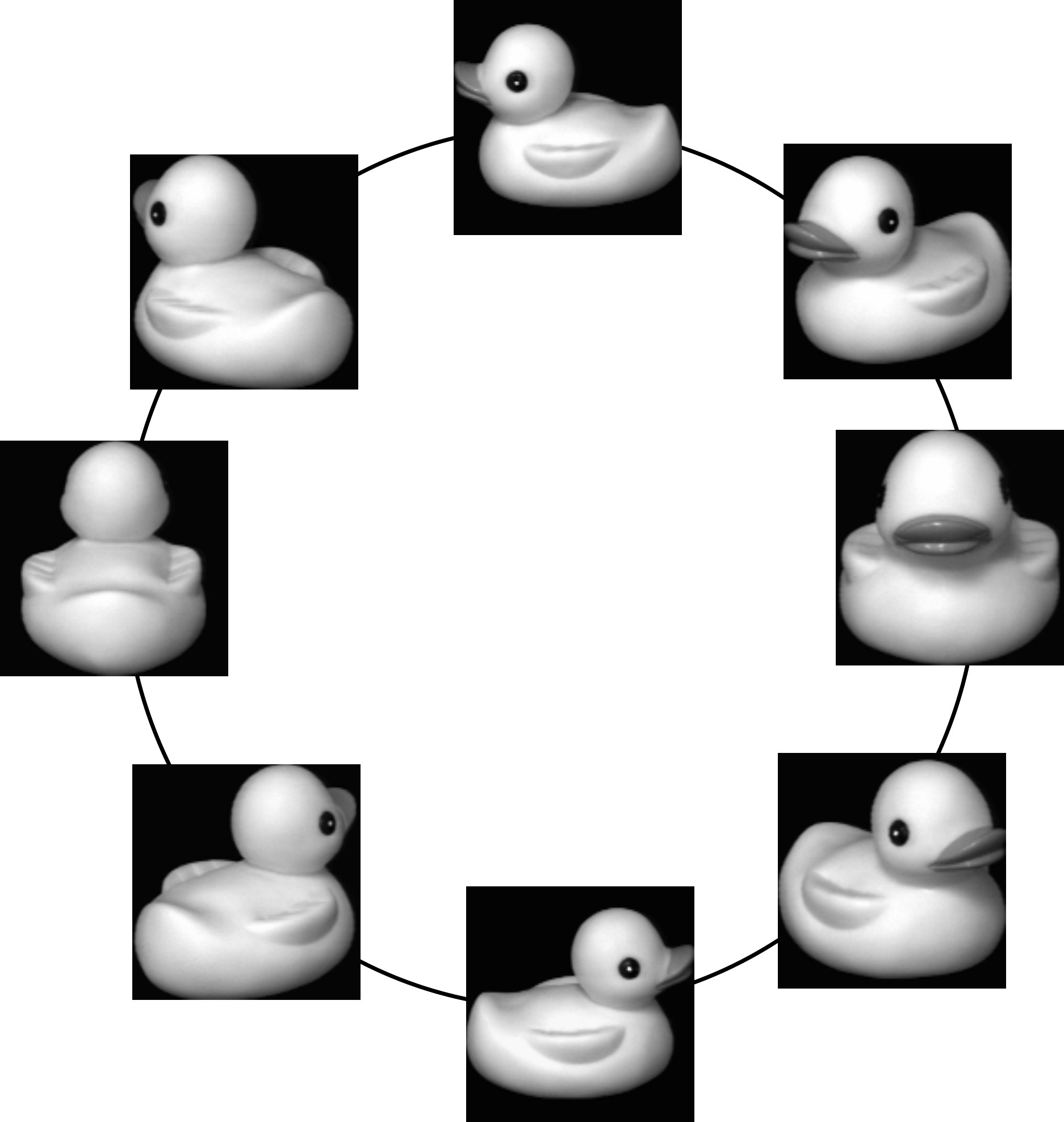} & \includegraphics[width=5cm]{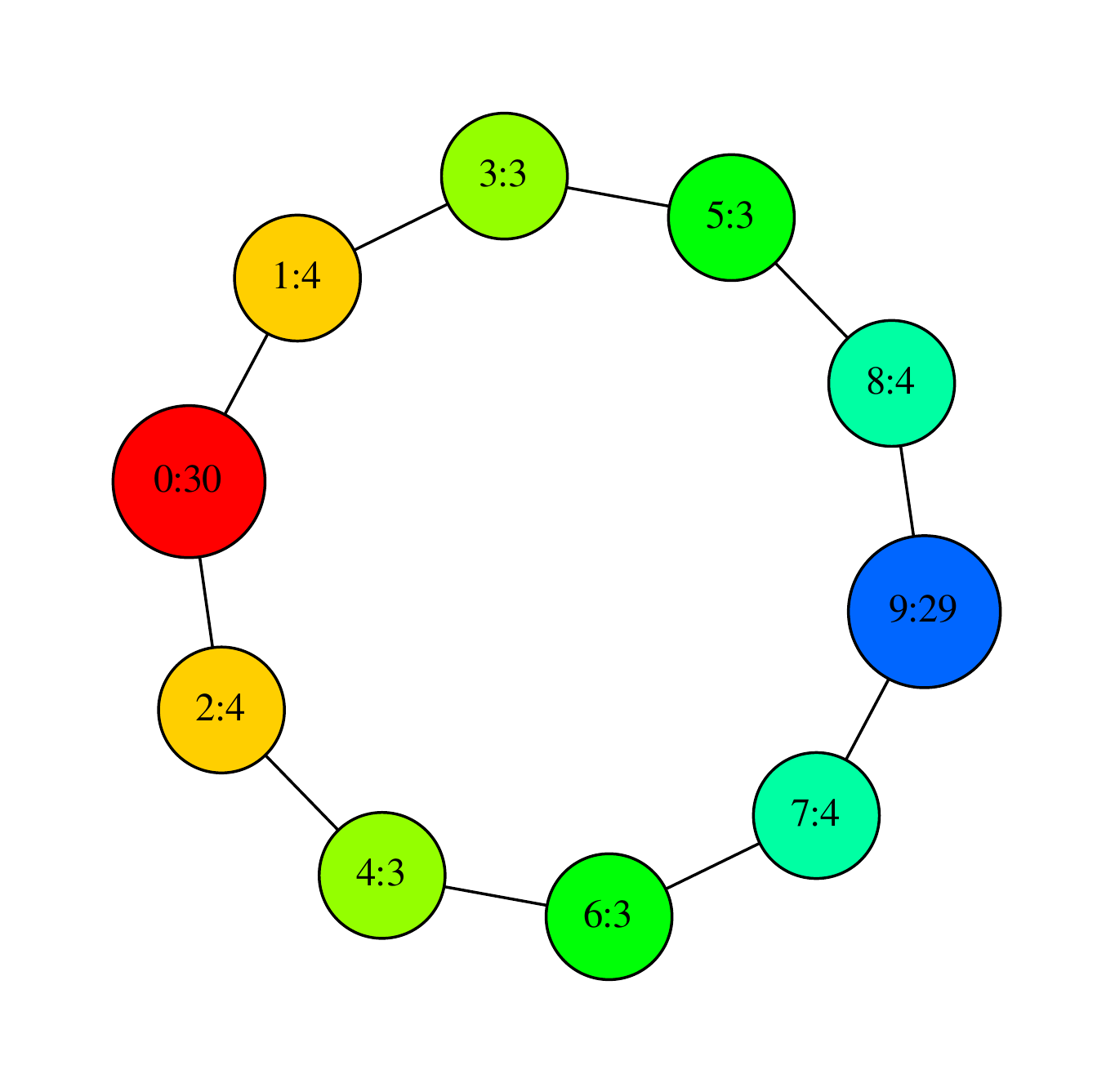} & \includegraphics[width=5cm]{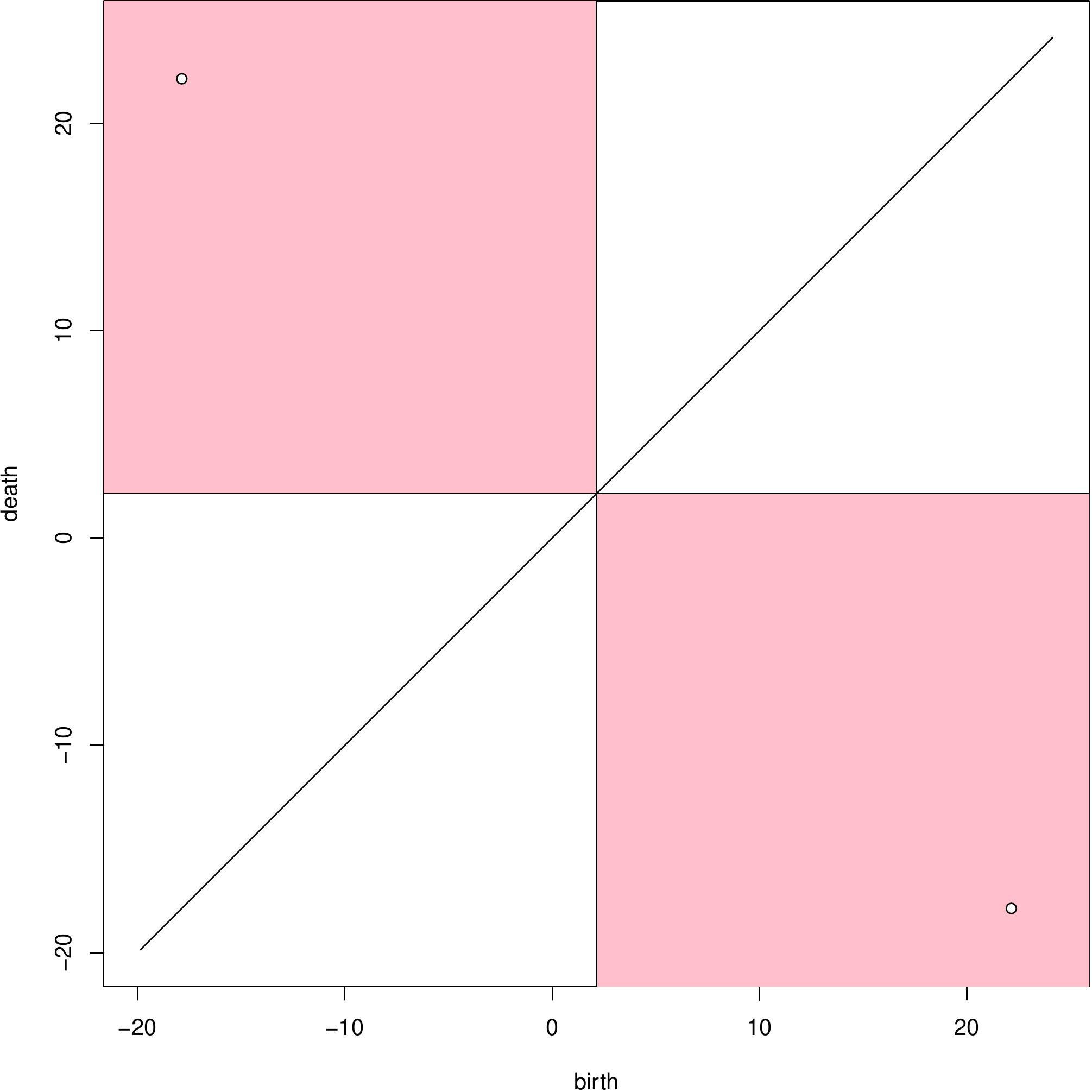} 
\end{tabular}
\caption{\label{fig:realapp} Mappers computed with automatic tuning (middle) and 85 percent confidence regions for their topological features (right) are provided for
 the Reaven-Miller dataset (first row) and the COIL dataset (second row). }

\end{figure}

\paragraph{Miller-Reaven dataset.} 
The first dataset comes from the Miller-Reaven diabetes study that contains 145 observations of patients suffering or not from diabete.
Observations were mapped into $\R^5$ by computing various medical features.
Data can be obtained in the ``locfit'' R-package. 
In~\cite{Reaven79}, the authors identified two groups of diseases
with the projection pursuit method, and in~\cite{Singh07}, the authors applied Mapper with hand-crafted parameters to get back this result.
Here, we normalized the data to zero mean and unit variance, and we obtained the two flares in the Mapper computed with the eccentricity function.
Moreover, these flares are at least 85 percent sure since the confidence squares on the corresponding
points in the extended persistence diagrams do not intersect the diagonal.
See the first row of Figure~\ref{fig:realapp}.

\paragraph{COIL dataset.}
The second dataset is an instance of the 16,384-dimensional COIL dataset~\cite{Nene96}. 
It contains 72 observations, each of which being a picture of a duck taken at a specific angle.
Despite the low number of observations and the large number of dimensions, we managed to retrieve the intrinsic 
loop lying in the data using the first PCA eigenfunction. However, the low number of observations made the bootstrap fail 
since the confidence squares computed around 
the points that represent this loop in the extended persistence diagram intersect the diagonal.
See the second row of Figure~\ref{fig:realapp}.

\subsection{Noisy data}

\begin{figure}
\begin{tabular}{ccc}
\includegraphics[width=6.5cm]{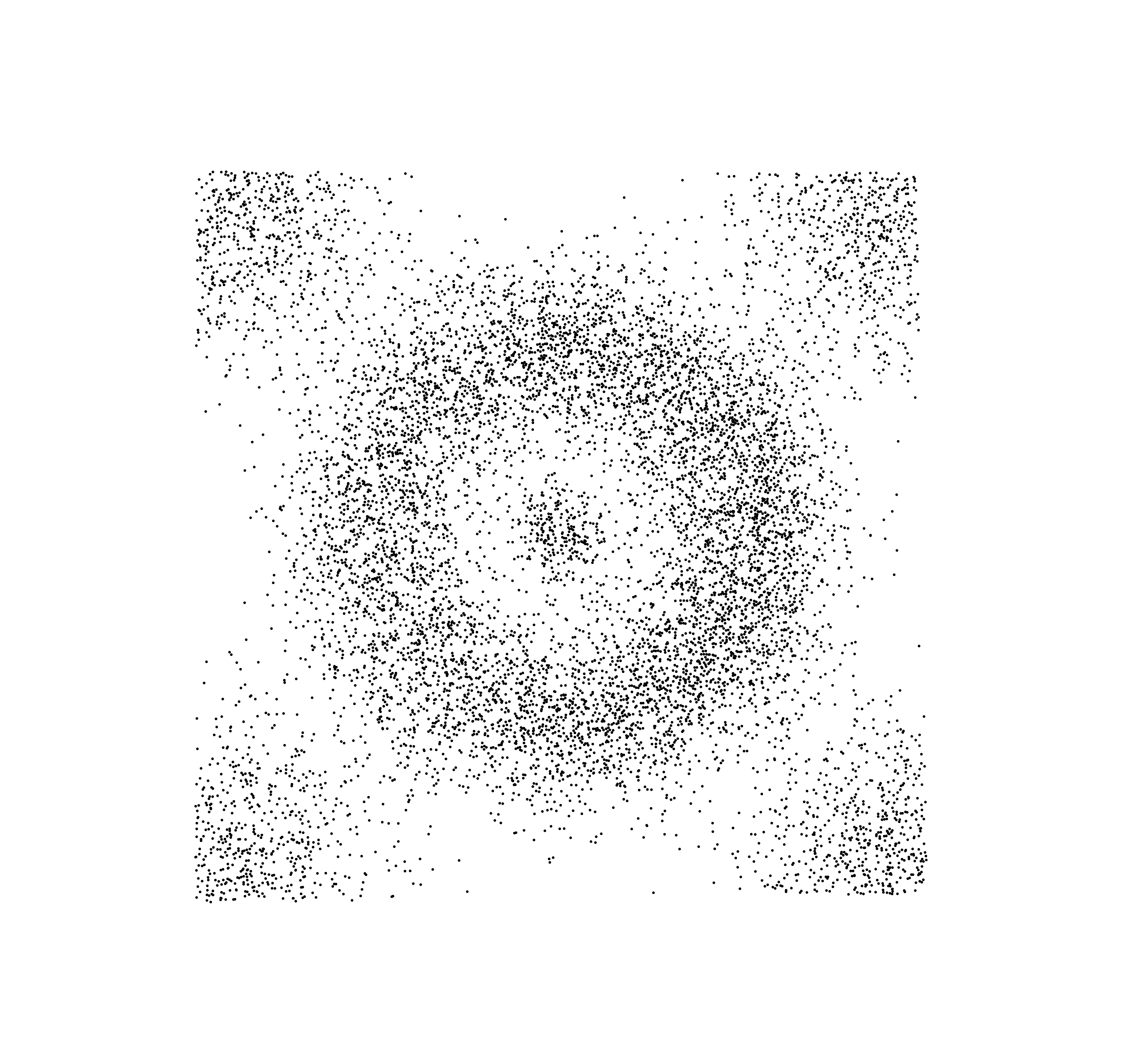} &\includegraphics[width=3.5cm]{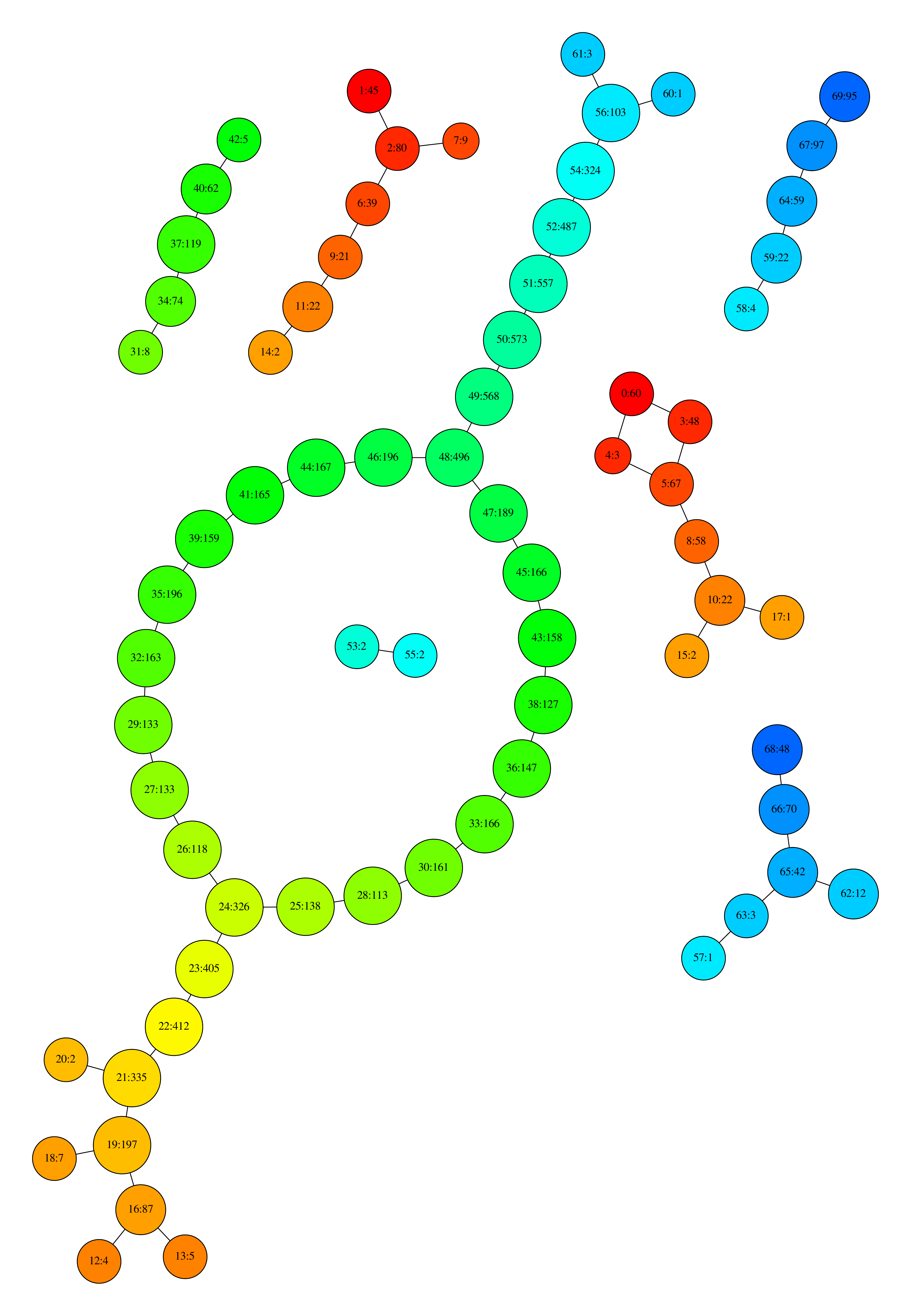} & \includegraphics[width=5cm]{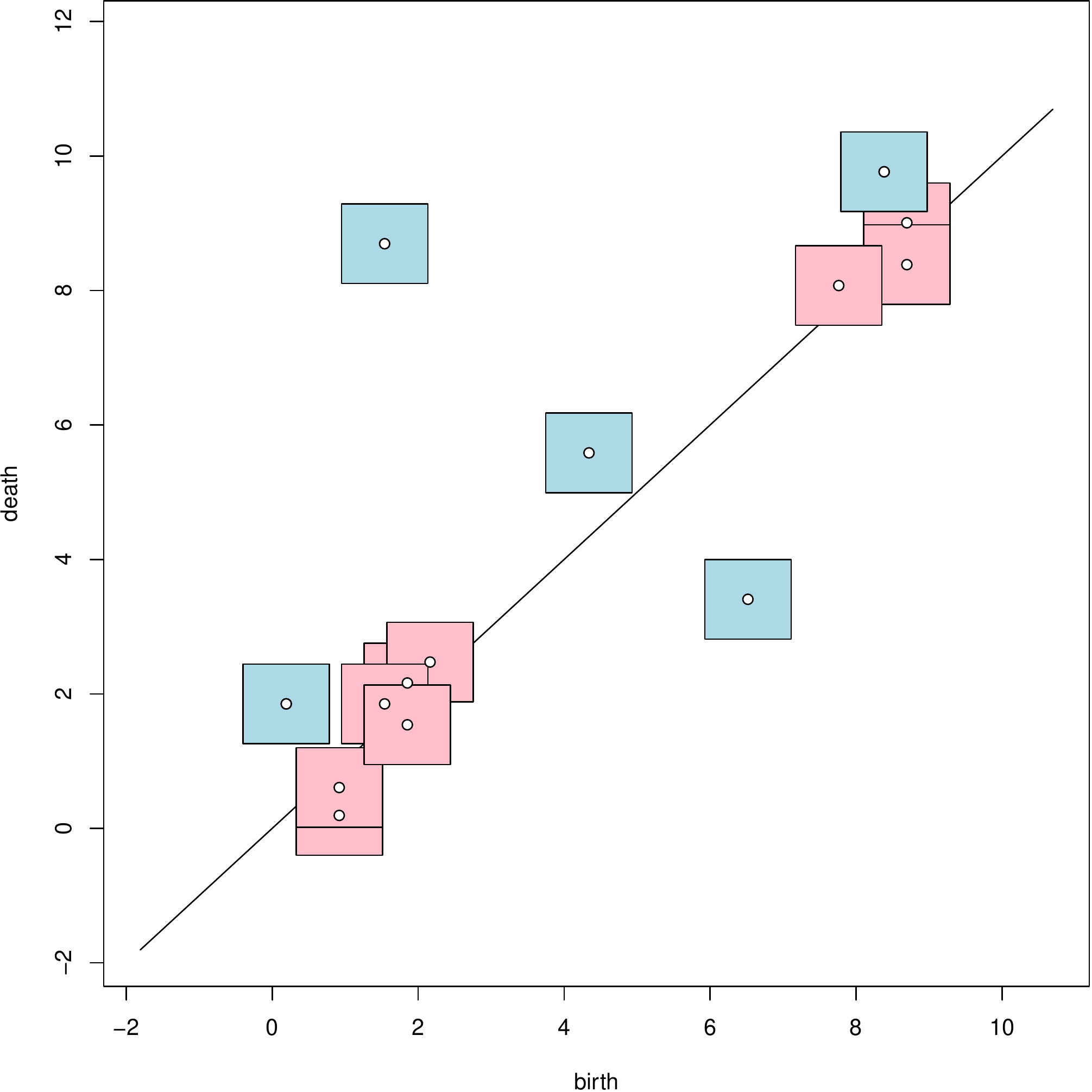} \\
\end{tabular}
\caption{\label{fig:noisy} Mappers computed with automatic tuning (middle) and 85 percent confidence regions for their topological features (right) are provided for
a a noisy crater in the Euclidean plane. }
\end{figure}

\paragraph{Denoising Mapper.}
An important drawback of Mapper is its sensitivity to noise and outliers.
See the crater dataset in Figure~\ref{fig:noisy}, for instance.
Several answers have been proposed for recovering the correct persistence homology from noisy data. 
The idea is to use an alternative filtration of simplical compexes instead of the Rips filtration. 
A first option is to consider the upper level sets of a density estimator rather then the 
distance to the sample (see Section  4.4 in \cite{Fasy14}). Another solution is to 
consider the sublevel sets of the DTM
and apply persistence homology inference in~\cite{Chazal14b}. 

\paragraph{Crater dataset.} To handle noise in our crater dataset, we simply smoothed the dataset by computing the empirical DTM with 10 neighbors on each point
and removing all points with DTM less than 40 percent of the maximum DTM in the dataset. Then we computed the Mapper with the height function.
One can see that all topological features in the Mapper that are most likely artifacts due to noise (like the small loops and connected components)
have corresponding confidence squares that intersect the diagonal in the extended persistence diagram. See Figure~\ref{fig:noisy}.

\section{Conclusion}

In this article, we provided a statistical analysis of the Mapper. Namely, we proved the fact that the Mapper is
a measurable construction in Proposition~\ref{prop:Measurability}, and we used 
the approximation Theorem~\ref{thm:geomineq} to show that the Mapper is a minimax optimal 
estimator of the Reeb graph in various contexts---see Propositions~\ref{prop:UpBdRestr},~\ref{prop:lecam} and~\ref{prop:subs}---and
that corresponding confidence regions can be computed---see Proposition~\ref{prop:conf1} and Section~\ref{sec:bootstrap}.
Along the way, we derived rules of thumb to automatically tune the parameters of the Mapper with Equation~(\ref{ref:coeff_ab_inconnus}). 
Finally, we provided few examples of our methods on various datasets in Section~\ref{sec:appli}.

\paragraph{Future directions.}
We plan to investigate several questions for future work.
\begin{itemize}

\item We will work on adapting results from~\cite{Chazal14b} to  
prove the validity of bootstrap methods for computing confidence regions on the Mapper,
since we only used bootstrap methods empirically in this article. 

\item We believe that using weighted Rips complexes~\cite{Buchet15} instead of the usual Rips complexes would
improve the quality of the confidence regions on the Mapper features,
and would probably be a better way to deal with noise that our current solution.

\item We plan to adapt our statistical setting to the question of selecting variables,
which is one of the main applications of the Mapper in practice.


\end{itemize}

\appendix

\section{Proofs}\label{sec:proofs}

\subsection{Preliminary results}

In order to prove the results of this article, we need
to state several preliminary definitions and theorems. 
All of them can be found, together with their proofs, in~\cite{Dey13a} and~\cite{Carriere17b}.
In this section, we let $\Xs_n\subset\mathcal X$ be a point cloud of $n$ points sampled on a smooth and compact  submanifold $\mathcal X$ embedded in $\R^D$,
with positive reach $rch$ and convexity radius $\rho$. 
Let $f:\mathcal X \rightarrow\R$ be a Morse-type filter function, 
$\I$ be a minimal open cover of the range of $f$
with resolution $r$ and gain $g$, $|\Rips_\delta(\Xs_n)|$ denote a geometric realization of the Rips
complex built on top of $\Xs_n$ with parameter $\delta$, and $\frips f:|\Rips_\delta(\Xs_n)|\rightarrow\R$ be the piecewise-linear
interpolation of $f$ on the simplices of $\Rips_\delta(\Xs_n)$.

\begin{defin}
Let $G=(\Xs_n,E)$ be a graph built on top of $\Xs_n$. Let $e=(X,X')\in E$ be an edge of $G$,
and let $I(e)$ be the open interval $(\min\{f(X),f(X')\},\max\{f(X),f(X')\})$. Then
%
$e$ is said to be {\em intersection-crossing} if there is a pair of consecutive
intervals $I,J\in\I$ such that $\emptyset\neq I\cap J \subseteq I(e).$
%
\end{defin}

\begin{thm}[Lemma 8.1 and 8.2 in~\cite{Carriere17b}]\label{th:Xcross}
Let $\Rips^1_\delta(\Xs_n)$ denote the 1-skeleton of $\Rips_\delta(\Xs_n)$.
If $\Rips^1_\delta(\Xs_n)$ has no intersection-crossing edges, then $\Map_{r,g,\delta}(\Xs_n,f(\Xs_n))$ and $\Map_{r,g}(|\Rips_\delta(\Xs_n)|,\frips f)$
are isomorphic as combinatorial graphs. 
\end{thm}


\begin{thm}[Theorem 7.2 and 5.1 in~\cite{Carriere17b}]\label{th:bottleReebMapp}
Let $f:\Xset\rightarrow\R$ be a Morse-type function. 
Then, we have the following inequality between extended persistence diagrams:
\begin{equation}\label{eq:stair}
d_\Delta(  \Dg( \Reeb_f(\mathcal X),\freeb f ),
\Dg( \Map_{r,g}(\mathcal X,f), \fmapp f ) )\leq r.
\end{equation}

Moreover, given another Morse-type function $\hat f:\Xset\rightarrow\R$, we have the following inequality:
\begin{equation}\label{eq:stab}
d_\Delta(  \Dg(\Map_{r,g}(\Xset,f), \fmapp f ) ,
\Dg( \Map_{r,g}(\Xset,\hat f), \fmapp{\hat f} ) )\leq r + \|f-\hat f\|_\infty.
\end{equation}
\end{thm}

\begin{thm}[Theorem 4.6 and Remark~2 in~\cite{Dey13a} and Theorem 8.3 in~\cite{Carriere17b}]\label{th:Reebapprox}
If $4 d_{\rm H}(\Xset,\Xs_n)\leq \delta\leq\min\{rch/4,\rho/4\}$, 
then:
$$d_\Delta(\Dg(\Reeb_f(\mathcal X),\freeb f),\Dg(\Reeb_{\frips f}(|\Rips_\delta(\Xs_n)|),\freeb{ \frips f } ))\leq 2\omega(\delta).$$
\end{thm}

Note that the original version of this theorem is only proven for Lipschitz functions in~\cite{Dey13a}, but it extends at no cost
to functions with modulus of continuity.

\subsection{Proof of Theorem~\ref{thm:geomineq} }
\label{sec:proofthgeomineq}



Let $|\Rips_\delta(\Xs_n)|$ denote a geometric realization of the Rips
complex built on top of $\Xs_n$ with parameter $\delta$. Moreover, let $\frips{f}:|\Rips_\delta(\Xs_n)|\rightarrow\R$ be the piecewise-linear
interpolation of $f$ on the simplices of $\Rips_\delta(\Xs_n)$, whose 1-skeleton is denoted by $\Rips^1_\delta(\Xs_n)$. 
Since $(|\Rips_\delta(\Xs_n)|,\frips{f})$ is a metric space, we also consider its Reeb graph $\Reeb_{\frips{f}}(|\Rips_\delta(\Xs_n)|)$,
with induced function $\freeb{\frips{f}}$, and its Mapper $\Map_{r,g}(|\Rips_\delta(\Xs_n)|,\frips{f})$, with induced function $\fmapp{\frips{f}}$.
See Figure~\ref{fig:recap}. Then, the following inequalities lead to the result:

\begin{figure}\centering
\includegraphics[width=15cm]{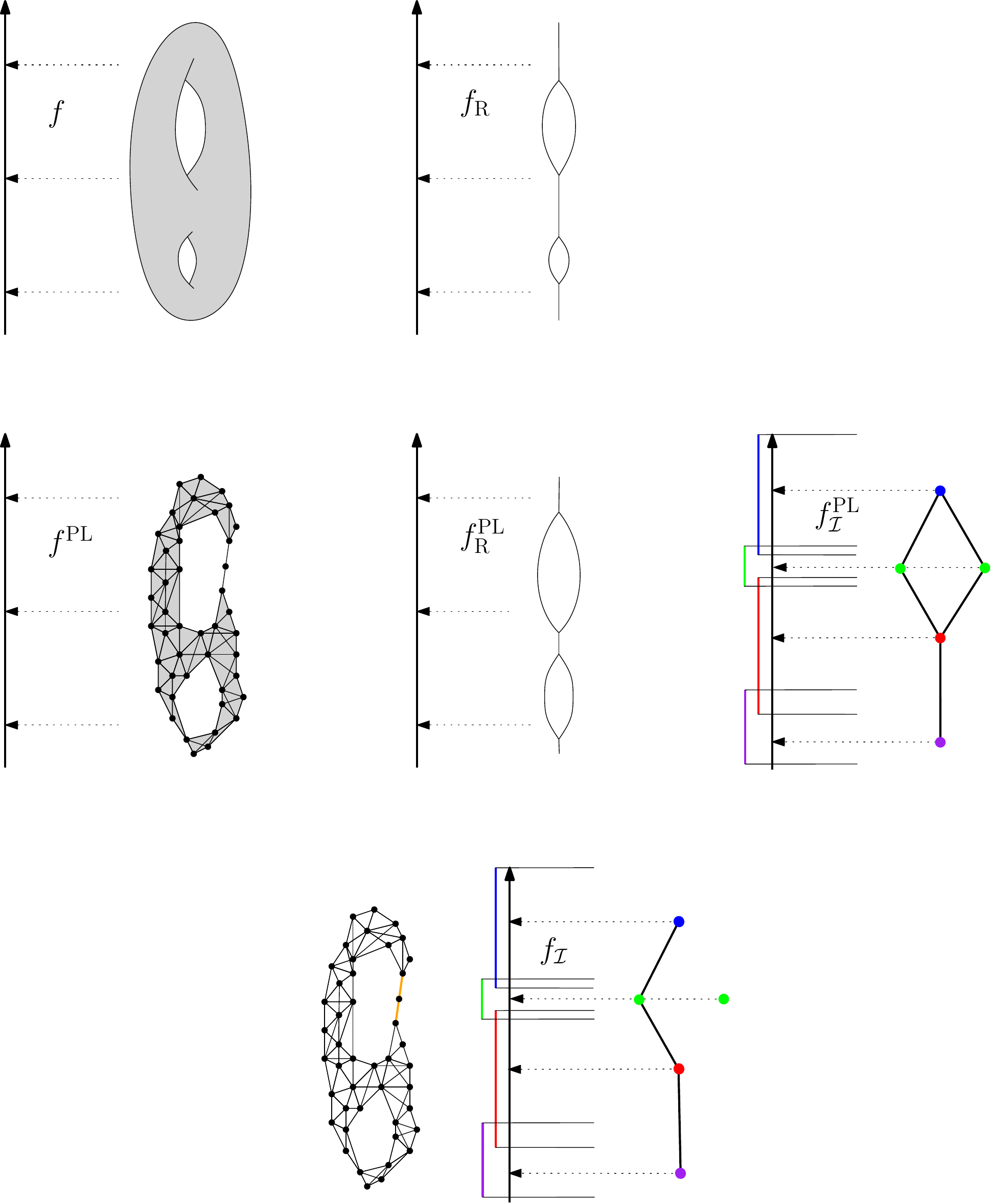}
\caption{\label{fig:recap}
Examples of the function defined on the original space (left column), its induced function defined on the Reeb graph (middle column) and 
the function defined on the Mapper (right column).
Note that the Mapper computed from the geometric realization of the Rips complex (middle row, right)
is not isomorphic to the standard Mapper (last row), since there are two intersection-crossing edges
in the Rips complex (outlined in orange).}
\end{figure}

\begin{align}
\dper&(\Reeb_f(\Xset),\Map_n)
=d_\Delta\left(\Dg(\Reeb_f(\Xset),\freeb{f}),\Dg(\Map_n,\fmapp{f})\right)\nonumber\\
&=d_\Delta\left(\Dg(\Reeb_f(\Xset),\freeb{f}),\Dg(\Map_{r,g}(|\Rips_\delta(\Xs_n)|,\frips f),\fmapp{\frips{f}})\right) \label{ineq:disc} \\
%
&\leq d_\Delta\left(\Dg(\Reeb_f(\Xset),\freeb{f})
,\Dg(\Reeb_{\frips{f}}(|\Rips_\delta(\Xs_n)|),\freeb{\frips{f}})
\right)\nonumber\\
&\ \ \ + d_\Delta\left(\Dg(\Reeb_{\frips{f}}(|\Rips_\delta(\Xs_n)|),\freeb{\frips{f}}),
\Dg(\Map_{r,g}(|\Rips_\delta(\Xs_n)|,\frips f),\fmapp{\frips{f}})\right)
\label{ineq:tri}\\
%
%
&\leq  2\omega(\delta)+ r .
\label{ineq:th}
\end{align}

Let us prove every (in)equality: \\

{\bf Equality~(\ref{ineq:disc}).}
Let $X_1,X_2\in \Xs_n$ such that $(X_1,X_2)$ is an edge of $\Rips^1_\delta(\Xs_n)$ 
i.e. $\|X_1-X_2\|\leq \delta$. Then, according to~(\ref{CdtB}): $|f(X_1)-f(X_2)| < gr$. 
Hence, there is no $s\in\{1,\dots,S-1\}$ such that $I_s\cap I_{s+1}\subseteq [\min\{f(X_1),f(X_2)\},\max\{f(X_1),f(X_2)\}]$.
It follows that there are no intersection-crossing edges in $\Rips^1_\delta(\Xs_n)$.
Then, according to Theorem~\ref{th:Xcross},
there is a graph isomorphism $i:\Map_n=\Map_{r,g,\delta}(\Xs_n,f(\Ys_n))\rightarrow\Map_{r,g}(|\Rips_\delta(\Xs_n)|,\frips f)$.
Since $\fmapp{f}=\fmapp{\frips{f}}\circ i$ by definition of $\fmapp{f}$ and $\fmapp{\frips{f}}$, the equality follows. \\

{\bf Inequality~(\ref{ineq:tri}).}
This inequality is just an application of the triangle inequality. \\

%


{\bf Inequality~(\ref{ineq:th}).}
According to~(\ref{CdtA}), we have 
$\delta \leq \min\{rch/4,\rho/4\}$. 
According to~(\ref{CdtD}), we also have $\delta\geq 4 d_{\rm H}(\Xset,\Xs_n)$.
Hence, we have 
$$d_\Delta(\Dg(\Reeb_f(\Xset),\freeb{f}),\Dg(\Reeb_{\frips{f}}(|\Rips_\delta(\Xs_n)|),\freeb{\frips{f}}))\leq 2 \omega(\delta),$$
according to Theorem~\ref{th:Reebapprox}.
Moreover, we have 
$$d_\Delta(  \Dg( \Reeb_{\frips{f}}(|\Rips_\delta(\Xs_n)|),\freeb{\frips{f}} ),
\Dg( \Map_{r,g}(|\Rips_\delta(\Xs_n)|,\frips f),\fmapp{\frips{f}} ) )\leq r,$$
according to Equation~(\ref{eq:stair}).

\subsection{Proof of Corollary \ref{cor:approxfilter}}

Let $|\Rips_\delta(\Xs_n)|$ denote a geometric realization of the Rips
complex built on top of $\Xs_n$ with parameter $\delta$. Moreover, let $\frips{f}:|\Rips_\delta(\Xs_n)|\rightarrow\R$ be the piecewise-linear
interpolation of $f$ on the simplices of $\Rips_\delta(\Xs_n)$, whose 1-skeleton is denoted by $\Rips^1_\delta(\Xs_n)$.
Similarly, let $\frips{\hat f}$ be the piecewise-linear
interpolation of $\hat f$ on the simplices of $\Rips^1_\delta(\Xs_n)$.
As before, since $(|\Rips_\delta(\Xs_n)|,\frips{f})$ and $(|\Rips_\delta(\Xs_n)|,\frips{\hat f})$ are metric spaces, we also consider 
their Mappers $\Map_{r,g}(|\Rips_\delta(\Xs_n)|,\frips{f})$ and $\Map_{r,g}(|\Rips_\delta(\Xs_n)|,\frips{\hat f})$. 
Then, the following inequalities lead to the result:

\begin{align}
\dper & (\Reeb_f(\Xset),\hat\Map_n)\leq \dper(\Reeb_f(\Xset),\Map_n) + \dper(\Map_n,\hat\Map_n)\text{ by the triangle inequality}\nonumber\\
&=\dper(\Reeb_f(\Xset),\Map_n) + \dper(\Map_{r,g}(|\Rips_\delta(\Xs_n)|,\frips{f}),\Map_{r,g}(|\Rips_\delta(\Xs_n)|,\frips{\hat f}))\label{Xcross} \\
&\leq r + 2\omega(\delta) + \dper(\Map_{r,g}(|\Rips_\delta(\Xs_n)|,\frips{f}),\Map_{r,g}(|\Rips_\delta(\Xs_n)|,\frips{\hat f}))\text{ by Theorem~\ref{thm:geomineq}}\nonumber\\
&\leq r + 2\omega(\delta) + r + \|\frips{f}-\frips{\hat f}\|_\infty \text{ by Equation~(\ref{eq:stab})}\nonumber\\
&= 2r + 2\omega(\delta) + \max\{|f(X)-\hat f(X)|\,:\,X\in\Xs_n\}\nonumber
\end{align}

Let us prove Equality~(\ref{Xcross}). By definition of $r$, there are no intersection-crossing edges for both $f$ and $\hat f$.
According to Theorem~\ref{th:Xcross}, $\Map_{r,g}(|\Rips_\delta(\Xs_n)|,\frips{f})$ and $\Map_n$ are isomorphic
and similarly for $\Map_{r,g}(|\Rips_\delta(\Xs_n)|,\frips{\hat f})$ and $\hat \Map_n$. See also the proof of Equality~(\ref{ineq:disc}). \\


\subsection{Proof of Proposition \ref{prop:Measurability} }\label{sec:proofMeasurability}

We check that not only the topological signature of the Mapper but
also the Mapper itself is a measurable object and thus can be seen as an
estimator of a target Reeb graph. This problem is
more complicated than for the statistical framework of persistence
diagram inference, for which the existing stability results give for
free that persistence estimators are measurable for adequate sigma
algebras.

Let $\bar\R=\R\cup\{
+\infty\}$ denote the extended real line.
Given a fixed integer $n\geq 1$, let $\mathcal{C}_{[n]}$ be the set of abstract simplicial complexes
over a fixed set of $n$ vertices. We see $\mathcal{C}_{[n]}$ as a
subset of the power set $2^{2^{[n]}}$, where $[n] = \{1, \cdots,
n\}$, and we implicitly identify $2^{[n]}$ with the set  $[2^n]$ via the map assigning to each subset 
$\{i_1, \cdots, i_k\}$ the integer $1+\sum_{j=1}^k 2^{i_j-1}$. Given a fixed parameter $\delta>0$, 
we define the application
$$
\Phi_{1}:   
 \left\{
\begin{array}{rcl} 
       (\R^D) ^n \times  \R ^n   &       \rightarrow    &    \mathcal{C}_{[n]} \times     \bar \R^{2^{[n]}}     \\[1ex]
 (\Xs_n,\Ys_n) & \mapsto &	(K,f_K)
 \end{array} \right.
$$ where $K$ is the abstract Rips complex of parameter $\delta$ over
the $n$ labeled points in $\R^D$, minus the intersection-crossing edges and their cofaces, 
and where $f_K$ is a function
defined 
by:
\[
f_K :
\left\{
\begin{array}{rcl} 
       2^{[n]}   &       \rightarrow &     \bar\R  \\[1ex]
\sigma  &  \mapsto &
 \left\{
\begin{array}{ll} 
 \max_{i \in \sigma}  \Ys_i  & \textrm{if $\sigma \in K$} \\[1ex]
 +\infty & \textrm{otherwise.}
 \end{array}
  \right.
  \end{array} 
\right.
\]
The space $(\R^D) ^n \times \R ^n$ is
equipped with the standard topology, denoted by $T_1$, inherited from $\R^{(D+1)n}$.
The space $\mathcal{C}_{[n]} \times \bar \R^{2^{[n]}}$ is
equipped with the product, denoted by $T_2$ hereafter, of the discrete topology on
$\mathcal{C}_{[n]}$ and the topology induced by the extended distance
$d(f,g)=\max\{|f(\sigma)-g(\sigma)|\,:\,\sigma\in 2^{[n]},\,f(\sigma){\rm\ or\ }g(\sigma)\neq +\infty\}$ on $\bar \R^{2^{[n]}}$. 
In particular, $K\neq K'\Rightarrow d(f_K,f_{K'})=+\infty$. 

Note that the map $ (\Xs_n, \Ys_n)   \mapsto K$ is piecewise-constant, 
with jumps located at the hypersurfaces defined by $\| X_i - X_j \|^2 = \delta^2$  
(for combinatorial changes in the Rips complex) or 
$Y_i=\mathrm{cst} \in \End((r,g))$ 
(for changes in the set of intersection-crossing edges) in 
$(\R^D) ^n \times \R^n$, where $\End((r,g))$ denotes the set of endpoints of elements of the gomic $(r,g)$. We can then define a finite measurable partition 
$(\mathcal C_\ell)_{\ell \in L} $ of $(\R^D) ^n\times \R^n$ whose boundaries are included in   
these hypersurfaces, and such that $(\Xs_n, \Ys_n)   \mapsto K$ is constant over each set $\mathcal C_\ell$.  
As a byproduct, we have that $(\Xs_n, \Ys_n)\mapsto f$ is  continuous over each set $\mathcal C_\ell$. 

We now define the operator
$$ 
\Phi_{2} :
\left\{
\begin{array}{rcl} 
     \mathcal{C}_{[n]}  \times   \bar\R^{2^{[n]}}   &   \rightarrow & \mathcal{A} \quad        \\[0.5em]
 (K,f)  & \mapsto &	(|K|,\frips f) 
 \end{array} 
\right.
 $$ 
where $\mathcal{A}$ denotes the class of topological spaces filtered
by Morse-type functions, and where $\frips{f}$ is the
piecewise-linear interpolation of $f$ on the geometric
realization $|K|$ of $K$.
For a fixed simplicial complex $K$, the extended persistence diagram of the lower-star filtration induced by~$f$
and of the sublevel sets of~$\frips f$ are identical---see e.g.~\cite{Morozov08}, therefore the map $\Phi_{2}$ is
distance-preserving (hence continuous) in the pseudometrics
$\distb$ on the domain and codomain. Since the topology~$T_2$ on
$\mathcal{C}_{[n]} \times \bar\R^{2^n}$ is a
refinement\footnote{This is because singletons are open balls in the
  discrete topology, and also because of the stability theorem for persistence diagrams~\cite{Chazal16a,Cohen07}
} of
the topology induced by $\distb$, the map $\Phi_2$ is also
continuous when $\mathcal{C}_{[n]} \times \bar\R^{2^{[n]}}$ is
equipped with $T_2$.

Let now $\Phi_3: \mathcal{A}\to \mathcal R$ map each Morse-type pair
$(\Xset,f)$ to its Mapper $\Map_{f}(\Xset,\I)$, where $\I=(r,g)$
is the gomic induced by $r$ and $g$. 
Note that, similarly to $\Phi_1$, the map $\Phi_3$ is piecewise-constant, 
since combinatorial changes in $\Map_{f}(\Xset,\I)$ 
are located at the regions $\Crit(f)\cap {\rm End}(\I)\neq\emptyset$. 
Hence, $\Phi_3$ is measurable in the pseudometric~$\distb$.
For more details on $\Phi_3$, we refer the reader to Definition 7.6 in~\cite{Carriere17b}.
 
Moreover, $\Map_{\frips f}(|K|,\I)$ is isomorphic to 
$\Map_{r,g,\delta}(\Xs_n, \Ys_n)$ 
by Theorem~\ref{th:Xcross} 
since all intersection-crossing edges were removed in the construction of $K$.
Hence, the map $\Phi$ defined by $\Phi=\Phi_3\circ\Phi_2\circ\Phi_1$ is a measurable map 
that sends $(\Xs_n,\Ys_n)$ to $\Map_{r,g,\delta}(\Xs_n,\Ys_n)$.

\subsection{Proof of Proposition \ref{prop:UpBdRestr} }\label{sec:proofdeter}


We fix some parameters $a >0$ and $b \geq 1$.  
First note that Assumption~\eqref{CdtB} is always satisfied by definition of $r_n$. 
Next, there exists  $n_0 \in \N$ such that for any $n \geq n_0$,  Assumption~\eqref{CdtA}  
is satisfied because $\delta_n\rightarrow 0$ and $\omega(\delta_n)\rightarrow 0$ as $n\rightarrow +\infty$.
Moreover, $n_0$ can be taken the same for all $f \in  \bigcup_{\p \in  \mathcal P(a,b)} \mathcal F(\p,\omega)$.

Let $\varepsilon_n = \disth(\Xset,\Xs_n)$.  
Under the $(a,\dimension)$-standard assumption, it is well known that (see for instance \cite{Cuevas04, Chazal15c}):
\begin{equation}\label{eq:cuevas}
\proba{ \varepsilon_n\geq u } \leq \min\left\{1, \frac{4^\dimension}{a u^\dimension}{\rm e}^{-a\left(\frac{u}{2}\right)^\dimension n}\right\}, \forall u > 0.
\end{equation}
In particular, regarding the complementary of~\eqref{CdtD} we have:
\begin{equation} \label{contrCondD}
\proba{ {\varepsilon_n > \frac{\delta_n}{4}} }\leq \min\left\{1,\frac{2^\dimension}{2{\log}(n)n}\right\}.
\end{equation}
Recall that ${\rm diam}(\mathcal{X}_\p)\leq L$.
Let $\bar C = \omega(L)$ be a constant that only depends on the parameters of the model. 
Then, for any $\p \in \mathcal P(a,b)$, we have:
\begin{equation} \label{bornetotpers}
\sup_{f \in \mathcal F(\p,\omega)} \dper \left(\Reeb_f(\Xset_\p),\Map_n\right) \leq \bar C .
\end{equation}
For $n \geq n_0$, we have :
\begin{equation*}
 \sup_{f \in \mathcal F(\p,\omega)}  \dper\left(\Reeb_f(\Xset_\p),\Map_n\right)   = 
 \sup_{f \in \mathcal F(\p,\omega)} \dper\left( \Reeb_f(\Xset_\p),\Map_n   \right)  \: \mathbb I_{\varepsilon_n  >  \delta_n / 4 }  
+\sup_{f \in \mathcal F(\p,\omega)} \dper\left( \Reeb_f(\Xset_\p),\Map_n  \right)  \: \mathbb I _{\varepsilon_n  \leq \delta_n / 4 }  
 \end{equation*}
and thus 
\begin{eqnarray}
\E \left[  \sup_{f \in \mathcal F(\p,\omega)}  \dper\left(\Reeb_f(\Xset_\p),\Map_n\right)   \right] & \leq & 
 \bar C \proba{ \varepsilon_n > \frac{\delta_n}{4} }  +  r_n + 2 \omega(\delta_n)   \notag \\
& \leq &  \bar C  \min\left\{1,\frac{2^\dimension}{2{\log}(n)n}\right\}  +  \left(\frac{1+2g}{g}\right)\omega(\delta_n)   \label{majdinf}
\end{eqnarray}
where we have used \eqref{bornetotpers}, Theorem~\ref{thm:geomineq} and the fact that $V_n(\delta_n)^+$ can be chosen less or equal to $\omega(\delta_n)$. 
For $n$ large enough, the first term in \eqref{majdinf} is of the order of $ \delta_n^b $,  which can be upper bounded by  $ \delta_n$ and thus by $\omega(\delta_n) $ (up to a constant) since $\omega(\delta) / \delta$ 
is non-increasing.
Since $\frac{1+2g}{g} < 6$ because $\frac13 < g < \frac 12$, we get that the risk is bounded by $ \omega(   \delta_n)   $ for $n\geq n_0$ up to a constant that only depends on the parameters of the model. 
The same inequality is of course valid for any $n$ by taking a larger constant, because $n_0$ itself only depends on the parameters of the model.

\subsection{Proof of Proposition~\ref{prop:lecam}}

The proof follows closely Section B.2 in~\cite{Chazal13c}.
Let $\Xset_0=[0,a^{-1/b}]\subset\R^D$.
Obviously, $\Xset_0$ is a smooth and compact submanifold of $\R^D$. Let $\mathcal U(\Xset_0)$ be the uniform measure on $\Xset_0$. 
Let $\mathcal P_{a,\dimension,\Xset_0}$ denote
the set of $(a,\dimension)$-standard measures whose support is included in $\Xset_0$.
Let $x_0=0\in\Xset_0$ and $\{x_n\}_{n\in\mathbb{N}^*}\in \Xset_0^{\mathbb{N}}$ such that $\|x_n-x_0\|=(an)^{-1/\dimension}$.
Now, let $$f_0:\left\{\begin{array}{ll} \Xset_0 & \rightarrow\R \\ x & \mapsto \omega(\|x-x_0\|) \end{array} \right.$$ 
By definition, we have $f_0\in\mathcal F(\mathcal U(\Xset_0),\omega)$ because $\Dg(\Xset_0,f_0)=\{(0,\omega(a^{-1/b}))\}$
since $f_0$ is increasing by definition of $\omega$.
Finally, given any measure $\p\in\mathcal{P}_{a,\dimension,\Xset_0}$, we let $\theta_0(\p)=\Reeb_{f_0|_{\Xset_\p}}(\Xset_\p)$.
Then, we have:
\begin{align}
\underset{\p\in\mathcal{P}_{a,\dimension}}{\rm sup}\  \mathbb{E}&\left[\sup_{f \in \mathcal F(\p,\omega) }\dper\left(\Reeb_f(\Xset_\p),{\hat \Reeb}_n \right)\right] \nonumber\\
&\geq \underset{\p\in\mathcal{P}_{a,\dimension,\Xset_0}}{\rm sup}\  \mathbb{E}\left[\sup_{f \in \mathcal F(\p,\omega) }\dper\left(\Reeb_f(\Xset_\p),{\hat \Reeb}_n \right)\right] \nonumber\\
&\geq \underset{\p\in\mathcal{P}_{a,\dimension,\Xset_0}}{\rm sup}\  \mathbb{E}\left[\dper\left(\Reeb_{f_0|_{\Xset_\p}}(\Xset_\p),{\hat \Reeb}_n \right)\right]
= \underset{\p\in\mathcal{P}_{a,\dimension,\Xset_0}}{\rm sup}\  \mathbb{E}\left[\rho\left(\theta_0(\p),{\hat \Reeb}_n \right)\right], \nonumber
\end{align}
where $\rho=\dper$. For any $n\in\mathbb{N}^*$, we let $\p_{0,n}=\delta_{x_0}$ be the Dirac measure on $x_0$ and 
$\p_{1,n}=(1 - \frac 1 n) \p_{0,n} + \frac 1 n \mathcal{U}([x_0, x_n])$.
As a Dirac measure, $\p_{0,n}$ is obviously in $\mathcal P_{a,\dimension,\Xset_0}$.
We now check that $\p_{1,n}\in\mathcal P_{a,\dimension,\Xset_0}$.
\begin{itemize}
\item Let us study $\p_{1,n}(B(x_0,r))$. 

Assume $r\leq (an)^{-1/b}$.
Then $$\p_{1,n}(B(x_0,r))=1-\frac 1 n + \frac 1 n \frac{r}{(an)^{-1/b}}\geq \left(1 -\frac 1 n + \frac 1 n\right)\left(\frac{r}{(an)^{-1/b}}\right)^b
\geq\left(\frac 1 2+\frac 1 n\right)anr^b\geq ar^b.$$

Assume $r > (an)^{-1/b}$. Then

$$\p_{1,n}(B(x_0,r))=1\geq \min\{ar^b\}.$$

\item Let us study $\p_{1,n}(B(x_n,r))$. 
Assume $r\leq (an)^{-1/b}$.
Then $$\p_{1,n}(B(x_n,r))=\frac 1 n \frac{r}{(an)^{-1/b}}\geq \frac 1 n \left(\frac{r}{(an)^{-1/b}}\right)^b=ar^b.$$

Assume $r > (an)^{-1/b}$. Then

$$\p_{1,n}(B(x_n,r))=1\geq \min\{ar^b\}.$$

\item Let us study $\p_{1,n}(B(x,r))$, where $x\in(x_0,x_n)$. 
Assume $r\leq x$. Then
$$\p_{1,n}(B(x,r))\geq \frac 1 n\frac{r}{(ab)^{-1/b}}\geq ar^b\text{ (see previous case)}.$$

Assume $r > x$. Then 
$\p_{1,n}(B(x,r))=1-\frac 1 n + \frac 1 n \frac{\left(x+\min\{r,\ (an)^{-1/b}-x\}\right)}{(an)^{-1/b}}$. If 
$\min\{r,\ (an)^{-1/b}-x\}=r$, then we have 
$$\p_{1,n}(B(x,r))\geq 1-\frac 1 n + \frac 1 n\frac{r}{(ab)^{-1/b}}\geq ar^b\text{ (see previous case)}.$$
Otherwise, we have
$$\p_{1,n}(B(x,r))=1\geq \min\{ar^b\}.$$
\end{itemize}

Thus $\p_{1,n}$ is in $\mathcal P_{a,\dimension,\Xset_0}$ as well.
Hence, we apply Le Cam's Lemma (see Section~\ref{sec:Lecam}) to get:
$$\underset{\p\in\mathcal{P}_{a,\dimension,\Xset_0}}{\rm sup}\  \mathbb{E}\left[\rho\left(\theta_0(\p),{\hat \Reeb}_n \right)\right]
\geq \frac 1 8 \rho(\theta_0(\p_{0,n}),\theta_0(\p_{1,n}))\left[1-{\rm TV}(\p_{0,n},\p_{1,n})\right]^{2n}.$$

By definition, we have:
$$\rho(\theta_0(\p_{0,n}),\theta_0(\p_{1,n}))=\dper\left(\Reeb_{f_0|_{\{x_0\}}}(\{x_0\}), \Reeb_{f_0|_{[x_0,x_n]}}(\mathcal{U}[x_0,x_n])\right).$$
Since $\Dg\left(\Reeb_{f_0|_{\{x_0\}}}(\{x_0\})\right)=\{(0,0)\}$ and $\Dg\left(\Reeb_{f_0|_{[x_0,x_n]}}(\mathcal{U}[x_0,x_n])\right)=\{(f(x_0),f(x_n))\}$
because $f_0$ is increasing by definition of $\omega$, it follows that 
$$\rho(\theta_0(\p_{0,n}),\theta_0(\p_{1,n}))=\frac 1 2 |f(x_n)-f(x_0)|=\frac 1 2 \omega\left((an)^{-1/b}\right).$$
It remains to compute ${\rm TV}(\p_{0,n},\p_{1,n})=\left|1-\left(1-\frac 1 n\right)\right|+\frac 1 n (an)^{-1/b}=\frac 1 n + o\left(\frac 1 n\right)$.
The Proposition follows then from the fact that $\left[1-{\rm TV}(\p_{0,n},\p_{1,n})\right]^{2n}\rightarrow {\rm e}^{-2}$.

\subsection{Proof of Proposition~\ref{prop:subs}}
Let $\p \in \mathcal P_{a,b}$ and $\omega$ a modulus of continuity for $f$. 
Using the same notation as in the previous section, we have
\begin{align}
\proba{\delta_n \geq u}& 
\leq\proba{\disth(\Xs_n,\Xset_\p)\geq\frac{u}{2}}+\proba{\disth(\Xs^{s_n}_{n},\Xset_\p)\geq\frac{u}{2}} \nonumber \\
& \leq\proba{\varepsilon_n\geq\frac{u}{2}}+\proba{\varepsilon_{s_n}\geq\frac{u}{2}}. \label{eq:subs}
\end{align}
Note that for any  $f \in \mathcal F(\p,\omega)$, according to \eqref{ApproxBound} and \eqref{bornetotpers}
\begin{equation} 
\label{DecompOmegan}
\dper\left(\Reeb_f(\Xset_\p),\Map_n\right)   \leq      \left[r + 2 \omega(\delta) \right] {\mathbb I}_{\Omega_n} + \bar C  \: {\mathbb I}_{\Omega_n^c}
\end{equation}
where $\Omega_n$ is the event defined by
$$
\Omega_n = 
 \left\{  4 \delta_n \leq \min \{  \kappa , \rho \}   \right\} 
  \cap\{ 4\varepsilon_{n}  \leq \delta_n \} .
$$
This gives
\begin{align}
\mathbb{E} \left[ \sup_{f \in \mathcal F(P,\omega)} \dper\left(\Map_n,\Reeb_f(\Xset)\right)\right]&\leq
\underbrace{\int_0^{\bar C}\proba{\omega(\delta_n)\geq\frac{g}{1+2g}\alpha}{\rm d}\alpha}_{(A)}
+\underbrace{\bar C\proba{\varepsilon_n\geq\frac{\delta_n}{4}}}_{(B)}\nonumber\\
&+
\underbrace{\bar C\proba{\delta_n\geq\min\left\{\frac{\kappa}{4},\frac{\rho}{4}\right\}}}_{(D)}.\nonumber
\end{align}

Let us bound the three terms $(A)$, $(B)$ and $(C)$. 
\begin{itemize}

\item {\bf Term $(C)$}.
It can be bounded using~(\ref{eq:subs}) then~(\ref{eq:cuevas}).

\item {\bf Term $(B)$}.
Let $t_n=2\left(\frac{2{\log}(n)}{a n}\right)^{1/\dimension}$ and $A_n=\{\varepsilon_n < t_n\}$.
We first prove that $\delta_n\geq 4\varepsilon_n$ on the event $A_n$, for $n$ large enough.
We follow the lines of the proof of Theorem~3 in Section~6 in~\cite{Fasy14}.  

Let $q_n$ be the $t_n$-{\em packing number} of $\Xset_\p$, i.e. the maximal number of Euclidean balls $B(X,t_n)\cap \Xset_\p$, where $X\in \Xset_\p$,
that can be packed into $\Xset_\p$ without overlap. It is known (see for instance Lemma 17 in~\cite{Fasy14}) that $q_n=\Theta(t_n^{-d})$, where $d$ is the (intrinsic) dimension of $\Xset_\p$.
Let ${\rm Pack}_n=\{c_1,\cdots,c_{q_n}\}$ be a corresponding packing set, i.e. the set of centers of a family of balls of radius $t_n$
whose cardinality achieves the packing number $q_n$. 
Note that $\disth({\rm Pack}_n,\Xset_\p) \leq 2t_n$. Indeed, for any $X\in \Xset_\p$, there must exist $c\in{\rm Pack}_n$ such that $\|X-c\|\leq 2t_n$, otherwise
$X$ could be added to ${\rm Pack}_n$, contradicting the fact that ${\rm Pack}_n$ is maximal.
By contradiction, assume $\epsilon_n < t_n$ and $\delta_n\leq 4\epsilon_n$. Then:
\begin{align*}
\disth(\Xs_n^{s_n},{\rm Pack}_n) & \leq \disth(\Xs_n^{s_n},\Xs_n) + \disth(\Xs_n,\Xset_\p) + \disth(\Xset_\p,{\rm Pack}_n) \\
&  \leq 5\disth(\Xs_n,\Xset_\p) + 2t_n \leq 7t_n.
\end{align*}

Now, one has $\frac{s_n}{q_n}=\Theta\left(\frac{n^{1-b/d}}{\log(n)^{1-b/d+\beta}}\right)$. Since $b\geq D\geq d$ by definition, it follows that $s_n=o(q_n)$.
In particular, this means that $\disth(\Xs_n^{s_n},{\rm Pack}_n)> 7t_n$ for $n$ large enough, which yields a contradiction. 

Hence, one has $\delta_n\geq 4\varepsilon_n$ on the event $A_n$. 
Thus, one has:
$$\proba{\varepsilon_n\geq\frac{\delta_n}{4}}\leq \underbrace{\proba{\varepsilon_n\geq\frac{\delta_n}{4}\ |\ A_n}}_{=0}\proba{A_n} + \proba{A_n^c} = \proba{A_n^c}.$$
Finally, the probability of $A_n^c$ is bounded with~(\ref{eq:cuevas}): 
$$\proba{A_n^c}\leq\frac{2^{\dimension}}{2{\log}(n)n}.$$ 

\item {\bf Term $(A)$}.
This is the dominating term. First, note that since $\omega$ is increasing, one has for all $u > 0$:
\begin{equation}\label{eq:omega}\proba{\omega(\delta_n)\geq u}=\proba{\delta_n\geq \omega^{-1}(u)}.\end{equation}

Then, using~(\ref{eq:subs}) and~(\ref{eq:omega}), we have:
$$(A)\leq \int_0^{\bar C} \proba{\varepsilon_n\geq\frac{1}{2}\omega^{-1}\left(\frac{g\alpha}{1+2g}\right)}{\rm d}\alpha
+ \int_0^{\bar C} \proba{\varepsilon_{s_n}\geq\frac{1}{2}\omega^{-1}\left(\frac{g\alpha}{1+2g}\right)}{\rm d}\alpha.$$
We only bound the first integral, but the analysis extends verbatim to the second integral when replacing $n$ by $s_n$.
Let $$\alpha_n=\frac{1+2g}{g}\omega\left[\left(\frac{4^\dimension{\log}(n)}{an}\right)^{1/\dimension}\right].$$
Since $x\mapsto\frac{\omega(x)}{x}$ is non-increasing, it follows that $x\mapsto\frac{\omega^{-1}(x)}{x}$ is non-decreasing, and 
\begin{equation}\label{eq:incrdecr}
\omega^{-1}(x)\geq\frac{x}{y}\omega^{-1}(y),\ \forall x\geq y >0.
\end{equation}

Taking inspiration from Section~B.2 in~\cite{Chazal13c} and using~(\ref{eq:cuevas}), we have the following inequalities:
\begin{align}
\int_0^{\bar C} &\proba{\varepsilon_n\geq\frac{1}{2}\omega^{-1}\left(\frac{g\alpha}{1+2g}\right)}{\rm d}\alpha
\leq \alpha_n + \frac{8^\dimension}{a}\int_{\alpha_n}^{\bar C}\frac{1}{\omega^{-1}\left(\frac{g\alpha}{1+2g}\right)^{\dimension}}
{\rm exp}\left[-\frac{an}{4^\dimension}\omega^{-1}\left(\frac{g\alpha}{1+2g}\right)^{\dimension}\right]{\rm d}\alpha \nonumber \\
&\leq \alpha_n + \frac{8^\dimension}{a}\int_{\alpha_n}^{\bar C}\frac{\alpha_n^\dimension}{\left[\alpha\omega^{-1}\left(\frac{g\alpha_n}{1+2g}\right)\right]^{\dimension}}
{\rm exp}\left[-\frac{an\alpha^\dimension}{(4\alpha_n)^\dimension}\omega^{-1}\left(\frac{g\alpha_n}{1+2g}\right)^{\dimension}\right]{\rm d}\alpha \nonumber \\
&\leq\alpha_n + \alpha_n\frac{2^\dimension 4n^{1-1/\dimension}}{\dimension a^{1/\dimension}  \omega^{-1}\left(\frac{g\alpha_n}{1+2g}\right)}
\int_{u\geq \frac{an}{4^\dimension}\omega^{-1}\left(\frac{g\alpha_n}{1+2g}\right)^\dimension} u^{1/\dimension-2}{\rm e}^{-u}{\rm d}u \nonumber \\
&= \alpha_n + \alpha_n\frac{ 2^\dimension n }{ \dimension{\log}(n)^{1/\dimension}}
\int_{u\geq {\log}(n)} u^{1/\dimension-2}{\rm e}^{-u}{\rm d}u \leq\left(1+\frac{2^\dimension}{\dimension\log(n)^2}\right)\alpha_n\text{ since }\dimension\geq 1\nonumber \\
&\leq C(\dimension)\alpha_n, \nonumber
\end{align}

where we used \eqref{eq:incrdecr} with $x = \frac{g\alpha}{1+2g}$ and $y=\frac{g\alpha_n}{1+2g}$ for the second inequality.
The constant $C(\dimension)$  only  depends on  $\dimension$.

Hence, since $\frac{1+2g}{g} < 6$, there exist constants $K,K'>0$ that depend only of the geometric parameters of the model such that:
$$(A)\leq  K\omega\left(\frac{K'{\log}(s_n)}{s_n}\right)^{1/\dimension}.$$

\end{itemize}

{\bf Final bound}.
Since $s_n = n{\log}(n)^{-(1+\beta)}$, by gathering all four terms, 
there exist constants $C,C'>0$ such that:
$$\mathbb{E}\left[\sup_{f \in \mathcal F(\p,\omega)} \dper\left(\Reeb_f(\Xset_\p),\Map_n\right)\right]\leq C\omega\left(\frac{C'{\log}(n)^{2+\beta}}{n}\right)^{1/\dimension}.$$

\subsection{Proof of Proposition~\ref{prop:conf1}}

We have the following bound by using \eqref{DecompOmegan} in the proof of Proposition~\ref{prop:subs}:
\begin{align}
&\proba{\dper(,\Reeb_f(\Xset_\p),\Map_n)\geq\eta}\nonumber \\
&\leq\proba{\omega(\delta_n)\geq \frac{g}{1+2g}\eta}+\proba{\varepsilon_n\geq\frac{\delta_n}{4}} 
+\proba{\delta_n\geq\min\left\{\frac{\kappa}{4},\frac{\rho}{4}\right\}} \nonumber \\
&\leq\proba{\varepsilon_n\geq \frac{1}{2}\omega^{-1}\left(\frac{g}{1+2g}\eta\right)}+
\proba{\varepsilon_{s_n}\geq \frac{1}{2}\omega^{-1}\left(\frac{g}{1+2g}\eta\right)} + o\left(\frac{1}{n{\log}(n)}\right). \nonumber
\end{align}
Following the lines of Section~6 in~\cite{Fasy14}, subsampling approximations give
$$\proba{\varepsilon_n\geq \frac{1}{2}\omega^{-1}\left(\frac{g}{1+2g}\eta\right)}\leq L_n\left(\frac{1}{4}\omega^{-1}\left(\frac{g}{1+2g}\eta\right)\right)+o\left(\frac{s_n}{n}\right)^{1/4},$$
and
$$\proba{\varepsilon_{s_n}\geq \frac{1}{2}\omega^{-1}\left(\frac{g}{1+2g}\eta\right)}\leq F_n\left(\frac{1}{4}\omega^{-1}\left(\frac{g}{1+2g}\eta\right)\right)+o\left(\frac{s^2_n}{s_n}\right)^{1/4}.$$
The result follows by taking  $s_n=n{\log}(n)^{-(1+\beta)}$.

\section{Le Cam's Lemma} \label{sec:Lecam}

The  version of Le Cam's Lemma given below is from \cite{Yu97} \citep[see also][]{Genovese12b}. 
Recall that  the total variation distance between two distributions $\p_0$ and $\p_1$ on a measured space $(\mathcal X, \mathcal B) $ 
is defined by $$ \TV(\p_0,\p_1)  = \sup_{B \in  \mathcal B} | \p_0(B) - \p_1(B) |.$$
Moreover, if $\p_0$ and $\p_1$ have densities $p_0$ and $p_1$
 for the same measure $\lambda$ on $\mathcal X$, then 
$$ \TV(\p_0,\p_1)    = \frac 1 2 \ell_1(p_0,p_1) = \int_{\mathcal X} |p_0-p_1|  d \lambda. $$

\begin{lem} \label{Lem:Lecam} Let $\mathcal P $ be a set of distributions. For $\p \in  \mathcal P$, let $\theta(\p)$ take values in a pseudometric space $(\X,\rho)$. 
Let $\p_0$ and $\p_1$ in $\mathcal P$ be any pair of distributions. Let $X_1,\dots,X_n$ be drawn i.i.d. from some $\p \in  \mathcal P$. Let $\hat 
\theta = \hat \theta(X_1,\dots,X_n) $ be any estimator of $\theta(\p)$, then
\begin{equation*}
 \sup_{ \p \in \mathcal P} \E _{\p^n} \left[ \rho( \theta , \hat \theta  )\right]  \geq   \frac 1 8    \rho \left( \theta(\p_0), \theta(\p_1) \right)   
\left[1 -  \TV(\p_0,\p_1) \right] ^{2 n } .
\end{equation*}
\end{lem}

\section{Extended Persistence}\label{sec:ExtPers}

Let $f$ be a real-valued function on a topological space $X$.
The family $\{X^{(-\infty, \alpha]}\}_{\alpha\in\R}$ of
  sublevel sets of $f$ defines a {\em filtration}, that is, it is
  nested w.r.t. inclusion: $X^{(-\infty, \alpha]}\subseteq
X^{(-\infty, \beta]}$ for all $\alpha\leq\beta\in\R$. The
  family $\{X^{[\alpha, +\infty)}\}_{\alpha\in\R}$ of superlevel sets
  of $f$ is also
  nested but in the opposite direction: $X^{[\alpha, +\infty)}\supseteq X^{[\beta, +\infty)}$
  for all $\alpha\leq\beta\in\R$. We can turn it into a filtration by
  reversing the real line. Specifically, let $\Rop=\{\tilde{x}\ |\ x\in\R\}$,
  ordered by $\tilde{x}\leq\tilde{y}\Leftrightarrow x\geq y$. We index
  the family of superlevel sets by $\Rop$, so now we have a
  filtration: $\{X^{[\tilde\alpha, +\infty)}\}_{\tilde\alpha\in\Rop}$, with
  $X^{[\tilde\alpha, +\infty)}\subseteq X^{[\tilde\beta, +\infty)}$ for all
  $\tilde\alpha\leq\tilde\beta\in\Rop$.

\begin{figure}[htb]
\begin{center}
\includegraphics[width=13cm]{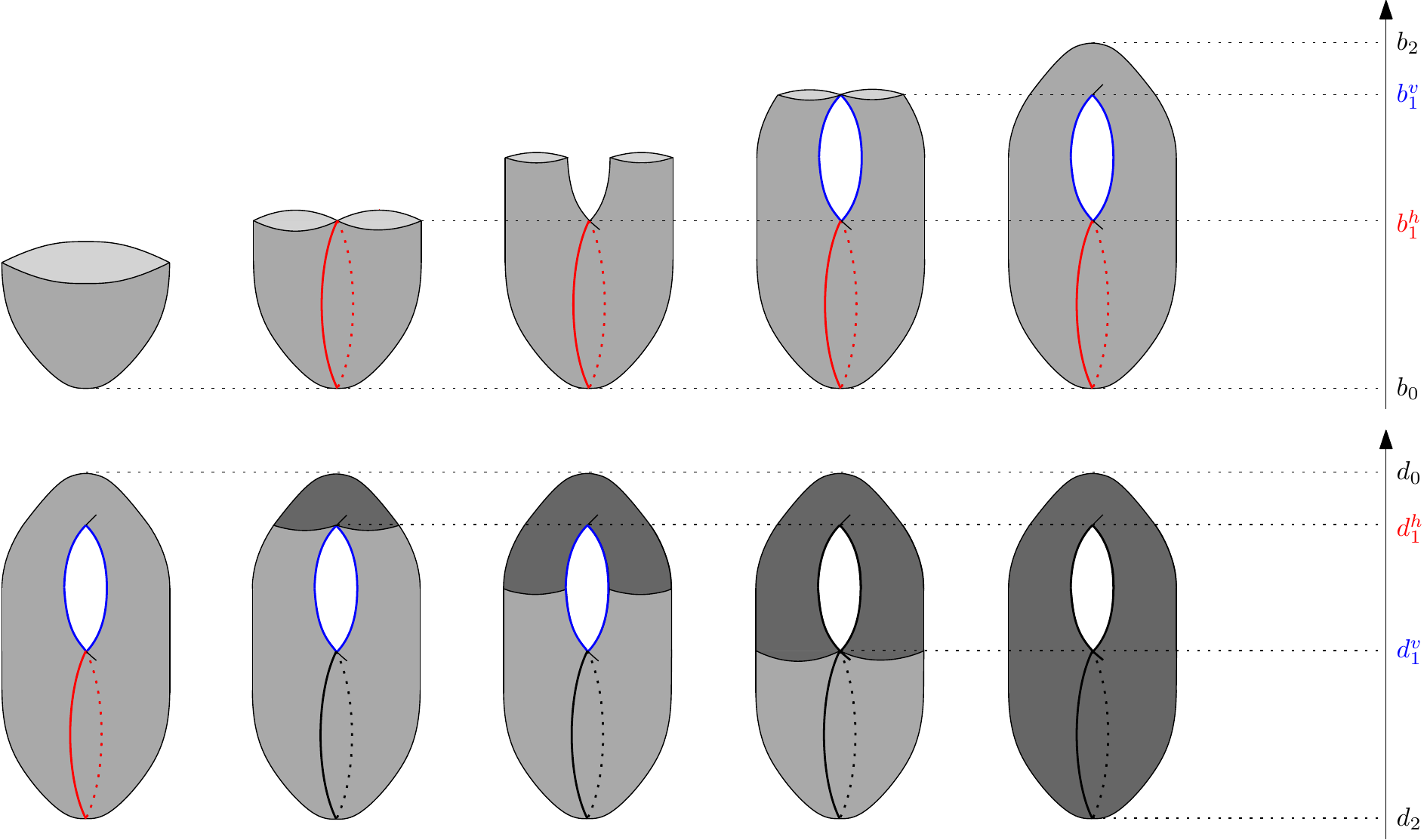}
\caption{\label{fig:ExFilt} The extended filtration of the
  height function on a torus.  The upper row displays the ordinary
  part of the filtration while the lower row displays the relative part.  The red and blue cycles both correspond to
  extended points in dimension 1.  The point corresponding to
  the red cycle is located above the diagonal ($d_1^h>b_1^h$), while
  the point corresponding to the blue cycle is located below the
  diagonal ($d_1^v>b_1^v$).}
\end{center}
\end{figure}

Extended persistence connects the two filtrations at infinity as
follows. Replace each superlevel set $X^{[\tilde\alpha, +\infty)}$ by
  the pair of spaces $(X, X^{[\tilde\alpha, +\infty)})$ in the second
    filtration. This maintains the filtration property since we have
    $(X, X^{[\tilde\alpha, +\infty)})\subseteq (X, X^{[\tilde\beta,
          +\infty)})$ for all
        $\tilde\alpha\leq\tilde\beta\in\Rop$. Then, let
        $\Rext=\R\cup\{+\infty\}\cup\Rop$, where the order is
        completed by $\alpha<+\infty<\tilde{\beta}$ for all
        $\alpha\in\R$ and $\tilde\beta\in\Rop$. This poset is
        isomorphic to $(\R, \leq)$. Finally, define the {\em extended
          filtration} of $f$ over $\Rext$ by:
\[
\begin{array}{llll}
F_\alpha &=& X^{(-\infty, \alpha]} & \mbox{for $\alpha\in\R$}\\[0.5em]
F_{+\infty} &=& X \equiv (X,\emptyset)\\[0.5em]
F_{\tilde\alpha} &=& (X, X^{[\tilde\alpha, +\infty)}) & \mbox{for $\tilde\alpha\in\Rop$},
\end{array}
\]
where we have identified the space $X$ with the pair of spaces $(X,
\emptyset)$. This is a well-defined filtration since we have
$X^{(-\infty, \alpha]}\subseteq X \equiv (X, \emptyset) \subseteq (X,
X^{[\tilde\beta, +\infty)})$ for all $\alpha\in\R$ and
  $\tilde\beta\in\Rop$.  The subfamily $\{F_\alpha\}_{\alpha\in\R}$ is
  called the \textit{ordinary} part of the filtration, and the
  subfamily $\{F_{\tilde\alpha}\}_{\tilde\alpha\in\Rop}$ is called the
  \textit{relative} part. See Figure~\ref{fig:ExFilt} for an
  illustration.

Applying the homology functor $H_*$ to this filtration gives the so-called \textit{extended persistence module} $\mathbb{V}$ of~$f$:
\[
\begin{array}{llll}
V_\alpha &=& H_*(F_\alpha)=H_*(X^{(-\infty, \alpha]}) & \mbox{for $\alpha\in\R$}\\[0.5em]
V_{+\infty} &=& H_*(F_{+\infty})=H_*(X) \cong H_*(X,\emptyset)\\[0.5em]
V_{\tilde\alpha} &=& H_*(F_{\tilde\alpha})=H_*(X, X^{[\tilde\alpha, +\infty)}) & \mbox{for $\tilde\alpha\in\Rop$},
\end{array}
\]
and where the linear maps between the spaces are induced by the
inclusions in the extended filtration.

For Morse-type functions, the extended persistence module can be
decomposed as a finite direct sum of half-open {\em interval
  modules}---see e.g.~\cite{Chazal16a}:
\[
\mathbb{V}\simeq\bigoplus_{k=1}^n \mathbb{I}[b_k, d_k),
\]
where each summand $\mathbb{I}[b_k, d_k)$ is made of copies of the
  field of coefficients at each index $\alpha\in [b_k, d_k)$, and of
    copies of the zero space elsewhere, the maps between copies of the
    field being identities.  Each summand represents the lifespan of a
    {\em homological feature} (cc, hole, void, etc.) within the
    filtration. More precisely, the {\em birth time} $b_k$ and {\em
      death time} $d_k$ of the feature are given by the endpoints of
    the interval.  Then, a convenient way to represent the structure
    of the module is to plot each interval in the decomposition as a
    point in the extended plane, whose coordinates are given by the
    endpoints. Such a plot is called the \textit{extended persistence
      diagram} of $f$, denoted $\Dg(f)$.  The distinction
    between ordinary and relative parts of the filtration allows to
    classify the points in $\Dg(f)$ in the following way:
\begin{itemize}
\item points whose coordinates both belong to $\R$  are called \textit{ordinary} points; 
they correspond to homological features being born and then dying in the ordinary part of the filtration;
\item points whose coordinates both belong to $\Rop$ are called \textit{relative} points; 
they correspond to homological features being born and then dying in the relative part of the filtration;
\item points whose abscissa belongs to $\R$  and whose ordinate belongs to $\Rop$ are called \textit{extended} points; 
they correspond to homological features being born in the ordinary part and then dying in the relative part of the filtration. 
\end{itemize}
Note that ordinary points lie strictly above the diagonal
$\Delta=\{(x,x)\ |\ x\in\R\}$ and relative points lie strictly below $\Delta$,
while extended points can be located anywhere, including on $\Delta$, e.g. cc 
that lie inside a single critical level.  It is common to
decompose $\Dg(f)$ according to this classification:
\[
\Dg(f)=\Ord(f)\sqcup\Rel(f)\sqcup\Ext^+(f)\sqcup\Ext^-(f),
\]
where by
convention $\Ext^+(f)$ includes the extended points located on the
diagonal~$\Delta$.

\bibliography{MapperStat_biblio}

\begin{thebibliography}{}

\bibitem[Biau and Mas, 2012]{Biau12}
Biau, G. and Mas, A. (2012).
\newblock {PCA-Kernel estimation}.
\newblock {\em Statistics and Risk Modeling with Applications in Finance and
  Insurance}, 29(1):19--46.

\bibitem[Blanchard et~al., 2007]{Blanchard07}
Blanchard, G., Bousquet, O., and Zwald, L. (2007).
\newblock {Statistical properties of kernel principal component analysis}.
\newblock {\em Machine Learning}, 66(2-3):259--294.

\bibitem[Buchet et~al., 2015]{Buchet15}
Buchet, M., Chazal, F., Oudot, S., and Sheehy, D. (2015).
\newblock {Efficient and Robust Persistent Homology for Measures}.
\newblock In {\em Proceedings of the 26th Symposium on Discrete Algorithms},
  pages 168--180.

\bibitem[Carri\`ere and Oudot, 2017a]{Carriere17a}
Carri\`ere, M. and Oudot, S. (2017a).
\newblock {Local Equivalence and Induced Metrics for Reeb Graphs}.
\newblock In {\em Proceedings of the 33rd Symposium on Computational Geometry}.

\bibitem[Carri\`ere and Oudot, 2017b]{Carriere17b}
Carri\`ere, M. and Oudot, S. (2017b).
\newblock {Structure and Stability of the 1-Dimensional Mapper}.
\newblock {\em Foundations of Computational Mathematics}.

\bibitem[Chazal et~al., 2011]{Chazal11}
Chazal, F., Cohen-Steiner, D., and M\'erigot, Q. (2011).
\newblock {Geometric Inference for Probability Measures}.
\newblock {\em Foundations of Computational Mathematics}, 11(6):733--751.

\bibitem[Chazal et~al., 2016a]{Chazal16a}
Chazal, F., de~Silva, V., Glisse, M., and Oudot, S. (2016a).
\newblock {\em {The Structure and Stability of Persistence Modules}}.
\newblock Springer.

\bibitem[Chazal et~al., 2014]{Chazal14b}
Chazal, F., Fasy, B., Lecci, F., Michel, B., Rinaldo, A., and Wasserman, L.
  (2014).
\newblock {Robust topological inference: distance to a measure and kernel
  distance}.
\newblock {\em CoRR}, abs/1412.7197.
\newblock Accepted for publication in Journal of Machine Learning Research.

\bibitem[Chazal et~al., 2015a]{Chazal15a}
Chazal, F., Fasy, B., Lecci, F., Michel, B., Rinaldo, A., and Wasserman, L.
  (2015a).
\newblock {Subsampling Methods for Persistent Homology}.
\newblock In {\em Proceedings of the 32nd International Conference on Machine
  Learning}, pages 2143--2151.

\bibitem[Chazal et~al., 2013]{Chazal13c}
Chazal, F., Glisse, M., Labru{\`e}re, C., and Michel, B. (2013).
\newblock {Optimal rates of convergence for persistence diagrams in Topological
  Data Analysis}.
\newblock {\em CoRR}, abs/1305.6239.

\bibitem[Chazal et~al., 2015b]{Chazal15c}
Chazal, F., Glisse, M., Labru{{\`e}}re, C., and Michel, B. (2015b).
\newblock Convergence rates for persistence diagram estimation in topological
  data analysis.
\newblock {\em Journal of Machine Learning Research}, 16:3603--3635.

\bibitem[Chazal et~al., 2016b]{Chazal16b}
Chazal, F., Massart, P., and Michel, B. (2016b).
\newblock Rates of convergence for robust geometric inference.
\newblock {\em Electronic Journal of Statistics}, 10(2):2243--2286.

\bibitem[Chen et~al., 2009]{Chen09}
Chen, X., Golovinskiy, A., and Funkhouser, T. (2009).
\newblock {A Benchmark for 3D Mesh Segmentation}.
\newblock {\em ACM Transactions on Graphics}, 28(3):1--12.

\bibitem[Cohen-Steiner et~al., 2007]{Cohen07}
Cohen-Steiner, D., Edelsbrunner, H., and Harer, J. (2007).
\newblock {Stability of Persistence Diagrams}.
\newblock {\em Discrete and Computational Geometry}, 37(1):103--120.

\bibitem[Cohen-Steiner et~al., 2009]{Cohen09}
Cohen-Steiner, D., Edelsbrunner, H., and Harer, J. (2009).
\newblock {Extending persistence using Poincar{\'e} and Lefschetz duality}.
\newblock {\em Foundation of Computational Mathematics}, 9(1):79--103.

\bibitem[Cuevas, 2009]{Cuevas09}
Cuevas, A. (2009).
\newblock {Set estimation: another bridge between statistics and geometry}.
\newblock {\em Bolet\'in de Estad\'istica e Investigaci\'on Operativa},
  25(2):71--85.

\bibitem[Cuevas and Rodr{\'\i}guez-Casal, 2004]{Cuevas04}
Cuevas, A. and Rodr{\'\i}guez-Casal, A. (2004).
\newblock {On boundary estimation}.
\newblock {\em Advances in Applied Probability}, pages 340--354.

\bibitem[DeVore and Lorentz, 1993]{DeVore93}
DeVore, R. and Lorentz, G. (1993).
\newblock {\em Constructive approximation}, volume 303.
\newblock Springer Science \& Business Media.

\bibitem[Dey and Wang, 2013]{Dey13a}
Dey, T. and Wang, Y. (2013).
\newblock {Reeb Graphs: Approximation and Persistence}.
\newblock {\em Discrete and Computational Geometry}, 49(1):46--73.

\bibitem[Fasy et~al., 2014]{Fasy14}
Fasy, B., Lecci, F., Rinaldo, A., Wasserman, L., Balakrishnan, S., and Singh,
  A. (2014).
\newblock {Confidence Sets for Persistence Diagrams}.
\newblock {\em The Annals of Statistics}, 42(6):2301--2339.

\bibitem[Friedman et~al., 2001]{Friedman01}
Friedman, J., Hastie, T., and Tibshirani, R. (2001).
\newblock {\em The Elements of Statistical Learning}.
\newblock Springer series in statistics Springer, Berlin.

\bibitem[Genovese et~al., 2012a]{Genovese12a}
Genovese, C., Perone-Pacifico, M., Verdinelli, I., and Wasserman, L. (2012a).
\newblock {Manifold estimation and singular deconvolution under Hausdorff
  loss}.
\newblock {\em The Annals of Statistics}, 40:941--963.

\bibitem[Genovese et~al., 2012b]{Genovese12b}
Genovese, C., Perone-Pacifico, M., Verdinelli, I., and Wasserman, L. (2012b).
\newblock {Minimax Manifold Estimation}.
\newblock {\em Journal of Machine Learning Research}, 13:1263--1291.

\bibitem[Hinks et~al., 2015]{Hinks15}
Hinks, T., Zhou, X., Staples, K., Dimitrov, B., Manta, A., Petrossian, T., Lum,
  P., Smith, C., Ward, J., Howarth, P., Walls, A., Gadola, S., and Djukanovic,
  R. (2015).
\newblock Innate and adaptive t cells in asthmatic patients: Relationship to
  severity and disease mechanisms.
\newblock {\em Journal of Allergy and Clinical Immunology}, 136(2):323--333.

\bibitem[Lum et~al., 2013]{Lum13}
Lum, P., Singh, G., Lehman, A., Ishkanov, T., Vejdemo-Johansson, M., Alagappan,
  M., Carlsson, J., and Carlsson, G. (2013).
\newblock {Extracting insights from the shape of complex data using topology}.
\newblock {\em Scientific Reports}, 3.

\bibitem[Morozov, 2008]{Morozov08}
Morozov, D. (2008).
\newblock {\em Homological Illusions of Persistence and Stability}.
\newblock {Ph.D.} dissertation, Department of Computer Science, Duke
  University.

\bibitem[Nene et~al., 1996]{Nene96}
Nene, S., Nayar, S., and Murase, H. (1996).
\newblock {Columbia Object Image Library (COIL-100)}.
\newblock Technical Report CUCS-006-96.

\bibitem[Nielson et~al., 2015]{Nielson15}
Nielson, J., Paquette, J., Liu, A., Guandique, C., Tovar, A., Inoue, T.,
  Irvine, K.-A., Gensel, J., Kloke, J., Petrossian, T., Lum, P., Carlsson, G.,
  Manley, G., Young, W., Beattie, M., Bresnahan, J., and Ferguson, A. (2015).
\newblock Topological data analysis for discovery in preclinical spinal cord
  injury and traumatic brain injury.
\newblock {\em Nature Communications}, 6.

\bibitem[Reaven and Miller, 1979]{Reaven79}
Reaven, G. and Miller, R. (1979).
\newblock An attempt to define the nature of chemical diabetes using a
  multidimensional analysis.
\newblock {\em Diabetologia}, 16(1):17--24.

\bibitem[Rucco et~al., 2015]{Rucco15}
Rucco, M., Merelli, E., Herman, D., Ramanan, D., Petrossian, T., Falsetti, L.,
  Nitti, C., and Salvi, A. (2015).
\newblock Using topological data analysis for diagnosis pulmonary embolism.
\newblock {\em Journal of Theoretical and Applied Computer Science},
  9(1):41--55.

\bibitem[Sarikonda et~al., 2014]{Sarikonda14}
Sarikonda, G., Pettus, J., Phatak, S., Sachithanantham, S., Miller, J., Wesley,
  J., Cadag, E., Chae, J., Ganesan, L., Mallios, R., Edelman, S., Peters, B.,
  and von Herrath, M. (2014).
\newblock Cd8 t-cell reactivity to islet antigens is unique to type 1 while cd4
  t-cell reactivity exists in both type 1 and type 2 diabetes.
\newblock {\em Journal of Autoimmunity}, 50:77--82.

\bibitem[Shawe-Taylor et~al., 2005]{Shawe05}
Shawe-Taylor, J., Williams, C., Cristianini, N., and Kandola, J. (2005).
\newblock {On the eigenspectrum of the Gram matrix and the generalization error
  of kernel-PCA}.
\newblock {\em IEEE Transactions on Information Theory}, 51(7):2510--2522.

\bibitem[Singh et~al., 2007]{Singh07}
Singh, G., M\'emoli, F., and Carlsson, G. (2007).
\newblock {Topological Methods for the Analysis of High Dimensional Data Sets
  and 3D Object Recognition}.
\newblock In {\em Symposium on Point Based Graphics}, pages 91--100.

\bibitem[{The GUDHI Project}, 2015]{gudhi}
{The GUDHI Project} (2015).
\newblock {\em {GUDHI} User and Reference Manual}.
\newblock {GUDHI Editorial Board}.

\bibitem[Yao et~al., 2009]{Yao09}
Yao, Y., Sun, J., Huang, X., Bowman, G., Singh, G., Lesnick, M., Guibas, L.,
  Pande, V., and Carlsson, G. (2009).
\newblock {Topological methods for exploring low-density states in biomolecular
  folding pathways}.
\newblock {\em Journal of Chemical Physics}, 130(14).

\bibitem[Yu, 1997]{Yu97}
Yu, B. (1997).
\newblock Assouad, {F}ano, and {L}e {C}am.
\newblock In {\em Festschrift for {L}ucien {L}e {C}am}, pages 423--435.
  Springer.

\end{thebibliography}
\bibliographystyle{apalike}

\end{document}